\documentclass[letterpaper, 12pt]{article}
\pdfoutput=1
\usepackage{amsfonts,amsthm,amsmath,amssymb,slashed}
\usepackage[textwidth = 500 pt, textheight = 630 pt]{geometry}
\usepackage{latexsym,amsfonts,amsmath,amssymb,theorem,dsfont,wasysym}
\usepackage{tabularx}
\usepackage{xcolor}
\usepackage{enumerate}
\usepackage{framed}
\usepackage{mdframed}
\usepackage{mathrsfs}
\usepackage[normalem]{ulem}
\usepackage{graphicx, rotating}
\usepackage{epsfig}
\usepackage{latexsym}
\usepackage{graphicx}
\usepackage{booktabs}
\usepackage{color}
\usepackage{amstext}
\usepackage{verbatim} 
\usepackage{fancyhdr}
\usepackage{fancybox}
\usepackage{pifont}
\usepackage{color}
\usepackage{ulem,bbold}
\usepackage{enumitem}
\usepackage{subfigure}
\usepackage{bbm}
\usepackage{parskip}
\usepackage{cite}
\usepackage{slashed}
\usepackage{hyperref}
\usepackage{amssymb}
\usepackage[T1]{fontenc}
\usepackage{epstopdf}
\usepackage[title]{appendix}
\usepackage[font=footnotesize,labelfont=]{caption}
\AppendGraphicsExtensions{.tif}
\definecolor{MidnightBlue}{RGB}{25, 25, 112}
\hypersetup{colorlinks=true, citecolor=MidnightBlue,
            linkcolor=MidnightBlue, urlcolor=MidnightBlue, linktocpage=true}

\def\beq{\begin{eqnarray}}
\def\eeq{\end{eqnarray}}

\def\mpl{M_{\rm pl}}

\def\d{{\rm d}}
\newcommand{\pvec}[1]{\vec{#1}\mkern2mu\vphantom{#1}}
\newcommand{\be}{\begin{equation}}
\newcommand{\ee}{\end{equation}}
\newcommand{\la}{\langle}
\newcommand{\ra}{\rangle}
\def\ea{\end{eqnarray}}
\def\ba{\begin{eqnarray}}

\def\beq{\begin{eqnarray}}
\def\eeq{\end{eqnarray}}

\def\mn{_{\mu \nu}}
\def\mpl{M_{\rm Pl}}

\def\la{\langle}
\def\ra{\rangle}

\newcommand{\lp}{\left (}
\newcommand{\rp}{\right )}
\def\BH{\text{\tiny BH}}
\def\DM{\text{\tiny DM}}

\def\lsim{\mathrel{\rlap{\lower3pt\hbox{\hskip0pt$\sim$}}
     \raise1pt\hbox{$<$}}}         
\def\gsim{\mathrel{\rlap{\lower4pt\hbox{\hskip1pt$\sim$}}
     \raise1pt\hbox{$>$}}}         
\def\lsim{\mathrel{\rlap{\lower3pt\hbox{\hskip0pt$\sim$}}
     \raise1pt\hbox{$<$}}}         
\def\gsim{\mathrel{\rlap{\lower4pt\hbox{\hskip1pt$\sim$}}
     \raise1pt\hbox{$>$}}}         

\begin{document}

\renewcommand{\thefootnote}{\fnsymbol{footnote}}

\makeatletter
\@addtoreset{equation}{section}
\makeatother
\renewcommand{\theequation}{\thesection.\arabic{equation}}

\rightline{}
\rightline{}
 

\begin{center}
{\Large \bf{Superfluid Dark Matter}}

 \vspace{1truecm}
\thispagestyle{empty} \centerline{\large  {Lasha  Berezhiani$^{1,2}$\footnote{\href{mailto:}{lashaber@mpp.mpg.de}}, Giordano Cintia$^{3,4}$\footnote{\href{mailto:}{giordano.cintia@univ-amu.fr}}, Valerio De Luca$^5$\footnote{\href{mailto:}{vdeluca@sas.upenn.edu}} and Justin Khoury$^5$\footnote{\href{mailto:}{jkhoury@sas.upenn.edu}}
}}

 \textit{$^1$Max-Planck-Institut f\"ur Physik, Boltzmannstra{\ss}e.~8, 85748 Garching, Germany\\
 \vskip 5pt
$^2$ Arnold Sommerfeld Center, Ludwig-Maximilians-Universit\"at, \\Theresienstra{\ss}e 37, 80333 M\"unchen, Germany\\ 
 \vskip 5pt
~$^3$ Galileo Galilei Institute for Theoretical Physics, Largo Enrico Fermi, 2, 50125 Firenze, Italy\\
 \vskip 5pt
~$^4$ Aix Marseille Universit\'e, Universit\'e de Toulon, CNRS, CPT, Marseille, France \\
 \vskip 5pt
~$^5$Center for Particle Cosmology, Department of Physics and Astronomy, \\ University of Pennsylvania, 209 South 33rd St, Philadelphia, PA 19104, USA 
 }

\end{center}   
\begin{abstract}
\noindent
The superfluid dark matter model offers an elegant solution to reconcile discrepancies between the predictions of the cold dark matter paradigm and observations on galactic scales. In this scenario, dark matter is composed of ultralight bosons with self-interactions that can undergo a superfluid phase transition in galactic environments. In this review, we explore the theoretical foundations of dark matter superfluidity, detailing the conditions required for the formation and stability of superfluid cores of astrophysical size. We examine the phenomenological consequences for galactic dynamics, including the impact on galaxy mergers, the formation of vortices, the behavior near supermassive black holes, modifications to dynamical friction, and the emergence of long-range interactions. By synthesizing theoretical developments with observational constraints, we aim to provide a comprehensive overview of the current status and future prospects of dark matter superfluidity as a viable extension of the standard cosmological model.
\end{abstract}

\newpage
\setcounter{page}{1}

\tableofcontents
\newpage
\renewcommand{\thefootnote}{\arabic{footnote}}
\setcounter{footnote}{0}

\linespread{1.1}
\parskip 4pt

\section{Introduction}

Cosmological and astrophysical data demonstrate that Cold Dark Matter (CDM) accounts for roughly a quarter of the total energy density of the Universe~\cite{Planck:2018vyg}.  Within the $\Lambda$CDM framework, dark matter (DM) is modeled as a pressureless cosmological fluid, leading to remarkable agreement with large-scale structure observations~\cite{Planck:2018nkj} and the temperature anisotropies of the cosmic microwave background~\cite{Planck:2018vyg}. Despite its success on cosmological scales, the microscopic nature of DM remains elusive. Moreover, several persistent anomalies in galactic dynamics lack a natural---or at least well-understood---explanation within the standard paradigm, prompting interest in models with richer microscopic properties that could impact astrophysical behavior. For comprehensive reviews, see Refs.~\cite{Bertone:2004pz, Ferreira:2020fam, Cirelli:2024ssz}.

A simple and intriguing extension of the standard picture, which has attracted much attention in the past decades, involves interactions within the dark sector, or between DM and ordinary matter (baryons), which can leave observable imprints on galactic dynamics. While these models face strong observational constraints, they still allow room for novel and potentially testable astrophysical phenomena. Among them, the idea that DM may form a superfluid phase on galactic scales has gained increasing attention. The original proposal was motivated in part to suppress dynamical friction, seemingly preferred by observations, and the possibility to support quasi-homogeneous, kiloparsec-scale cores at zero temperature~\cite{Goodman:2000tg}. The idea was revitalized ten years ago by including the emergence of a phonon-mediated force between baryons~\cite{Berezhiani:2015bqa, Berezhiani:2015pia}, which arises from the long-range coherence intrinsic to the superfluid phase. This force can offer a tantalizing explanation for the tight empirical correlation between the observed baryon mass distribution and orbital velocities in galaxies.
Since then, the model has undergone significant development, accompanied by a growing body of observational and theoretical constraints, which we will review in our discussion of its current status.

This review provides a comprehensive overview of the superfluid dark matter (SDM) paradigm, highlighting both its theoretical foundations and observational implications. We begin by discussing the conditions required for the superfluid phase transition and the formation of superfluid cores in the central regions of DM halos. We then examine various observational features, including impact on galaxy mergers, the emergence of vortices, the formation of DM spikes around supermassive black holes (BHs), and the suppression of dynamical friction for objects moving within the core. Finally, we explore the idea of phonon-mediated long-range forces between baryons and their phenomenological implications on galactic scales. 

Before turning to the details of the superfluid model itself, we first outline the main challenges faced by the standard CDM paradigm on small (galactic) scales, which motivate the exploration of alternatives such as SDM.

\subsection{Cold dark matter and small scale challenges}

In the standard~$\Lambda$CDM model, the collapse of the primordial fluctuations leads to the formation of virialized halos. These describe overdense DM regions with enclosed mass 
\begin{equation}
{M}_\text{\tiny halo}=\frac{4 \pi}{3} R_\text{\tiny V}^3 \rho_{\rm m} \delta \,,
\label{eq:HaloMass}
\end{equation}
where~$R_\text{\tiny V}$ is the virial radius, which conventionally defines the size of the halo, while~$\delta$ is the overdensity parameter that determines the mean density of the halo relative to the background cosmological density~$\rho_{\rm m}$. The mass identification inherently involves some arbitrariness due to the choice of the parameter~$\delta$. Conventionally, the choice~$\delta \sim 200$ is made, in such a way that the enclosed mass in Eq.~\eqref{eq:HaloMass} matches the theoretical virialized overdensity of an idealized spherical collapse of a DM region~\cite{1998ApJ...495...80B}.

The non-linear nature of the halo's collapse makes it extremely challenging to analyze the dynamics analytically. The problem is usually tackled numerically, where N-body simulations of collisionless DM reveal a nearly power-law density profile in the inner region of halos, of the form~\cite{Navarro:2008kc}
\begin{equation}
    \rho(r)\sim r^{-\gamma} \qquad \text{with} \qquad \gamma= 0.8-1.4\,,
\end{equation}
and a steeper fall-off in the outer regions. The main outcome of simulations is the quite universal behavior of the density distribution of halos across all mass ranges. Different density profiles, such as the Navarro-Frank-White (NFW)  profile~\cite{Navarro:1996gj} or the Einasto profile~\cite{1965TrAlm...5...87E}, have been proposed as empirical formulas to fit the simulated virialized overdensities~\cite{Dutton:2014xda, Moore:1999gc, 1993ApJ...417..450K, Jaffe:1983iv}.

However, numerical simulations have encountered challenges in reproducing astrophysical observations on galactic scales. While many of these small-scale issues are complicated by the uncertainties inherent in modeling baryonic physics and observational systematics, certain discrepancies have emerged over the past years as significant and persistent puzzles for the~$\Lambda$CDM paradigm.

The first discrepancy concerns low surface brightness galaxies and late-type dwarfs, which are dominated by DM throughout their halos, including the central regions. This issue, commonly referred to as the \textit{cusp-core problem}, highlights the mismatch between the central density profiles predicted by simulations and those inferred from observations. Specifically, numerical simulations typically produce halos with dense, cuspy centers, in contrast to the faint, cored density distributions inferred from rotation curve measurements~\cite{Flores:1994gz,Moore:1994yx,RC1,RC2,RC3,RC4,RC5,RC6,RC7,RC8,RC9,RC10,RC11,RC12}.
In comparison, high surface brightness objects, such as spiral galaxies, present additional complexities due to the dominance of baryonic matter at small radii. Even among dwarf galaxies, a surprising diversity in rotation curves is observed~\cite{Oman:2015xda}, with some systems exhibiting cuspy profiles and others showing cored distributions. See Fig.~5 of~\cite{Oman:2015xda}.

Addressing this discrepancy within the CDM framework requires incorporating the effects of baryonic physics. In particular, feedback from collective baryonic processes can significantly modify the density profiles of halos compared to predictions from DM-only simulations. Hydrodynamical simulations have demonstrated the critical role of these effects in shaping the central regions of halos. For instance, energy injection from supernovae~\cite{2013MNRAS.429.3068T, 2014ApJ...789L..17M, Onorbe:2015ija, Read:2015sta} and density fluctuations induced by star formation~\cite{2010Natur.463..203G, 2012MNRAS.422.1231G, 2012MNRAS.424.1275B} can drive the formation of cores, emphasizing the importance of baryonic feedback. These processes operate by ejecting baryonic matter from the central regions, thereby reducing the gravitational potential and allowing the DM halo to expand, resulting in a shallower central density profile. Significant deviations from standard CDM predictions are found in halos with efficient star formation, while halos with little to no baryonic content generally follow the evolution expected from DM-only simulations. The main implication of these effects is the emergence of a correlation between the cumulative stellar mass formed in the halo and the properties of the resulting core~\cite{DiCintio:2013qxa, Burkert:1995yz}. However, explaining the dynamics of low surface brightness galaxies through baryonic feedback alone remains challenging due to their intrinsically low baryonic content.

The overprediction of the number of halos, commonly known as the {\it missing satellites} problem, has long represented a significant challenge for~$\Lambda$CDM. However, this issue has largely been alleviated in recent years. In hierarchical structure formation models, DM halos are expected to host a population of smaller subhalos, with the halo mass function quantifying their abundance within specific mass intervals. This yields a robust prediction within the~$\Lambda$CDM framework: as the mass of subhalos increases, their number decreases monotonically across all observable mass ranges. In particular, simulations predict that Milky Way-like galaxies should host approximately a thousand subhalos with masses exceeding~$10^7 M_\odot$~\cite{Springel:2008cc, 2016ApJ...818...10G}---masses large enough to potentially host visible galaxies. Initially this prediction appeared to be in stark contrast with the relatively small number of satellites observed around the Milky Way. However, beginning in the mid-2000s, improved observational campaigns have led to the discovery of numerous faint satellite galaxies in both the Milky Way and Andromeda, resolving halos with masses as small as approximately~$300 \,M_\odot$~\cite{DES:2015zwj,DES:2016jjg}. The number of detected satellites is expected to continue increasing with future surveys.

Baryonic physics appears to play an important role in softening the discrepancy between the predicted and observed number of subhalos. A key factor is the inefficiency of low-mass subhalos in forming stars, rendering many of them effectively dark and thus undetectable. This idea is supported by semi-analytic models of galaxy formation, which show that supernova feedback and the ultraviolet (UV) background from reionization can strongly suppress star formation in low-mass halos~\cite{Benson:2001at, 2009ApJ...696.2179K, Munoz:2009kp, 2010MNRAS.401.2036L, 2010AIPC.1240..355M, 2011MNRAS.413..101G}. When coupled to the galactic medium, supernova-driven outflows can expel baryonic matter from the inner regions of subhalos, regulating or even quenching subsequent star formation~\cite{1974MNRAS.169..229L, 1986ApJ...303...39D}. Simultaneously, the UV background generated during the epoch of reionization can heat the baryonic gas, preventing it from cooling and collapsing into stars~\cite{Efstathiou:1992zz, Thoul:1996by, Barkana:1999apa, Gnedin:2000uj, Hoeft:2005jn, Okamoto:2008sn}. As a result, the ``missing satellites'' issue can be reinterpreted as a challenge in determining how many subhalos host visible galaxies, rather than an overproduction of substructures. The increasing discovery of faint satellites~\cite{Kim:2017iwr, Perivolaropoulos:2021jda} continues to reduce this discrepancy. Interestingly, recent observations~\cite{Homma:2016fzg,10.1093/pasj/psx050, Homma:2019cmf, Homma:2023ppu, 2024arXiv240505303K} suggest that we may even be approaching a new tension: an apparent ``too many satellites'' problem.

The third shortcoming is known as the \textit{too-big-to-fail} problem~\cite{2012MNRAS.422.1203B, 2016A&A...591A..58P}. It was observed that the most massive satellites of the Milky Way are significantly less dense than the most massive subhalos predicted by DM-only N-body simulations. Based on our earlier discussion, resolving this discrepancy would require that star formation occurs preferentially in lower-mass halos---those within our detection capabilities---while being suppressed in the most massive subhalos, which appears counterintuitive within the standard paradigm. This puzzling situation is illustrated, for example, in Fig.~10 of Ref.~\cite{Bullock:2017xww}.

Several possible resolutions to the too-big-to-fail problem have been investigated. One possibility is that the mass of the Milky Way’s halo has been overestimated, which would imply that its largest satellites are naturally lighter, bringing simulations into better agreement with observations~\cite{Wang:2012sv, Dutton:2015nvy, Cirelli:2024ssz}. Alternatively, scatter in halo masses could mean that the Milky Way simply represents a statistical downward fluctuation~\cite{Wang:2012sv, Dutton:2015nvy, Cirelli:2024ssz}. Moreover, the discrepancy appears less severe in simulations that include baryonic physics, where feedback processes can transform central density cusps into cores, effectively reducing the inferred halo masses---similar to the mechanisms discussed for the cusp-core problem. Recent simulations suggest that the satellites of the Milky Way and Andromeda may no longer exhibit a significant too-big-to-fail issue~\cite{Dutton:2015nvy, Cirelli:2024ssz}, although the problem may persist for isolated dwarf galaxies in the Local Group.

These issues constitute some of the most prominent small-scale challenges to the~$\Lambda$CDM model. For a more detailed discussion, we refer the interested reader to~\cite{Cirelli:2024ssz}.

\subsection{Beyond the cold dark matter paradigm}
\label{Intro-models}

Although baryonic physics is expected to play a critical role in shaping DM structures on small scales, the unresolved issues discussed above motivate the exploration of alternative and complementary approaches. One such approach, which we focus on in this review, suggests that the traditional CDM paradigm may be incomplete when describing dynamics on galactic scales, and that additional dark sector features may be required. For instance, relaxing the assumption that DM behaves as a purely pressureless fluid below kiloparsec scales could significantly impact halo formation, potentially alleviating the small-scale challenges outlined earlier.

One of the simplest modifications to the DM paradigm that addresses small-scale discrepancies is {\it warm} DM~\cite{Colin:2000dn}. In this model, DM particles possess a small but non-negligible mass, typically in the range of \( m \sim \mathrm{keV} \), leading to a residual thermal velocity dispersion that suppresses the growth of density perturbations below a characteristic free-streaming scale. As a result, warm DM behaves similarly to CDM on large cosmological scales while exhibiting distinct dynamics on subgalactic scales. Even with modest velocities, DM particles can  erase primordial fluctuations by free-streaming out of potential wells, suppressing the formation of small-scale structures and mitigating several associated problems~\cite{Bode:2000gq, Viel:2005qj, Lovell:2013ola, Viel:2013fqw, Irsic:2017ixq}.
A well-studied realization of warm DM is a thermal relic fermion that decoupled while still relativistic and later became non-relativistic as the universe expanded. Its free-streaming length, and thus the suppression of small-scale structure, depends primarily on the particle mass and decoupling temperature: a thermal relic with \( m \sim (1 \div 3)~\mathrm{keV} \) produces a cutoff in the matter power spectrum at comoving scales \(\sim (0.1\div1)~\mathrm{Mpc} \). Non-thermal scenarios---such as sterile neutrinos $\nu_s$ produced via the Dodelson--Widrow mechanism~\cite{Dodelson:1993je}---yield similar effects, though with distinct momentum distributions. Current bounds from the Lyman-\(\alpha\) forest constrain the thermal relic mass to \( m \gtrsim (3 \div 5)~\mathrm{keV} \)~\cite{Viel:2013fqw, Irsic:2017ixq, Rogers:2020ltq, Garzilli:2019qki}, while X-ray searches for the radiative decay \( \nu_s \to \nu \gamma \) impose complementary bounds on sterile neutrino models~\cite{Boyarsky:2018tvu}. Despite these stringent limits, warm DM remains an appealing minimal extension of CDM capable of mitigating small-scale structure tensions.

Another model motivated by these discrepancies is {\it self-interacting} DM~\cite{Carlson:1992fn, Spergel:1999mh} in which, unlike warm DM, the astrophysical phenomenology is primarily controlled by the DM scattering cross section per unit mass~$\sigma/m$, rather than the particle mass itself. 
Self-interactions redistribute DM from the inner to the outer regions of halos, softening the central cusp of the NFW profile and suppressing small-scale structure formation~\cite{Peter:2012jh, Rocha:2012jg,Kamada:2016euw, Creasey:2017qxc, Ren:2018jpt}. 
This model can reproduce the observed diversity in rotation curve shapes of dwarf galaxies~\cite{Kamada:2016euw, Creasey:2017qxc, Ren:2018jpt} provided that the required amplitude for self-interactions  is typically $\sigma/m\sim\mathcal{O}(0.1 \div 10)\ \mathrm{cm^2/g}$ on dwarf galactic scales (with relative velocities $v\sim \mathcal{O}(10) \, \mathrm{km/s}$) whereas, to remain consistent with larger cluster-scale probes, the cross section has to be $\sigma/m\lesssim (0.1 \div 1)\ \mathrm{cm^2/g}$ at $v\sim 10^3\ \mathrm{km/s}$,  motivating velocity-dependent interactions~\cite{Randall:2008ppe, Robertson:2016xjh, Harvey:2015hha}. Particle physics realizations that naturally produce this velocity dependence include Yukawa forces mediated by light scalar or vector (dark photon) fields~\cite{Loeb:2010gj, Kaplinghat:2015aga}, giving rise to models  spanning DM masses in the MeV to TeV range.
Numerical simulations of self-interacting DM including baryonic feedback confirm that such models can mitigate small-scale tensions of the CDM paradigm~\cite{Kamada:2016euw, Creasey:2017qxc, Ren:2018jpt, Kaplinghat:2015aga}, while being compatible with existing astrophysical and cosmological constraints~\cite{Tulin:2017ara}.

Further well-motivated class of DM candidates are {\it axions} and axion-like particles (ALPs), which arise as pseudo Nambu-Goldstone bosons associated with the spontaneous breaking of a global $U(1)$ symmetry~\cite{Peccei:1977hh, Weinberg:1977ma, Wilczek:1977pj}. The QCD axion was originally proposed to solve the strong CP problem, but it can also constitute all or part of the DM in the universe. ALPs generalize this idea to a broader class of light pseudoscalars, which need not solve the strong CP problem but can share similar properties, such as extremely weak couplings to Standard Model (SM) particles and long lifetimes~\cite{Jaeckel:2010ni, Marsh:2015xka}. Both are typically very light, with masses ranging from $m_a \sim 10^{-12}\ \mathrm{eV}$ to $10^{-2}\ \mathrm{eV}$, depending on the symmetry-breaking scale $f_a$. In the simplest scenario, they are produced non-thermally via the vacuum misalignment mechanism~\cite{Preskill:1982cy, Abbott:1982af, Dine:1982ah}, in which the initial field displacement from the minimum of its potential leads to coherent oscillations that behave as CDM. For QCD axions, the observed abundance is achieved for $f_a \sim (10^{11} \div 10^{12})\ \mathrm{GeV}$, corresponding to $m_a \sim 10^{-5}\ \mathrm{eV}$~\cite{Wantz:2009it, Borsanyi:2016ksw}, whereas ALPs can span a broader range of masses and couplings, potentially allowing for heavier sub-eV candidates~\cite{Hui:2016ltb, Marsh:2015xka}. These candidates are characterized by a rich phenomenology, which spans from ultralight quantum-suppressed structures on galactic scales to cold and coherent fields that behave like conventional CDM, potentially accessible through experimental, astrophysical, and cosmological observables~\cite{Marsh:2015xka, Hui:2016ltb}. Further constraints on axion DM models will be discussed later in the Review.

Yet another scenario that has gained significant attention in recent years is {\it fuzzy} DM, due to its rich phenomenology on small scales~\cite{Hu:2000ke, Hui:2016ltb}. In this model, DM consists of ultralight bosons with masses lighter than~$10^{-20} \, {\rm eV}$. The extremely small mass invalidates a classical fluid description and allows DM to form a condensate on galactic scales through the balance between quantum pressure and gravity. This wave-like behavior gives rise to several distinctive phenomena, including the suppression of substructures below the de Broglie wavelength and the formation of cores in the inner regions of halos, with characteristic sizes set by quantum pressure and gravity~\cite{May:2021wwp, May:2022gus, Liu:2022rss}. On cosmological scales, however, fuzzy DM behaves similarly to CDM, preserving the large-scale successes of the standard model.
Fuzzy DM can be naturally produced via non-thermal mechanisms in the early universe, giving rise to Bose-Einstein condensates with extremely long coherence lengths, leading to macroscopic wave interference effects in galactic halos. These quantum features lead to characteristic observational signatures, including granular density fluctuations, soliton-halo transitions, and modifications to the halo mass function at dwarf-galaxy scales.
Observational constraints on fuzzy DM come from multiple probes. Lyman-$\alpha$ forest observations place a lower bound of~$m>2\times 10^{-20}$~eV at 95\% confidence level~\cite{Irsic:2017yje,Armengaud:2017nkf,Kobayashi:2017jcf,Rogers:2020ltq}. Additional bounds arise from dwarf galaxy counts, stellar streams, and rotation curve measurements, which constrain the soliton core sizes and halo substructure~\cite{Marsh:2018zyw, Safarzadeh:2019sre, Dalal:2022rmp, Bar:2018acw}. We defer the reader to recent reviews on fuzzy and Bose-Einstein condensate DM models for a discussion of further constraints on this scenario~\cite{Suarez:2013iw,Urena-Lopez:2019kud, Ferreira:2020fam,Hui:2021tkt,Eberhardt:2025caq,Chavanis:2025qcg}.

\begin{figure} 
\centering
\includegraphics[scale=0.4]{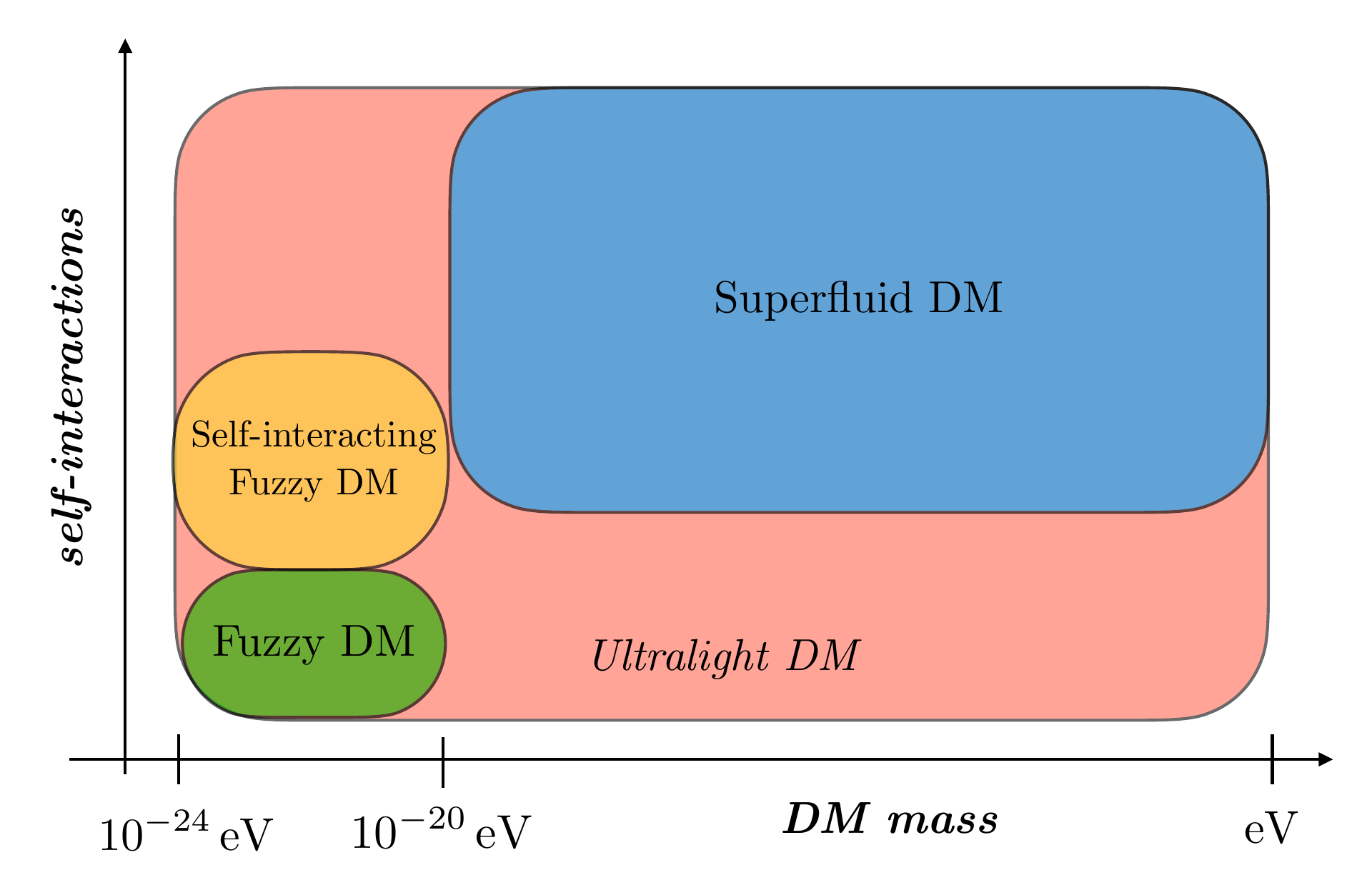}
\caption{Schematic illustration of ultralight DM models in the mass/self-interactions plane. This review focuses on the Superfluid DM model.}
	\label{figDM}
\end{figure}

The extremely light mass range considered in fuzzy DM models requires that gravitational collapse be counterbalanced by quantum pressure. However, one can envision alternatives involving heavier DM candidates, extending up to the~eV mass range. A prime example is the superfluid DM model, which postulates sub-eV DM particles with self-interactions capable of undergoing a superfluid phase transition in galactic halos.\footnote{The formation of superfluids in galaxies can also be achieved through fermions  with attractive interactions via Bardeen-Cooper-Schrieffer pairing. Such models typically involve a phase transition from a relativistic fermion fluid in the early Universe to a non-relativistic condensate at late times, resulting in a mildly evolving effective equation of state and characteristic astrophysical signatures, including core-like halo structures. The presence of a fermionic condensate can relax the Tremaine–Gunn lower bound \cite{Tremaine:1979we} on the DM particle mass and can be observationally distinguished from ultralight bosonic DM through phenomena analogous to superconductivity, such as Andreev reflection and superconducting vortices. For developments in this direction, see Refs.~\cite{Alexander:2016glq,Alexander:2018fjp,Alexander:2020wpm, Carena:2021bqm, Tong:2023krn, Garani:2022quc, Liang:2024xww, Alexander:2024qml}.} This framework allows for the formation of coherent superfluid cores, stabilized by the balance between self-interactions and gravity~\cite{Goodman:2000tg, Berezhiani:2015bqa, Berezhiani:2015pia}. It offers a distinct approach to addressing the small-scale challenges mentioned earlier and admits, in principle, microscopic realizations.\footnote{Another ultralight DM model, named self-interacting fuzzy DM, shares the same mass range as standard fuzzy DM but incorporates weak self-interactions~\cite{Gavrilik:2020mjy, Mocz:2023adf, Painter:2024rnc, Capanelli:2025nrj}. The main difference with SDM lies in the coherence length of the system. Although both frameworks involve self-interacting ultralight particles, self-interactions in the fuzzy case are not efficient enough to bring the system to equilibrium. As a result, the coherence length remains approximately set by quantum pressure, as in the non-interacting fuzzy DM scenario, in contrast to SDM, where self-interactions govern the macroscopic coherence properties.} For earlier reviews on the SDM scenario, see Refs.~\cite{Suarez:2013iw,Ferreira:2020fam}. A schematic summary of the DM models discussed is shown in Fig.~\ref{figDM}.

In this review, we focus on the SDM scenario, summarizing its distinctive phenomenology on galactic scales and its potential to address small-scale challenges. 
Section~\ref{Sec: Superfluidity} introduces the phenomenon of superfluidity, outlining its theoretical foundations within the frameworks of effective field theory and hydrodynamics. We also discuss finite-temperature effects, the interaction of the superfluid with external perturbations, and the role of gravity.
Section~\ref{Sec: SDMgalaxies} focuses on DM superfluidity in galactic environments, with particular attention to the formation of superfluid halos and their observable phenomenology.
Section~\ref{sec: cosmo}
 explores its cosmological aspects, including possible production mechanisms and the subsequent evolution of the condensate. 
Section~\ref{sec: pheno} provides an overview of the phenomenological implications of the SDM framework in both astrophysical and cosmological contexts, with particular emphasis on the dynamics of merging galaxy clusters, galaxy rotation curves and galactic relations, portals to the SM, vortex formation within halos and the development of DM spikes around BHs, while
Section~\ref{sec:DynamicalFr} examines the phenomenon of dynamical friction and its suppression in the presence of a superfluid, in contrast to the expectations from standard CDM. We present the theoretical foundations of this effect using both a quasiparticle description and a hydrodynamical approach. An application to the evolution of BH binaries is also discussed.
Section~\ref{sec:Forces} explores the possibility of long-range forces arising between baryons immersed in the DM superfluid, by coupling them to the constituent degrees of freedom. We present two illustrative examples capable of generating such interactions.
Finally, Section~\ref{sec:conclusions} summarizes the main results of the review and provides a brief comparison between this model and other DM candidates.

Throughout the paper we adopt natural units,~$\hbar = c = k_{\rm B} =1$, and mostly positive metric signature.

\section{Superfluidity and its effective description}
\label{Sec: Superfluidity}

The phenomenon of superfluidity refers to a set of remarkable hydrodynamic properties exhibited by certain fluids when cooled below a critical temperature. One of its hallmark features is frictionless flow, first observed in 1937 by Kapitza~\cite{Kapitza1938}, and independently
by Allen and Misener~\cite{Allen1938}, in liquid helium. They observed that, when cooled
below the~$\lambda$-point, 2.17~K, the frictionless flow allows liquid helium to move seamlessly through narrow
capillaries without energy dissipation.

As originally derived by Landau~\cite{Landau1941}, what distinguishes a superfluid from other substances is the fact that the only energetically accessible, low-energy degrees of freedom of the system are~\textit{phonons}. These are sound waves satisfying the dispersion relation
\begin{equation}
    \omega_k= c_s k+\dots\,; \qquad c^2_s=\frac{\partial P}{\partial \rho}\,,
    \label{eq:dispersionLandau}
\end{equation}
where~$c_s$ is the sound speed of the superfluid, expressed in terms of its pressure~$P$ and density~$\rho$.\footnote{As usual,~$\omega_k$ and~$k$ denote
the energy and momentum of the wave, respectively.}

The frictionless behavior of a superfluid is summarized in Landau's criterion of superfluidity. Consider an external perturbation ({\it e.g.}, a body) moving through the fluid. This perturbation can  transfer energy to the medium by exciting phonons with dispersion relation $\omega_k$. Landau's criterion states that, as long as the relative velocity between the superfluid and the perturber is smaller than the superfluid sound speed, then no dissipation can take place. To illuminate this, denote the relative velocity by~$\vec{v}_s$. In the rest frame of the perturber, the fluid carries an energy
\begin{equation}
E_\text{\tiny fluid}=E_\text{\tiny kin}+\omega_k+\vec{k}\cdot \vec{v}_s\,,
\label{eq:EnergyFluid}
\end{equation}
where~$E_\text{\tiny kin}$ is the kinetic energy of the superfluid.\footnote{In the rest frame of the unperturbed fluid, the total energy is determined solely by the perturbation energy~$\omega_k$. When transforming back to the laboratory frame via a Galilean boost, the energy acquires a contribution from the kinetic energy~$E_\text{\tiny kin}$ of the moving fluid, one from the intrinsic energy $\omega_k$ of the perturbations in the comoving frame, and a Doppler term, $\vec{k}\!\cdot\!\vec{v}_s$, arising from boosting the perturbation wave packet to the laboratory frame.} For the superfluid to be slowed down by the passage of the perturber, a certain amount of kinetic energy has to be damped by excited phonons. This is possible if the sum of the second and third terms of Eq.~\eqref{eq:EnergyFluid} is negative. By plugging in the phonon spectrum in Eq.~\eqref{eq:dispersionLandau}, we have dissipation if~$c_s k+\vec{k}\cdot \vec{v}_s\,<0$, which requires
\begin{equation}
    v_s>c_s\,.
    \label{eq:LandauCr}
\end{equation}
As a result, subsonic inertial motion is dissipationless, in contrast to normal fluids where the sonic barrier serves as an additional source of friction on top of ordinary dissipation channels. 

Although phonons can explain the superfluid phenomenology, their fundamental origin comes from the microscopic properties of the system. In particular, superfluidity is 
deeply connected to the phenomenon of Bose-Einstein condensation. From the effective field theory (EFT) point of view, a superfluid can be described as a finite density configuration of bosons, which spontaneously breaks the~$U(1)$ global symmetry of the system responsible for particle number conservation.\footnote{Although superfluidity seems to occur only in systems of globally charged bosons, non-relativistic superfluids can also form from real bosonic fields, as number-changing processes are suppressed in this regime. We will derive this result later  by adopting the Gross-Pitaevskii formalism.}
In the simplest realization, the phonon field emerges as the Nambu-Goldstone  boson of the spontaneously broken internal symmetry, and the underlying~$U(1)$ invariance is now realized non-linearly by the phonon as a shift symmetry. The rest of this Section is devoted to reviewing the EFT description of superfluidity.

\subsection{Simplest example: superfluidity in a quartic theory}
\label{sec:Phi4}
As a starting point, it is useful to study the low-energy description of the simplest superfluid, described by a condensate of bosons with two-body contact interactions. Consider a complex scalar field~$\Phi$ with quartic potential
\begin{equation} 
    \mathcal{L}=-\partial_\mu \Phi^* \partial^\mu \Phi-m^2|\Phi|^2-\frac{\lambda_4}{2}|\Phi|^4\,.
    \label{eq:LagrPhi4}
\end{equation}
The quartic coupling~$\lambda_4$ is of course assumed positive, to ensure stability. This corresponds to \textit{repulsive self-interactions}, which is fundamental for viable superfluid phenomenology.

The classical field solution describing a homogeneous configuration is given by  
\begin{equation}
\Phi_0(t)=v\,{\rm e}^{{\rm i}\mu t}\,; \qquad \mu=\sqrt{m^2+\lambda_4 v^2}\,.
\label{sect2:fieldconf}
\end{equation}
Here,~$v$ describes the vacuum expectation value of the field, while~$\mu$ is the relativistic chemical potential of the system, which is chosen in such a way that the field configuration \eqref{sect2:fieldconf} satisfies the equation of motion. In the non-relativistic limit, generically obtained by assuming $\lambda_4 v^2 \ll m^2$~\cite{Berezhiani:2020umi}, we can expand~$\mu$ to obtain\footnote{This is because, upon restoring the speed of light $c$, the chemical potential satisfies $\mu^2=m^2c^4+\lambda_4 v^2 c^2$. Expanding this expression in the non-relativistic (large-$c$) limit yields Eq.~\eqref{muRel}.}
\begin{equation}
    \mu
    \simeq m+\frac{\lambda_4 v^2}{2m}= m+\mu_\text{\tiny NR}\,,
    \label{muRel}
\end{equation}
where in the last step we have identified the non-relativistic chemical potential~$\mu_\text{\tiny NR} = \frac{\lambda_4 v^2}{2m}$. We stress that the order parameter~$v$ of the solution is directly related to the number density of zero-momentum particles comprising the condensate. In particular, if we evaluate the~$U(1)$ charge for the configuration~\eqref{sect2:fieldconf}, we find the number density
\begin{equation}
\label{numberdensity}
n\simeq 2m v^2\,,
\end{equation}
where we have taken the non-relativistic limit.

The field configuration~\eqref{sect2:fieldconf} spontaneously breaks the global~$U(1)$ symmetry of the theory, and phonons emerge as the Nambu-Goldstone bosons for this breaking. To derive their spectrum, one should therefore perturb the homogeneous solution as
\begin{equation}
\Phi(\vec{x},t)=\big(v+h(\vec{x},t)\big){\rm e}^{{\rm i}\left(\mu t+\pi(\vec{x},t)\right)}\,.
\label{perturbedSol}
\end{equation}
Substituting in the original Lagrangian~\eqref{eq:LagrPhi4}, we formulate the theory in terms of the radial~$h(\vec{x},t)$ and angular~$\pi(\vec{x},t)$ fields:
\begin{equation}
\mathcal{L}=\dot{h}^2-\big(\vec{\nabla} h\big)^2+\left(\dot{\pi}^2-\left(\vec{\nabla} \pi\right)^2+2\mu \dot{\pi} -\frac{\lambda_4}{2}\left(v+h\right)^2+\lambda_4 v^2\right)\left(v+h\right)^2\,.
\label{eq:Lagrpert}
\end{equation}
In contrast with the standard spontaneous symmetry breaking of the vacuum, here the radial and  angular modes are kinetically mixed by the relativistic chemical potential~$\mu$. The main implication of this mixing is that canonically normalized degrees of freedom are a mixture of the angular and radial modes, and have therefore a quite cumbersome general expression~\cite{Creminelli:2023kze,Hui:2023pxc}. 

Several approximations can be considered to simplify the analysis. First, since we aim to study non-relativistic superfluids, we can drop the~$\dot{h}^2$ and~$\dot{\pi}^2$ terms.\footnote{This assumption is justified by the fact that those contributions are suppressed by an explicit factor of $c^{-2}$. The kinetic mixing term, however, is not suppressed, $2\mu c^{-2} \dot{\pi} h\sim 2m \dot{\pi}h$, and therefore survives in the non-relativistic limit.} This implies that the typical energy scales of interest are smaller than the mass~$m$ of the scalar. Second, we can consider a leading-order approximation in spatial derivatives of~$h$ by dropping the gradient term~$\big(\vec{\nabla} h\big)^2$. This is possible if we are interested in energy scales smaller than~$\mu_\text{\tiny NR}$. In the context of DM superfluidity, the shortcomings of this approximation were discussed for dissipative processes~\cite{Berezhiani:2020umi} and in the analysis of the gravitational stability of the superfluid phase~\cite{Berezhiani:2021rjs}.

With these approximations, the Lagrangian~\eqref{eq:Lagrpert} becomes an algebraic polynomial function of the radial field~$h$, which can be integrated out exactly by solving its equation of motion. The final Lagrangian, written in terms of the angular field~$\pi$, reads
\begin{equation}
    \mathcal{L}_\pi=\frac{1}{2\lambda_4} \left(2 m \dot{\pi}-\big(\vec{\nabla} \pi\big)^2+\lambda_4 v^2\right)^2\,.
    \label{eq:LagrNoncanonical}
\end{equation}
Notice that we have approximated the relativistic chemical potential with the mass of a constituent particle,~$\mu\simeq m$, as in Eq.~\eqref{muRel}, consistent with the non-relativistic approximation. Equation~\eqref{eq:LagrNoncanonical} provides the low energy theory of long-wavelength fluctuations of a homogeneous condensate, for a quartic theory. 

If we expand the Lagrangian and keep up to quadratic terms in~$\pi$, the angular mode reveals itself as the phonon field:
\begin{equation}
\label{Lp}
    \mathcal{L}_\pi=\frac{1}{2}\dot{\pi}^2-\frac{1}{2}c_s^2\big(\vec{\nabla} \pi\big)^2+\dots\,; ~~~\qquad c_s^2=\frac{\lambda_4 v^2}{2m^2}\,,
\end{equation}
where we have canonically normalized the phonon field and identified the sound speed of the superfluid.  As predicted, the low energy theory is described by a gapless sound mode, which in turn leads us to Landau's criterion for superfluidity.

Lastly, we comment on the requirement of repulsive self-interactions for superfluidity. If the theory had negative quartic coupling,~$\lambda_4 < 0$, the long-wavelength fluctuations of the homogeneous solution would have an imaginary sound speed (see Eq.~\eqref{Lp}) and thus would be unstable. One may therefore consider the simple quartic potential with repulsive self-interactions as in~\eqref{eq:LagrPhi4}. However, such realizations are typically sensitive to sizable radiative corrections, which can destabilize the tiny mass required for superfluidity on astrophysical scales.

This radiative instability can be avoided if the potential is protected by an almost exact symmetry.
A well-known way to stabilize it is to endow the scalar field with a shift symmetry. This naturally arises if the scalar is the Nambu-Goldstone boson associated with the spontaneous breaking of a global symmetry. The shift symmetry restricts the scalar to appear only through derivative interactions. Consequently, a small explicit breaking of the global symmetry is required to introduce non-derivative terms, generating a mass and self-interactions. In this case, all such couplings are proportional to the parameter controlling the explicit symmetry breaking, the \textit{spurion}, which also ensures the radiative stability of the potential as in~\eqref{eq:LagrPhi4}.

However,  generating a repulsive potential through this mechanism is  non-trivial. For instance, in the well-known case of the QCD axion, where the shift symmetry is broken by instanton effects, the resulting potential is a periodic cosine. Expanding it in powers of the scalar field, the quartic term carries an attractive coupling~\cite{Vafa:1983tf} (see also Sec.~\ref{sec: axions} for further details on the QCD axion):
\begin{equation}
V(a)=f_a^2 m_a^2 \left[1-\cos\left(\frac{a}{f_a}\right)\right]\sim\frac{m_a^2}{2}a^2-\frac{m_a^2}{4!f_a^2}a^4+\ldots
\end{equation}
Here, $f_a$ is the decay constant of the axion, while $m_a$ is the spurion.
As a result, such particles cannot produce long-range correlations due to their attractive self-interactions~\cite{Guth:2014hsa}. More generally, pseudo-Nambu-Goldstone bosons arising from the breaking of a compact group typically have periodic trigonometric potentials, and any single cosine or sine term inevitably leads to attractive interactions.

The challenges of constructing pseudo-Nambu-Goldstone boson models with repulsive self-interactions are discussed in~\cite{Fan:2016rda}, where various scenarios are analyzed. The author shows that, in most cases, repulsive self-interactions cannot be achieved. On the other hand, they also show a working example in which the repulsive interactions arise from an axion originating as the remnant of a five-dimensional $U(1)$ gauge symmetry. 

These limiting considerations, however, arise from naturalness arguments for the model parameters, analogous to the so-called hierarchy problem for the Higgs boson mass. Although this is an interesting issue worthy of further investigation, at the level of theoretical consistency the construction of repulsively self-interacting massive bosons requires no more effort than the Higgs mechanism --- the key difference being that no tachyonic mass term is needed.

\subsection{The effective field theory of superfluidity:~$P(X)$ approach}
\label{sec:PX}

Although the example considered above was specific to the case of a quartic potential, certain features of the result are universal and completely determined by the symmetries of the problem.  To isolate these general features, it is convenient to write the Lagrangian of Eq.~\eqref{eq:LagrNoncanonical} in the following form
\begin{equation}
\mathcal{L}_\pi=\frac{2m^2}{\lambda_4}X^2\,; \qquad X=\dot{\pi}-\frac{(\vec{\nabla} \pi)^2}{2m}+m c_s^2\,,
\end{equation}
where we introduced the perturbed, non-relativistic chemical potential~$X$.
While the functional dependence of the Lagrangian on~$X$ is determined by the specific choice of the potential, the explicit form of~$X$ is completely fixed by symmetry principles and, as such, remains the same in the general EFT of superfluidity. 

To show this, we can build the EFT as usual by identifying the degrees of freedom and symmetries of the theory~\cite{Greiter:1989qb,Son:2002zn}. This approach has proven successful in capturing the low-energy properties of many condensed matter systems, such as fluids, solids, and other media~\cite{Leutwyler:1996er,Son:2005ak,Dubovsky:2005xd,Gabadadze:2007si,Berezhiani:2010db,Dubovsky:2011sj, Hoyos:2012dh,Nicolis:2013lma,Nicolis:2015sra,Babichev:2018twg}. In the EFT of superfluidity, the only low energy degree of freedom is the phonon field, which we identify with the phase~$\theta$ of the order parameter responsible for the symmetry breaking. Concerning symmetries, the spontaneously broken global~$U(1)$ symmetry is non-linearly realized as the shift invariance of the Nambu-Goldstone boson
\begin{equation}
    \theta\rightarrow\theta+c\,,
\end{equation}
where~$c$ is an arbitrary constant.
Therefore, the low energy EFT governing the dynamics of the system must be written in terms of a shift-invariant Lagrangian of a single degree of freedom, which at lowest order in derivatives is
\beq
\mathcal{L}= 
P(X_{\rm R})\,;\qquad {\rm with} \quad X_{\rm R}\equiv  
-\partial_\mu \theta\partial^\mu \theta\,.
\label{PX}
\eeq
The object~$X_{\rm R}$ is the only function of~$\theta$ that satisfies Lorentz invariance, exhibits shift invariance, and involves the minimum number of derivatives.
We emphasize that the low energy nature of the description~\eqref{PX} stems from this last property: the truncation of the derivative expansion of the effective theory at leading order in derivatives.
We will elaborate on the regime of validity of such a formulation when needed.

The second step is to determine the explicit form of~$X_{\rm R}$ by deriving the Noether current associated with the shift symmetry.
This is readily given by
\beq
J^\mu=-P'(X_{\rm R})\partial^\mu \theta\,,
\label{Jc}
\eeq
where primes are derivatives with respect to~$X_{\rm R}$.
If one chooses~$P(X_{\rm R})$ such that the unperturbed system corresponds to a constant background for~$\theta$, then one is dealing with a Lorentz invariant theory of the Nambu-Goldstone boson. Indeed, as one can easily see from Eq.~\eqref{Jc}, this choice corresponds to a vanishing background charge, which in turn implies a vanishing charge density of the underlying~$U(1)$ group.\footnote{A perfect example of such a realization is the~$U(1)$-Goldstone model, in which the low energy EFT of the Nambu-Goldstone boson of the spontaneously broken~$U(1)$, at tree level, reduces to
\beq
\mathcal{L}=-\frac{1}{2}(\partial\pi)^2+\frac{1}{4\Lambda^4}(\partial\pi)^4\,.
\eeq
Here~$\Lambda$ is the UV cutoff of the theory,  determined by the mass and interaction coupling constant of the scalar mode which has been integrated out.
}

Instead, for superfluids we are interested in a finite charge configuration. This implies that the background value of the phase~$\theta$ must be linear in time
\begin{equation}
    \theta=\mu t+\pi\,,
\end{equation}
where we have introduced the dynamical perturbation of the phase~$\pi$, which carries no expectation value. The functional dependence of~$\theta$ is fixed by the absence of the current~$\vec{J}$, consistent with homogeneity, and by the time-independence of~$J^0$. The field~$\pi$ is the same perturbed phase introduced earlier in the explicit field solution of Eq.~\eqref{perturbedSol}. Following this decomposition, the explicit form of the relativistic perturbed chemical potential becomes $X_{\rm R} = \mu^2 + 2 \mu \dot{\pi} - \partial_\mu \pi \partial^\mu \pi$.

The arbitrary value of~$\mu$ is rooted in the fact that it represents the chemical potential of the theory, and therefore is determined by the density~$J^0$ of the substance. In other words, it is fixed for a definite charge density according to the relation
\beq
n\equiv J^0=P'\left(\mu^2\right)\mu\,.
\eeq
Lastly, the non-relativistic limit of~$X_{\rm R}$ reads
\begin{equation}
    X_{\rm R}\simeq m^2+2m\left(\mu_\text{\tiny NR}+\dot{\pi}-\frac{(\vec{\nabla}\pi)^2}{2m}\right)=m^2+2m X\,.
\end{equation}
Therefore, in the non-relativistic limit,~$X_{\rm R}$ reduces to the perturbed, non-relativistic chemical potential~$X$, which we already identified in the special case of a superfluid with quartic contact self-interactions. We have therefore proven what we stated in the first part of this Section: in the non-relativistic limit and at leading order in derivatives, the theory of a non-relativistic superfluid is only a function of the perturbed non-relativistic chemical potential~$X$.

At this point we can delve in the study of the function~$P$. Examining the background energy-momentum tensor of the theory,
\beq
T\mn={\rm diag}\Big[2\mu^2 P'\big(\mu^2\big)-P\big(\mu^2\big),\,P\big(\mu^2\big),\,P\big(\mu^2\big),\,P\big(\mu^2\big)\Big]\,,
\eeq
we see that~$P$ reveals itself as the pressure of the system, evaluated on the background chemical potential. 
Furthermore, in the case of a non-relativistic equation of state, we would have 
\beq
\rho\equiv T_{00}\gg P \qquad \Longrightarrow \qquad T_{00}\simeq 2 \mu^2 P'\left(\mu^2\right)\,.
\eeq
Notice that this relation is trivially equivalent to specifying the equation of state~$P(\rho)$, which results from eliminating the chemical potential~$\mu$ in terms of the density~$\rho$. 

Couplings between the superfluid and external probes can be introduced following the same logic. For example, to describe an external particle~$\chi$ coupled to the number density of the fluid, one would consider the Lagrangian
\begin{equation}
    \mathcal{L}=P(X)+g\, n(X) \mathcal{O}(\chi)\,,
    \label{eq:Lagrwithprobe}
\end{equation}
where~$n(X)$ is the number density operator of the superfluid, and the operator~$\mathcal{O}(\chi)$ is a general function of~$\chi$. This class of operators can arise if the probe particle interacts with the superfluid constituents, for instance, through a contact interaction, or via a massive mediator which can be integrated out. This formalism has been used in Refs.~\cite{Acanfora:2019con, Caputo:2019ywq, Caputo:2020sys} to study the response of superfluid helium-4 to external sub-GeV DM particles~\cite{Knapen:2017xzo}, which couple to the bulk of the fluid substance. Such interactions are expected to arise from couplings between the dark sector and the Standard Model, in particular by coupling to quarks. The results obtained in the~$P(X)$ framework agree with those obtained using the dynamic structure function~\cite{Schutz:2016tid, Knapen:2016cue}, validating the theoretical formulation described above. 

Lastly, one can compute the spectrum of the excitations~$\pi$. Expanding the perturbed phase~$\theta=\mu t+\pi$ around the background value, the quadratic theory reads
\beq
\mathcal{L}^{(2)}_\pi=\Big(P'\big(\mu^2\big)+2\mu^2P''\big(\mu^2\big)\Big) \dot{\pi}^2-P''\big(\mu^2\big)\big(\vec{\nabla} \pi\big)^2\,.
\eeq
For a stable spectrum,\footnote{Stability of the background requires the following conditions  to be satisfied
\beq
P'(\mu^2)+2\mu^2P''(\mu^2)>0\, \qquad {\rm and} \qquad P''(\mu^2)>0\,.
\eeq} 
the sound wave nature of fluctuations straightforwardly follows from the dispersion relation
\beq
\omega_k^2=c_s^2k^2\,; \qquad {\rm with} \qquad c_s^2\equiv\frac{P''(\mu^2)}{P'(\mu^2)+2\mu^2P''(\mu^2)}\,.
\eeq
As long as the theory, in conjunction with the background, accommodates~$0<c_s^2\ll 1$, the system at hand represents a non-relativistic superfluid.

\subsection{Higher-order corrections and hard phonons as single particle excitations}
\label{sec:HigherOrder}

The regime of validity of the EFT discussed above is determined by its cutoff, which sets the scale beyond which the assumptions underlying its derivation break down. 
The~$P(X)$ theory assumes a leading-order approximation in spatial derivatives and, due to this assumption, the theory of superfluid phonons is constrained to be described by the equation of state of the system, evaluated on the perturbed chemical potential~$X$. Nevertheless, this assumption becomes inadequate whenever phonons with typical momentum larger than the cutoff~$m c_s$ are excited. 

In this review, we refer to excitations with momentum~$k\gg  m c_s$ as \textit{hard phonons}. We now show how higher-order derivative corrections can be consistently introduced to describe hard phonons, as long as contact self-interaction potentials are considered. This will allow us to extend the theory of phonons up to the scale~$m$, above which the non-relativistic approximation breaks down. 

For simplicity, let us focus again on the theory of a scalar field with contact two-body interactions, described by the Lagrangian \eqref{eq:LagrPhi4}. By following the steps of Sec.~\ref{sec:Phi4}, we derive again the Lagrangian for the radial and angular perturbations. This time, however, we retain the term with spatial gradients of the radial field~$h$, to obtain
\begin{equation}
\mathcal{L}=-\big(\vec{\nabla} h\big)^2+\left(-\big(\vec{\nabla} \pi\big)^2+2 m \dot{\pi} -\frac{\lambda_4}{2}\left(v+h\right)^2+\lambda_4 v^2\right)\left(v+h\right)^2\,.
\label{eq:Lagrpert2}
\end{equation}
The advantage of this theory is that it is now valid beyond the scale~$mc_s$, up to order~$k/m$ corrections to the dynamics. The downside is that we cannot integrate out exactly the radial field~$h$ as we did at leading order in derivatives, since its equation of motion is no longer an algebraic function of~$v+h$. However, we can solve the equation of motion for the radial field perturbatively in powers of the angular fluctuation~$\pi$. We begin by solving the linearized equation of motion for the radial field~$h$, whose formal solution reads
\begin{equation}
    h(\vec{x},t)=\frac{2v m \dot{\pi}}{-\Delta+2\lambda_4 v^2}\,,
\end{equation}
where~$\Delta$ is the spatial Laplacian. Substituting this in~\eqref{eq:Lagrpert2}, we obtain the Lagrangian of the phonon field,
\begin{flalign}
\mathcal{L}_\pi=\frac{1}{2}\left(\dot{\pi}^2 + c_s^2\,\pi\Delta\pi-\frac{\left({\Delta}\pi\right)^2}{4 m^2}\right)
+ \dots
\label{eq:Lagr1}
\end{flalign}
where we have performed the canonical rescaling~$\pi\rightarrow \frac{\sqrt{-\Delta+4\mu^2 c_s^2}}{2\sqrt{2}\mu v} \pi$.
Ellipses denote cubic and higher-order terms in~$\pi$,
while the sound speed~$c_s^2$ is determined as in the analysis at leading order in derivatives, Eq.~\eqref{eq:LagrNoncanonical}. 

By inspection of~\eqref{eq:Lagr1}, the dispersion relation of the phonon field now reads
\begin{equation}
    \omega_k^2=c_s^2 k^2+\frac{k^4}{4 m^2}\,.
    \label{eq:dispersionHO}
\end{equation}
The first term is the sound mode part of the spectrum, which matches its analog obtained at leading order in derivatives. A perturbation with a dispersion relation dominated by this contribution can be interpreted as a collective excitation of the condensate. The second term,~$\frac{k^4}{4 m^2}$, is a new contribution called single particle term, and it dominates the dispersion relation for momenta~$k>2 m c_s$. This is exactly the scale at which the~$P(X)$ EFT breaks down and higher-order corrections must be introduced. An excitation with a dispersion relation dominated by the second term can be interpreted as a constituent particle of the condensate, with enough energy to win over collective self-interactions and propagate in the superfluid bulk as a free particle, albeit with a mass slightly heavier than $m$ due to the presence of collective self-interactions.

It should be stressed that contact self-interactions are fundamental for this interpretation. Indeed, for local interactions, both the sound term~$c_s^2 k^2$ and the single particle contribution~$\frac{k^4}{4 m^2}$ are universal~\cite{Bernstein:1990kf}. More generally, non-local interactions provide a richer superfluid phenomenology. For example, new degrees of freedom beyond single particle excitation can appear above the scale~$m c_s$. This is the case of superfluid helium, where long-range dipole interactions induce the existence of additional degrees of freedom, such as maxons and rotons~\cite{1949ZhETF..19..637L, Landau:1949ixg, terHaar:1965zry,Nicolis:2017eqo}, beyond the typical phonon spectrum. 

So far, we have derived the interpretation of hard phonons as propagating constituents by inspecting the phonon dispersion relation. However, the analysis of scattering processes involving phonons reveals the same conclusions. Specifically, as shown in greater details in Ref.~\cite{Berezhiani:2020umi} and Appendix~\ref{sec:Scattering}, the derivative coupling between a probe and the phonon field allows the former to dissipate energy as phonon radiation.  The interaction rate associated to this process, described in Appendix~\ref{sec:Scattering}, exactly matches the scattering rate of a constituent off a moving perturber, confirming the expectation that exciting a hard phonon is essentially equivalent to kicking out a constituent particle from the condensed phase.


\subsection{Gross–Pitaevskii equation and the hydrodynamical formulation} \label{sec:Gross}

The EFT of superfluidity relies on the symmetries of the system to infer the low-energy theory of phonons. As shown above, starting from the relativistic framework, one obtains a consistent non-relativistic theory for the phase of the condensate wavefunction. In this subsection, we provide this derivation in reverse order, by first deriving the non-relativistic theory and then delving into the theory of fluctuations.

The non-relativistic approach is based on the \textit{Gross–Pitaevskii equation}. Let us start from the equation of motion of a complex scalar field with quartic interactions, derived from Eq.~\eqref{eq:LagrPhi4}:
\begin{equation}
    -\Box \Phi+m^2\Phi+\lambda_4 |\Phi|^2\Phi=0\,,
\label{eq:eomPhi}
\end{equation}
where~$\Box$ is the d'Alembertian. To study the low-energy limit of the theory, we introduce the following field decomposition
\begin{equation}
    \Phi(\vec{x},t)=\frac{1}{\sqrt{2 m}} \Psi(\vec{x},t) {\rm e}^{-{\rm i} m t}\,.
    \label{eq:NRdecomposition}
\end{equation}
Substituting back into~\eqref{eq:eomPhi}, and ignoring  quadratic in time derivatives, we obtain the Schr\"odinger-like equation,
\begin{equation}
    {\rm i} \frac{\partial}{\partial t} \Psi+\frac{\Delta}{2 m } \Psi-\frac{\lambda_4}{4 m^2}|\Psi|^2 \Psi-V_\text{\tiny ext}\Psi=0\,.
    \label{eq:GPeq}
\end{equation}
Note that we have generalized the equation to include a coupling to an external potential~$V_\text{\tiny ext}$. 
This is the Gross–Pitaevskii equation, which describes the non-relativistic dynamics of the ground state wavefunction~$\Psi$ of a system of many bosons. It is mathematically equivalent to the non-linear Schrödinger equation of a particle subject to an external potential~$V_\text{\tiny ext}$, with the non-linear part describing the self-interactions of the condensate.\footnote{The most general derivation of the Gross–Pitaevskii equation is obtained by considering the non-relativistic Hamiltonian of a dilute many-body system. In this context, two-body interactions, characterized by the potential~$V(\vec{r}-\vec{r}\,')$, represent the dominant interaction channel. At low energies, the scattering length~$a$ emerges as the sole parameter to approximate this potential, leading to~$V(\vec{r}-\vec{r}\,')\sim \frac{4 \pi a}{m} \delta^{(3)}(\vec{r}-\vec{r}\,')+\ldots$ Consequently, the Hamiltonian simplifies to that of a theory featuring contact self-interactions. By deriving the equation of motion and applying the Hartree-Fock method, Eq.~\eqref{eq:GPeq} follows~\cite{fetter2003quantum, pethick2008bose}.} 

Before proceeding, let us mention that the derivation of Eq.~\eqref{eq:GPeq} relied on further few  assumptions outlined in Refs.~\cite{Salehian:2020bon,Salehian:2021khb}. In particular, the decomposition~\eqref{eq:NRdecomposition} allows to rewrite the  relativistic field $\Phi$ in terms of its nonrelativistic component~$\Psi$, which is a slowly varying function of time and space
(compared to the dominant frequency of the system given by the scalar field mass $m$). Due to the quartic nonlinearities
involved in the relativistic system~\eqref{eq:eomPhi}, small amplitude, high-frequency oscillations sourced nonlinearly by $\Psi$ would in turn backreact on $\Psi$ itself, affecting its dynamics. This effect can be taken care of by integrating out the fast modes, thus inducing relativistic corrections to the Gross–Pitaevskii equation, which we neglect in this section and postpone to Sec.~\ref{Sect:stability}.
Furthermore, in an expanding background, there would be additional terms in  Eq.~\eqref{eq:GPeq} proportional to both the metric and matter perturbation sectors, which we also neglect as small corrections in the late
universe, when the fields are essentially nonrelativistic.

To show that there is a sound mode in the condensate, it is convenient to rewrite the above equation in the form of hydrodynamical equations. To do so, perform the Madelung decomposition of the wavefunction as~\cite{1927ZPhy...40..322M}\footnote{Evidently, the decomposition~\eqref{eq:MadelungTr} is ill-defined in the limit of vanishing density. In this case, one must resort to the Gross-Pitaevskii equation~\eqref{eq:GPeq} to capture the correct behavior.}
\begin{equation}
\Psi(\vec{x},t)=\sqrt{\frac{\rho(\vec{x},t)}{m}} {\rm e}^{{\rm i} \theta(\vec{x},t)}\,,
\label{eq:MadelungTr}
\end{equation}
where~$\rho(\vec{x},t)= m |\Psi(\vec{x},t)|^2$ is the mass density distribution of the superfluid phase, and~$\theta(\vec{x},t)$ is the phase distribution. The latter is related to the velocity field of the fluid via
\begin{equation}
    \vec{v}(\vec{x},t)=\frac{\vec{\nabla}\theta(\vec{x},t)}{m}\,.
\end{equation}
Since the velocity is the gradient of a scalar, the flow is irrotational.

Substituting these into the Gross–Pitaevskii equation and following Ref.\cite{Boehmer:2007um}, we derive the Madelung equations\footnote{Although this decomposition has been applied to systems of ultralight particles with negligible self-interactions, it has been shown to inadequately capture the interference patterns resulting from the particles' wave-like behavior~\cite{Hui:2020hbq}.}
\begin{align}
\nonumber
    \frac{\partial}{\partial t} \rho+\vec{\nabla} \cdot (\rho \,\vec{v})& =0\,;\\
     \frac{\partial}{\partial t} \vec{v}+\big(\vec{v}\cdot \vec{\nabla}\big) \vec{v} & =-\frac{1}{\rho}\vec{\nabla} P+\frac{1}{2 m} \vec{\nabla} \left(\frac{\Delta \sqrt{\rho}}{\sqrt{\rho}}\right)\,.
     \label{eq:MadelungEq}
\end{align}
These are recognized, respectively, as the continuity and Euler's equations, and can be interpreted as the quantum hydrodynamical equations of the system. The function~$P=\frac{\lambda_4 \rho^2}{2 m^2}$ is the equation of state of a two-body interacting system. The form of Euler's equation remains the same if we consider any polytropic equation of state,~$P=P(\rho)$. The quantum nature of this system is represented by the last term of the velocity equation, commonly known as the quantum pressure contribution, which captures the effect of the wave nature of particles on the dynamics of the system.\footnote{In fact, this term arises from an anisotropic stress tensor contribution, and not from a pressure term, as one can show by rewriting it as~$\partial_i \left(\frac{\Delta \sqrt{\rho}}{\sqrt{\rho}}\right)=-\rho^{-1} \partial_j\sigma_{ij}$, with~$\sigma_{ij}=-\frac{\rho}{4m^2}\partial_i \partial_j \ln{\rho}$. See Ref.~\cite{Hui:2016ltb} for further details.} It describes the reluctance of particles with velocity dispersion to be confined within a certain volume, per Heisenberg's uncertainty principle. Notice that this term becomes negligible as we increase the particle mass, since wave properties are only important below the de Broglie wavelength~$\lambda_{\rm dB} = 2 \pi/m v$. 

The continuity and Euler's equations can be combined into a single wave equation, at the linearized level. To do this, perturb the density field as
\begin{equation}
\rho(\vec{x},t)=\rho_0\big(1+\alpha(\vec{x},t)\big)\,,
\label{alphaDF}
\end{equation}
where~$\rho_0$ is the mean density of the system, assumed to be homogeneous. Expanding Eqs.~\eqref{eq:MadelungEq} at first order in~$\alpha$, the resulting equations can be combined into the following wave equation
\begin{equation}
    \ddot{\alpha}-c_s^2 \Delta 
    \alpha +\frac{\Delta^2}{4m^2} \alpha=0\,,
    \label{eq:EOMphononGP}
\end{equation}
where~$c_s$ is the adiabatic sound speed of Eq.~\eqref{eq:dispersionLandau}. One can appreciate that this equation for the density inhomogeneity~$\alpha(\vec{x},t)$ is equivalent to the linearized equation of motion for phonons, obtained in~\eqref{eq:Lagr1}. In particular, the sound and quantum terms of the dispersion relation match precisely. Thus gradient corrections are automatically accounted for in this formulation.

Let us also mention that the motion of an external heavy perturber coupled to the superfluid constituents, within the Gross–Pitaevskii theory, matches the results obtained using the EFT approach. In particular, energy dissipation is still prohibited if the impurity moves with subsonic velocity~\cite{2004PhRvA..70a3608A}.

We conclude our discussion of the Gross–Pitaevskii and hydrodynamical formulations with a remark. Although the above derivation was based on a complex scalar field, it can be readily extended to real scalar fields. This implies that a global~$U(1)$ symmetry is not strictly necessary in the relativistic theory to obtain phonon excitations. At first glance, a real scalar field theory at finite density---lacking a~$U(1)$ symmetry---would seem unable to support a Goldstone boson in its fluctuation spectrum. This is consistent with the fact that such configurations are typically unstable at the quantum level, as there is no symmetry to prevent the annihilation of constituent particles, leading to depletion of the background over time. Nevertheless, in the non-relativistic limit, an effective~$U(1)$ symmetry emerges due to particle number conservation, thereby allowing for the appearance of phonons.\footnote{However, this approximation breaks down over long timescales. In the absence of a globally conserved charge, number-changing processes---driven by self-interactions---gradually erode the condensate. For example, in a scalar field theory with a quartic coupling, four zero-momentum particles in the condensate can annihilate into two relativistic quanta, resulting in a damping of the condensate amplitude~\cite{Boyanovsky:1995pf,Dvali:2017ruz,Berezhiani:2021gph,Dvali:2022vzz}. These cumulative effects ultimately define the timescale beyond which the mean-field description fails and the emergent~$U(1)$ symmetry of the non-relativistic theory is spoiled by quantum effects.}

\subsection{Finite-temperature effects}
\label{sec: finitetemperature}

Our discussion thus far has focused on the effective description of a superfluid at zero temperature. On the other hand, in environments like galaxies and for some specific mass ranges of constituents, DM particles can have non-negligible velocity dispersion, and hence a non-zero effective temperature. In this subsection we extend the~$P(X)$ formalism, valid at zero temperature, to finite temperature, following~\cite{Nicolis:2011cs}. We discuss various theoretical aspects of finite-temperature effects in superfluids, which will be relevant for our discussion of DM superfluidity. 

According to Landau's phenomenological model~\cite{Landau:1941vsj}, at finite temperature and below the critical point~$T_{\rm c} = \frac{2 \pi}{m} \left( \frac{n}{\zeta(3/2)}\right)^{2/3}$, the system 
behaves as a two-component fluid: an inviscid superfluid component, and a ``normal'' component characterized by viscosity and entropy transport.
Hence, finite temperature effects can be incorporated in the low energy EFT approach by introducing a second fluid component, with its own degrees of freedom and symmetries, and interacting with the superfluid component. A detailed discussion on the emergence of the two-fluid model from a microscopic, relativistic field description can be found in Refs.~\cite{Alford:2012vn, Alford:2013koa}.

We begin by fixing the symmetries and degrees of freedom of the finite temperature theory. The superfluid component is described as before by~$\theta$, the phase of the ground state wavefunction, which enjoys a shift invariance. For the normal fluid, as in the EFT of regular fluids~\cite{Dubovsky:2005xd, Endlich:2010hf}, we introduce three scalar fields~$\phi^I(\vec x,t)$,~$I=1,2,3$, describing the three Lagrangian coordinates of the normal fluid. They enjoy the following symmetries: 
\begin{flalign}
&\phi^I\rightarrow \phi^I+a^I\,; \nonumber \\
 &\phi^I\rightarrow R_{\,J}^I\,\phi^J\,; \nonumber \\
    &\phi^I\rightarrow f^I\big(\phi^J\big)\qquad \text{with} \qquad \text{det}\,\frac{\partial f^I}{\partial\phi^J}=1\,.
\end{flalign}
These are respectively a consequence of homogeneity, rotational invariance and incompressibility of the fluid. The latter is mathematically imposed as an invariance of the system under volume-preserving reparametrizations. To realize these symmetries, the three scalars can only appear in the combination 
\begin{equation}
    J_\text{\tiny N}^\mu=\frac{1}{3!}\epsilon^{\mu \nu \alpha\beta} \epsilon_{IJK}\partial_\nu  \phi_I\partial_\alpha \phi_J\partial_\beta \phi_K\,.
\end{equation}
Thus, the building blocks of the finite-temperature effective theory are~$J_\text{\tiny N}^\mu$ and~$\partial_\mu \theta$. From these we can form three scalar invariants. The first invariant is still~$X$, as in the zero-temperature case. Then, we can build an invariant involving only the fluid four-vector~$J_\text{\tiny N}^\mu$ as 
\begin{equation}
   B=\sqrt{-J_\text{\tiny N}^\mu J_{\text{\tiny N} \, \mu}}\,.
\end{equation}
 The last invariant is built by contracting~$J_\text{\tiny N}^\mu$ and~$\partial_\mu \theta$, and is conventionally written as
 \begin{equation}
     Y=\frac{1}{B}J_\text{\tiny N}^\mu \partial_\mu \theta\,.
 \end{equation}
Therefore, the most general Lagrangian density describing the low-energy physics of a finite-temperature superfluid is given by
\begin{equation}
\label{FXYB}
\mathcal{L}_T=F(X,Y,B)\,.
\end{equation}
This matches the Carter-Langlois equation of state derived from the partition function for a thermal gas of phonons using thermodynamic and hydrodynamic methods~\cite{Carter:1995if}. 
See also Ref.~\cite{Favero:2024ryn} for a recent analysis of the non-relativistic limit of~\eqref{FXYB}.

A complementary derivation of finite temperature corrections to the superfluid equation of state can be obtained following the recent discussions in Refs.~\cite{Slepian:2011ev,Sharma:2018ydn}. This approach is based on the Hartree-Fock-Bogoliubov approximation and the introduction of two different chemical potentials at~$T \leq T_{\rm c}$~\cite{Yukalov_2008}---one for the condensed phase and another for the normal phase---that become equal at the critical temperature.\footnote{This approach enforces the condition of a gapless spectrum~\cite{Hugenholtz:1959zz}, allowing to overcome the Hohenberg-Martin dilemma~\cite{HOHENBERG1965291}, and ensures that the self-consistency condition for the mean field expansion is satisfied, avoiding complex corrections in the equation of state.} 

For a quartic superfluid, the most general form of the equation of state is derived in~\cite{Sharma:2018ydn}. Since there is no closed form expression, we only report certain limits of the result. At sub-critical temperatures,~$T \leq T_{\rm c}$, the finite temperature equation of state reduces to
\begin{align}
\nonumber
P\left(n,T \leq T_{\rm c}\right) &=\frac{\lambda_4 n^2}{4m^2}\left( 1 + \frac{n_1^2}{n^2} -\frac{1}{2}\left(1-\frac{n_1 + \sigma}{n}\right)^2\right) -  \frac{8m^{3/2}\Delta_2^{5/2} }{15\pi^2} \\
&-~T\int \frac{\text{d}^3k}{(2\pi)^3}\text{log}\left[1 - {\rm e}^{-\frac{1}{T}\sqrt{\frac{k^2}{2m}\left(\frac{k^2}{2m} + 2\Delta_2 \right) }}\right]\,.
\label{condensed phase pressure}
\end{align}
where~$n$ is the total number density, while~$n_0$ and~$n_1$ denote respectively the ground state and phonon number densities.
We have also introduced 
\begin{equation}
\Delta_2=\frac{\lambda_4}{4m^2}\left( n_0 +\sigma\right)\,,
\end{equation}
where the so-called anomalous average~$\sigma$ is defined as
\begin{equation}
    \sigma=\frac{1}{V}\sum_{k\neq 0} \langle a_{\textbf{k}} a_{-\textbf{k}}\rangle\,,
\end{equation}
with~$a_{\textbf{k}}$ being the ladder operator of the constituent field and $V$ the volume of the system. Although the anomalous average vanishes above the critical temperature, it does not in the presence of a condensate.\footnote{This quantity allows to overcome the Hohenberg-Martin dilemma at finite, non-critical temperatures. In fact, the anomalous average is usually ignored in the Popov approximation~\cite{doi:10.1142/9789814340960_0010}, but this is a reasonable assumption only near the critical temperature.}

At temperatures larger than the critical one,~$T \geq T_{\rm c}$, the condensate fraction~$n_0$ and anomalous average~$\sigma$ both vanish, such that the equation of state reduces to\footnote{The polylogarithm of order~$n$ is defined as~$\text{Li}_{n}(z) = \sum_{k =1}^\infty \frac{z^k}{k^n}$.}
\begin{equation}
P\left(n,T \geq T_{\rm c}\right) = \frac{\lambda_4 n^2}{4m^2} + \frac{\sqrt{2}\Gamma\left(5/2\right)}{3\pi^2}T^{5/2}m^{3/2}\text{Li}_{5/2}\left({\rm e}^{\frac{1}{T}\left(\mu - \frac{\lambda_4n}{2m^2}\right)}\right)\,,
\label{normal phase pressure}
\end{equation}
where the chemical potential~$\mu$ can be replaced by the number density to obtain a standard equation of state~$P = P(n,T)$.
We refer the reader to~\cite{Sharma:2018ydn} for further details on finite temperature corrections, as well as a derivation of
the equation of state of superfluids with 3-body interactions.

\subsection{Superfluidity and gravity}

We complete this preliminary discussion on the EFT of superfluidity by discussing the role of gravity, which turns out to be fundamental in the context of DM superfluidity. In particular, we will discuss the role of a minimal coupling between the superfluid constituents and gravity. For simplicity, we focus here on the situation where the system can be approximated as having zero temperature.

In this case, following Ref.~\cite{Son:2005rv}, the most general non-relativistic Lagrangian of a zero-temperature superfluid in an external gravitational potential~$\Phi_\text{\tiny N}$, at lowest order in spatial derivatives, is still determined by a~$P(X)$ theory, with~$X$ now given by\footnote{This result is obtained by noting that the most general Lagrangian for a non-relativistic superfluid in an external gauge field~$A_\mu$ is a~$P(X)$ theory with 
\begin{equation}
   X= D_t \theta-\frac{D_i\theta D^i\theta }{2m}\,,
\end{equation}
with~$D_t \theta=\partial_t \theta-A_0$ and~$D_i\theta =\partial_i\theta-A_i$~\cite{Son:2005rv}. 
By identifying the external field~$A_0$ with the gravitational potential energy~$m\Phi_\text{\tiny N}$, the above result follows.}
\begin{equation}
  X=\dot{\theta}-\frac{\partial_i \theta \partial^i \theta}{2m}-m\Phi_\text{\tiny N}\,; \qquad \theta=\mu_\text{\tiny NR} t +\pi\,.
\end{equation}
In other words, the leading order description of a superfluid in the presence of gravity is determined by an  effective chemical potential 
$\mu_\text{eff}=\mu_\text{\tiny NR}-m \Phi_\text{\tiny N}$. Expanding the Lagrangian in~$\pi$, the theory of phonons follows as before. 

It is, however, important to also keep track of gradient corrections to the~$P(X)$ theory, which can be important to characterize the gravitational stability of self-gravitating superfluids, or to describe the motion of a supersonic perturber through the inviscid condensate~\cite{Berezhiani:2020umi}. To properly keep track of gradient corrections, it is simplest to start from the complex scalar field theory. For concreteness, we will consider a massive scalar field minimally coupled to gravity, with quartic potential: 
\begin{equation}
    \mathcal{L}=\sqrt{-g}\left\{\frac{R}{16\pi G}-\partial_\mu \Phi^*\partial^\mu \Phi -m^2 |\Phi|^2-\frac{\lambda_4}{2} |\Phi|^4\right\}\,,
    \label{eq:minimalCoupling}
\end{equation}
where~$R$ is the Ricci scalar,~$g_{\mu\nu}$ is the space-time metric, and~$G$ is Newton's constant. We approach the problem in two complementary ways: using the quasi-particle and hydrodynamical descriptions.

Before proceeding to the details, let us briefly comment on possible effects arising from quantum gravity. It is widely expected that gravity violates global symmetries: indeed, fundamental theories of gravity (such as string theory) appear incompatible with exact continuous global symmetries~\cite{Banks:2010zn}. Hence, even if the starting point of the above discussion is a scalar field endowed with a fundamental $U(1)$ invariance, one should generically expect explicit $U(1)$-breaking mediated by operators suppressed by powers of the Planck scale.

However, as discussed in Sec.~\ref{sec:Gross}, the non-relativistic superfluid description remains valid even in the absence of a global $U(1)$ symmetry because the effective non-relativistic dynamics automatically restores particle-number conservation. Hence, gravity-mediated effects will merely generate effective quartic, sextic, and higher-order operators in the non-relativistic Lagrangian~\eqref{eq:GPeq}, suppressed by powers of the Planck mass. One can then expect to absorb this entire class of operators into a redefinition of the effective potential and, correspondingly, of the superfluid equation of state.

Substantial deviations would, however, arise when the relativistic modes of the condensate (or fast modes, in the terminology of~\cite{Salehian:2020bon,Salehian:2021khb}) are taken into account, since relativistic corrections can induce a slow evaporation of the condensate, as we will discuss in Sec.~\ref{Sect:stability}. 

\subsubsection{Quasi-particle description}
\label{sec:JeansQuasiP}

We begin by generalizing the derivation of Sec.~\ref{sec:HigherOrder} to include gravity. We provide only the main steps and refer the reader to~\cite{Berezhiani:2019pzd} for a more detailed discussion. 

Since we are ultimately interested in non-relativistic superfluids, in regimes where Newtonian gravity provides a good description of the dynamics of the system,
it suffices to approximate the gravitational action as Newtonian gravity~$\mathcal{L}_{\rm grav} =\frac{1}{8\pi G} \Phi_\text{\tiny N} \Delta \Phi_\text{\tiny N}$.
Substituting the perturbed superfluid solution~\eqref{perturbedSol} into the Lagrangian~\eqref{eq:minimalCoupling}, we obtain, in the Newtonian approximation,
\begin{equation}
    \mathcal{L}=\frac{1}{8\pi G} \Phi_\text{\tiny N} \Delta \Phi_\text{\tiny N} -(\vec{\nabla} h)^2+2 m (v+h)^2\left(\mu_\text{\tiny NR}- m \Phi_\text{\tiny N}+\dot{\pi}-\frac{\big(\vec{\nabla}\pi\big)^2 }{2 m}\right)-\frac{\lambda_4}{2}(v+h)^4\,.
    \label{eq:LagrangianFluctuationsGr}
\end{equation}
This is the generalization of Eq.~\eqref{eq:Lagrpert2}.

We can now integrate out the radial field to derive a theory for the phonon~$\pi$. First, let us notice that, if we ignored the gradient term~$(\vec{\nabla} h)^2$, we could exactly solve the equation of motion of~$h$ as in the case without gravity. Substituting the solution back into the Lagrangian~\eqref{eq:LagrangianFluctuationsGr}, the resulting theory would be a function of
\begin{equation}
    X= \mu_\text{\tiny NR}- m \Phi_\text{\tiny N} +\dot{\pi}-\frac{\big(\vec{\nabla}\pi\big)^2 }{2 m}\,.
\end{equation}
This matches the results obtained in the~$P(X)$ theory~\cite{Son:2005rv}. 

However, as emphasized earlier, it is important to keep track of higher-order gradient corrections, which means that the~$(\vec{\nabla} h)^2$ term cannot be ignored.  
Our strategy is to solve the equation of motion for~$h$, perturbatively in powers of~$\pi$. For instance, at leading order in~$\pi$, the solution is
\begin{equation}
    h(\vec{x},t)=\frac{2m v}{-\Delta+4 m^2 c_s^2} \big(\dot{\pi}-m \Phi_\text{\tiny N}\big)\,.
\end{equation}
Substituting this back into~\eqref{eq:LagrangianFluctuationsGr}, we obtain
\begin{equation}
   \mathcal{L}_\pi=\frac{1}{8\pi G} \Phi_\text{\tiny N} \Delta \Phi_\text{\tiny N} - \frac{\rho}{2m^2} \big(\vec{\nabla}\pi\big)^2 + 2\rho \big(\dot{\pi}-m \Phi_\text{\tiny N}\big)\frac{1}{-\Delta+4 m^2 c_s^2}\big(\dot{\pi}-m \Phi_\text{\tiny N}\big)\,,
   \end{equation}
where we introduced the superfluid density~$\rho = m n = 2 m^2v^2$. Integrating out the Newtonian potential, we obtain, to quadratic order in the phonon field,
\begin{equation}
    \mathcal{L}_\pi=\frac{1}{2} \dot{\pi}^2+\frac{1}{2}\pi\left(4 \pi G \rho+ c_s^2 \Delta-\frac{\Delta^2}{4m^2}\right) \pi\,,
    \label{eq:LagrGRphonon}
\end{equation}
where we have redefined~$\pi$ to canonically normalize the kinetic term. Note that, in deriving Eq.~\eqref{eq:LagrGRphonon}, we assumed that the combination~$\mu_\text{\tiny NR}- m \Phi_\text{\tiny N}$ is homogeneous on the background, which will turn out to be a reasonable approximation in the context of DM superfluidity. 

The upshot is that, in the weak-field approximation, gravity affects superfluid phonons by changing their dispersion relation in the infrared. Indeed, Eq.~\eqref{eq:LagrGRphonon} implies the phonon dispersion relation
\begin{equation}
    \omega_k^2=-4\pi G\rho+c_s^2 k^2+\frac{k^4}{4m^2}\,.
    \label{eq:phononGR}
\end{equation}
From the first term,$-4\pi G\rho$, we see that long-wave phonons are now unstable. This instability bears resemblance to the familiar Jeans instability in fluids under the influence of gravity~\cite{2008gady.book.....B, Chavanis:2011zi}. In other words, large-scale fluctuations of the condensate grow due to gravity, which cannot be counterbalanced by repulsive self-interactions on those scales.   See also Ref.~\cite{Celoria:2017bbh} for a generalization on a cosmological background.

We conclude by discussing the implications for spectator particles, minimally coupled to gravity, moving in the superfluid medium. For concreteness, let us introduce an external probe particle of mass~$M$, as the quantum excitation of a massive scalar field~$\chi$ with Lagrangian:
\begin{equation}
    \mathcal{L}_\chi=\sqrt{-g}\left(\frac{1}{2}\partial_\mu \chi \partial^\mu \chi-\frac{1}{2}M^2\chi^2\right)\,.
\end{equation}
The superfluid interacts with the probe via gravity, such that, once the Newtonian potential is integrated out, a cubic interaction vertex appears in the phonon Lagrangian
\begin{equation}
    \mathcal{L}_{\chi^2 \pi}=-4 \pi G  \sqrt{\rho}\,\frac{\dot{\pi}}{\sqrt{\Delta\Big(4\pi G\rho+c_s^2\Delta-\frac{\Delta^2}{4m^2}\Big) }}\, \chi^2\,.
\end{equation}
An interesting feature of this operator is that, although its origin is purely gravitational, it depends on the self-interacting nature of the condensate through the sound speed. We will discuss this feature in more details in Sec.~\ref{sec:DynamicalFr}, where dynamical friction in models of SDM will be reviewed. 

\subsubsection{Hydrodynamical description}

A complementary approach to describe the coupling to gravity is based on the hydrodynamical treatment, similar to Sec.~\ref{sec:Gross}, following Refs.~\cite{Ostriker:1998fa, Chavanis:2011zi}. The starting point is to substitute the field decomposition of Eq.~\eqref{eq:NRdecomposition} into the Lagrangian~\eqref{eq:minimalCoupling}.
Upon dropping terms with two time-derivatives, we obtain the non-relativistic equation of motion of the system in the presence of gravity\footnote{For completeness, the non-relativistic Lagrangian from which the Gross-Pitaevskii equation is derived reads  
\begin{equation}
    \mathcal{L}=\frac{1}{8 \pi G} \Phi_\text{\tiny N} \Delta \Phi_\text{\tiny N}+\frac{{\rm i}}{2}\big(\Psi^* \partial_t \Psi+\Psi \partial_t \Psi^*\big)-\frac{|\vec{\nabla}\Psi|^2}{2m}-\frac{\lambda_4}{8 m^2}|\Psi|^4-m \Phi_\text{\tiny N} |\Psi|^2 \,.
    \label{eq:GPgravity}
\end{equation}}
\begin{equation}
     {\rm i} \frac{\partial}{\partial t} \Psi+\frac{\Delta}{2 m } \Psi - \frac{\lambda_4}{4 m^2}|\Psi|^2 \Psi - m \Phi_\text{\tiny N} \Psi = 0\,.
\end{equation}
This is supplemented with Poisson's equation for the Newtonian potential
\begin{equation}
    \Delta \Phi_\text{\tiny N}=4\pi G\big(\rho_\text{\tiny ext}(r)+\rho(r)\big)\,,
\end{equation}
where~$\rho(r)= m |\Psi|^2$ is the superfluid mass density, and~$\rho_\text{\tiny ext}$ is any external mass distribution. This system of equations is commonly referred to as the \textit{Gross-Pitaevskii-Poisson system}, or alternatively, the \textit{Schrödinger-Poisson system} when the special case~$\lambda_4=0$ is considered.

Substituting the Madelung decomposition~\eqref{eq:MadelungTr} and repeating the derivation of Sec.~\ref{sec:Gross}, we obtain the hydrodynamical equations of the condensate in the presence of gravity. The continuity equation remains unchanged, while Euler's equation is of course affected by the inclusion of the Newtonian potential, resulting in the following modification to Eq.~\eqref{eq:MadelungEq}:
\begin{flalign}
     \frac{\partial}{\partial t} \vec{v}+\big(\vec{v}\cdot \vec{\nabla}\big) \vec{v}=-\frac{1}{\rho}\vec{\nabla} P-\vec{\nabla} \Phi_\text{\tiny N}+\frac{1}{2 m^2} \vec{\nabla} \left(\frac{\Delta \sqrt{\rho}}{\sqrt{\rho}}\right)\,.
     \label{eq:MadelungEqGr}
\end{flalign}
If we set the velocity field to zero and drop the quantum pressure term, this equation reduces to the one governing the hydrostatic equilibrium of a fluid with a polytropic equation of state. In the case of a self-gravitating fluid, for which~$\rho_\text{\tiny ext}=0$, the equation can be rewritten into a Lane–Emden form~\cite{2008gady.book.....B} 
\begin{equation}
\frac{1}{\xi^2}\frac{{\rm d}}{{\rm d} \xi}\left(\xi^2\frac{{\rm d}}{{\rm d} \xi} \Xi\right)=-\Xi^n\,,~~~ \text{with}~~~ \Xi=\left(\frac{\rho}{\rho_0}\right)^{\frac{1}{n}}\,; \quad \xi=\frac{r}{r_\text{\tiny LE}}\,,
\label{eq:LaneEmden}
\end{equation}
where~$\rho_0$ is the central density,~$n$ is the index of a general polytropic equation of state 
\be
P= K \rho^{1+\frac{1}{n}}\,,
\label{eos polytrope}
\ee
and we introduced the characteristic scale~$r_\text{\tiny LE}^2={\frac{n+1}{4\pi G} K \rho_0^{\frac{1}{n}-1}}$. 
The solution to this equation gives the density profile of a stable self-gravitating polytropic gas, in which the balance between the gravitational potential and the pressure is established at every point of the system. The differential equation is solved by imposing the boundary conditions~$\Xi(0)=1$ and~$\Xi'(0)=0$.  In the context of DM superfluidity, we will study in detail this equation for the polytropic indices~$n=1$ and~$n=\frac{1}{2}$.

Lastly, let us mention that also the Gross-Pitaevskii-Poisson system can be converted into a wave equation for the density perturbation~$\alpha(\vec x,t)$ in the presence of gravity. This reads
\begin{equation}
    \ddot{\alpha}-c_s^2 \Delta 
    \alpha-4\pi G\rho_0 \,\alpha +\frac{\Delta^2}{4m^2} \alpha=4\pi G \rho_\text{\tiny ext}\,,
    \label{eq:EOMphononGPGr}
\end{equation}
where we have assumed again a homogeneous background. This expression reproduces the result from the EFT approach of Eq.~\eqref{eq:LagrGRphonon}.

\subsection{Quantum stability of superfluids and evaporation}
\label{Sect:stability}

The discussion above summarizes the semiclassical physics of non-relativistic superfluids, a regime well suited to describe SDM. A key element in constructing the EFT of superfluidity is the presence of a microscopic $U(1)$ symmetry that is spontaneously broken. However, non-relativistic condensates can also arise from a real scalar field theory, where the symmetry emerges in the non-relativistic limit. This occurs because the coarse-graining procedure over long time scales, employed to recover the Schrödinger equation from the real Klein-Gordon field (see {\it e.g.} \cite{Hui:2016ltb} for this derivation in the context of fuzzy DM), effectively freezes the number-changing processes typical of theories without a conserved charge, leading to an emergent particle-number conservation. However, this emergent conservation is only approximate:  number-changing interactions can appear through subleading secular corrections to the time-averaging procedure, ultimately causing the EFT of superfluidity for a real scalar to break down over sufficiently long timescales. In this Section, we quantify the magnitude of this effect. 

To do so, we must examine the quantum dynamics of the superfluid. In particular, 
the exact time evolution of a quantum system is determined by the Hamiltonian dynamics of its corresponding quantum state. For coherent configurations, such as the condensates discussed above, the system can be represented, for example, by a coherent state or by a number eigenstate containing a large number of particles. A detailed treatment of these quantum states lies beyond the scope of this review. However, in Section~\ref{sec:Caveat}, we provide a brief overview in the context of cosmological structure formation.
Here, instead, we focus on outlining the key aspects of quantum time evolution that are relevant for such systems. 

An important distinction arises depending on whether the underlying field is real or complex.
In the latter case, one can show that superfluid configurations remain stable in the full quantum theory. The key point is that, because the relativistic Hamiltonian for the superfluid fluctuations~\eqref{eq:Lagrpert} is time independent, the theory admits a vacuum that is exactly time-translation-invariant. This vacuum is precisely what one identifies with the superfluid state. The very existence of such a time-translation invariant vacuum for the fluctuations is what prevents a complex scalar field from decaying in time.
Aspects of the quantum dynamics of complex scalar superfluids are discussed in Refs.~\cite{Kovtun:2020udn, Nicolis:2023pye,Berezhiani:2025tkp} for the background evolution and stability, and in Refs.~\cite{Alford:2012vn,Alford:2013koa,Sharma:2018ydn,Sharma:2022jio} for the theory of fluctuations.

However, reiterating the derivation for a real scalar field shows that the theory of fluctuations becomes time independent only after applying the temporal smoothing procedure. Without this averaging, no time-translation invariant vacuum can be defined, and the superfluid fluctuations necessarily evolve nontrivially, inducing a corresponding nontrivial backreaction on the classical superfluid background. It is precisely this backreaction that drives the slow, gradual evaporation of the superfluid through number-changing processes among its internal constituents.

We now provide the depletion timescale of the real scalar condensate.  This can be done by following the procedure outlined in~\cite{Baacke:1996se, Berezhiani:2021gph}. Let us consider the Heisenberg field equation of motion for a real scalar field with quartic interactions, where we consider a normalized potential~$\frac{\lambda_4}{4!} \hat{\Phi}^4$. Next, we introduce the following field redefinition
\begin{equation}
    \hat{\Phi}(\vec{x},t)=v(t)+\hat{\psi}(\vec{x},t)\,, \qquad \langle \hat{\psi}(\vec{x},t)\rangle=0\,.
\end{equation}
Notice that, although it resembles the standard semiclassical expansion employed in the previous section, here the expansion is not organized around the classical solution to the equation of motion.
By definition, here $v(t)$ is the one-point function of the field operator on the state that describes the condensate, and has to be determined. The time evolution of this function provides information about the evolution of the condensate.  However, we have the freedom to choose the following initial conditions for this function
\begin{equation}
    \langle \hat{\Phi}(\vec{x},0)\rangle=v_0\,,\qquad \langle \partial_t\hat{\Phi}(\vec{x},0)\rangle=0\,,
    \label{eq:initial1p}
\end{equation}
which can be set by choosing the initial state of the system accordingly.
To find the evolution of the one-point function, and hence of the quantum background, we need to solve the following system of differential equations
\begin{flalign}
(\partial_t^2+m^2)v(t)+\frac{\lambda_4}{3!}v(t)^3=-v(t)\frac{\lambda_4}{2}\langle \psi^2(x)\rangle\,;\\
\left(\Box_x+m^2+\frac{\lambda_4}{2}v^2(t)\right)\langle \hat{\psi}(x)\hat{\psi}(y)\rangle=0\,.
\end{flalign}
The derivation of this system, as well as its numerical solution, is provided in Refs.~\cite{Baacke:1996se, Berezhiani:2021gph}.

This system is one-loop exact, and hence contains information on the leading non-trivial quantum correction to the dynamics of the condensate. The first equation represents the equation of motion for the background, corrected by one-loop quantum corrections on its right-hand side. Without these corrections, it would reduce to the standard classical equation of motion. 
The second equation represents the equation of motion for the two-point function of the fluctuation field. As we see, it is similar to the equation of motion of a scalar field in the vacuum, albeit corrected with a time and background-dependent mass. In order to solve this system, we use the initial conditions~\eqref{eq:initial1p}, and the fact that $\hat{\psi}$ behaves as a real scalar field with an effective mass $m_\text{eff}=m^2+\frac{\lambda_4}{2}v_0^2$ at time $t=0$. 

The time evolution reveals that, while classically $v(t)$ is an oscillatory function describing a stationary configuration, at the one-loop level its amplitude is damped, reflecting the gradual loss of condensate particles through evaporation. The timescale at which the amplitude becomes half-damped ({\it i.e.}, when an order-one fraction of the condensate has been depleted) matches the parametric estimate~\cite{Dvali:2017ruz, Berezhiani:2021gph}
\begin{equation}
    t^{-1}_\text{dep}\sim \lambda_4  \left(\frac{\lambda_4 \rho}{m^4}\right)^3 m\,.
\label{eq:depletiontime}
\end{equation}
 This is the scale setting the survival time of the condensate.

Moreover, as a result of this evaporation, relativistic particles are produced through the conversion of the zero-momentum constituents of the condensate into a radiation component.
It is important to emphasize that this particle production is fundamentally different from the standard parametric resonance mechanism described by the Mathieu equation~\cite{Kofman:1994rk, Lozanov:2019jxc}, which occurs only when the field lies within an instability band. By contrast, the process described above is generically present and does not rely on such conditions. Moreover, it is much slower, as it does not involve any exponential growth of mode functions. The standard exponential amplification can still occur, however, if the initial conditions are appropriately chosen~\cite{Baacke:1996se}.

\section{Superfluid dark matter in galaxies}
\label{Sec: SDMgalaxies}

The idea of DM existing in the form of condensates at typical galactic densities
can be summarized in two different but complementary proposals, which mainly differ by the
impact that DM self-interactions have on the dynamics of the condensed phase. To be as general as possible, we will consider a homogeneous configuration
of real scalar bosons with repulsive quartic and hexic self-interactions, minimally coupled to
gravity:
\begin{equation}
\mathcal{L}=-\frac{1}{2}\partial_\mu \Phi \partial^\mu \Phi-\frac{1}{2}m^2\Phi^2-\frac{\lambda_4}{4!}\Phi^4-\frac{\lambda_6}{6!}\Phi^6\,.
\label{eq:LagrangianDM}
\end{equation}
From this Lagrangian, two-body and three-body superfluids are simply obtained by setting~$\lambda_6=0$ and~$\lambda_4=0$, respectively. No interactions between the dark and baryonic sectors will be considered in this Section.

According to the discussion of the previous Section, the hydrodynamical description of this system is provided by the Euler equation~\eqref{eq:MadelungEqGr},
together with Poisson's equation for the Newtonian potential, and the equation of state of the system. The latter is given by, in the two extreme cases of two-body~($\lambda_6=0$) and three-body~($\lambda_4=0$) interactions, 
\begin{align}
\nonumber
P &=K_4 \rho^2\,;\qquad K_4=\frac{\lambda_4}{16m^4} \qquad ~~(\lambda_6 = 0)\\
P &= K_6 \rho^3\,; \qquad K_6=\frac{\lambda_6}{144 m^6} \qquad (\lambda_4 = 0) \,.
\label{eos 2 cases}
\end{align}
Comparing with~\eqref{eos polytrope}, these correspond to polytropic indices of~$n = 1$ and~$1/2$, respectively. The pressure is positive for repulsive interactions~($\lambda_4,\lambda_6 > 0$). 

As shown above, the effect of gravity is balanced by two contributions. The first is the gradient of the polytropic equation of state,~$-\frac{1}{\rho}\vec{\nabla} P$, which encodes the role of self-interactions in the evolution of the system. The second is the quantum pressure contribution,~$\frac{1}{2 m^2} \vec{\nabla} \left(\frac{\Delta \sqrt{\rho}}{\sqrt{\rho}}\right)$, which arises from the wave-like nature of the light bosonic particles. Depending on which dominates at the characteristic scales of interest---typically $\sim {\rm kpc}$ scales in astrophysical contexts---different dynamical regimes emerge. If quantum pressure is the dominant contribution, the system behaves according to the fuzzy DM model. In the opposite limit, where self-interactions dominate, the dynamics are governed by the SDM model, which is the primary focus of this review.

We begin by reviewing the formation and phenomenological implications of a DM superfluid phase on the evolution of galaxies, originating from bosonic self-interactions. In this Section, we neglect the coupling to ordinary matter, which will be the subject of Sec.~\ref{sec:Forces}. 

\subsection{Hydrodynamics of the dark matter superfluid}

The first model of SDM was proposed by J. Goodman in the seminal paper~\cite{Goodman:2000tg}. In this work, the author introduced DM as a scalar boson with repulsive two-body interactions, which is degenerate at typical galactic densities and interacts through repulsive short-range interactions. The important implication of the original model is that
the ground state pressure sourced by self-interactions can sustain self-gravitating, zero-temperature configurations. We refer to these structures as non-topological solitons. This stands in contrast to self-interacting DM models, where cores are prone to gravothermal collapse upon cooling (see the discussion at the end of Sec.~\ref{sec:FormationSF})~\cite{Spergel:1999mh,Balberg:2002ue}, which was one of the original motivations behind the introduction of the model.

To better understand this point, it is convenient to recall the dispersion relation of the fluctuations of a self-interacting scalar condensate, minimally coupled to gravity. According to Eq.~\eqref{eq:phononGR}, long-wavelength fluctuations are unstable. In particular, length scales larger than 
\begin{equation}
 \lambda_{\rm J}=\frac{\sqrt{2\pi^2}}{m c_s}\left(-1+\sqrt{1+\frac{4 \pi G \rho}{m^2 c_s^4}}\right)^{-1/2}
    \label{eq:Js}
\end{equation}
are deemed to grow and form structures. This scale is known as the Jeans scale, above which fluctuations of the system are unstable due to gravity~\cite{2008gady.book.....B}.
For two-body interacting superfluids, it was derived in Ref.~\cite{Goodman:2000tg} in the limit of vanishing quantum pressure, while the derivation of the full expression using the Gross–Pitaevskii formalism can be found in~\cite{Chavanis:2011zi}.  

It is immediately apparent that Eq.~\eqref{eq:Js} admits two distinct regimes, depending on the magnitude of the dimensionless parameter\footnote{The dimensionless parameter in Eq.~\eqref{xidef} is denoted by~$\xi$ in~\cite{Berezhiani:2021rjs}. We chose to use~$\zeta$ in this review to avoid confusion with the dimensionless radial variable in the Lane-Emden equation~\eqref{eq:LaneEmden}.}  
\begin{equation}
    \zeta= \frac{ m^2 c_s^4} {4 \pi G \rho}\,.
\label{xidef}
\end{equation}
This measures the relative strength of self-interactions and gravity at the scale of the instability~$\lambda_{\rm J}$. This parameter characterizes the nature of pressure that will sustain the aforementioned configurations. If self-interactions are negligible,~$\zeta\ll 1$, then the Jeans scale reduces to 
\begin{equation}
\label{JeansscaleQ}
   \lambda_{\rm J} \simeq \left(\frac{\pi^3}{G \rho m^2}\right)^{1/4} \qquad  (\zeta \ll 1)\,.
\end{equation}
This matches the Jeans scale obtained in fuzzy DM models. We refer to this regime as \textit{degeneracy pressure} case, as it describes condensates where the quantum pressure counteracts gravity within the soliton~\cite{Hu:2000ke}. 

The opposite regime,~$\zeta\gg 1$, corresponds to condensates where self-interactions are the main contribution to the pressure of the system. In this case, the Jeans scale reduces to
\begin{equation}
\label{IPJS}
    \lambda_{\rm J}\simeq \sqrt{\frac{\pi c_s^2}{G \rho}} \qquad  (\zeta \gg 1)\,.
\end{equation}
We refer to this regime as the \textit{interaction pressure} case, wherein the system behaves effectively as a superfluid. Indeed, if instead~$\zeta\ll 1$, then the collapse of the fluid would result in the formation of substructures in which all modes are dominated by the quadratic part of the phonon dispersion relation,~$\omega_k \sim \frac{k^2}{2m}$. In this situation, there would be no superfluid phenomenology because Landau's criterion for dissipation would always be satisfied.

Thanks to the Jeans instability, long-wavelength fluctuations of an infinitely homogeneous fluid are prone to collapse and rearrange themselves into a collection of self-gravitating stable configurations, with a size comparable to the Jeans scale~$\lambda_{\rm J}$. Their density profile, under the assumption of self-gravitation, is a solution of the Lane-Emden equation~\eqref{eq:LaneEmden}, where the polytropic index depends on the underlying theory. For two-body and three-body interacting superfluids, we will consider polytropic indices~$n=1$ and~$n=1/2$, respectively.

With~$n=1$, the density profile can be derived analytically:
\be
\label{eq:2bodyprofile}
\rho_{n=1}(r)=\rho_0 \frac{\sin \frac{\pi r}{R}}{ \frac{\pi r}{R}} \,,\qquad \text{with}\qquad R=\sqrt{\frac{\pi K_4}{2G}} = \sqrt{\frac{\pi \lambda_4}{32 Gm^4}} \,.
\ee
Here,~$\rho_0$ is the soliton central density, and~$K_4$ is the coefficient of the polytropic equation of state in Eq.~\eqref{eos 2 cases}. 
Notice that the size of the soliton is density-independent and completely fixed by the parameter of the microscopic theory. This can be understood from the fact that, as we increase the number of particles in the condensate, the gravitational energy increases as the total energy provided by self-interactions. Another noteworthy property is that, since~$c_s^2 = 2K_4\rho$ in this case,  the soliton diameter is precisely set by the Jeans scale:~$2R = \sqrt{\frac{\pi c_s^2}{G\rho_0}} = \lambda_{\rm J}$.  

For~$n=1/2$, the Lane-Emden equation does not admit an analytic solution. However, an approximate density profile is provided by 
\begin{equation}
   \rho_{n=\frac{1}{2}}(r)\simeq \rho_0 \cos^{1/2}\left(\frac{\pi}{2}\frac{r}{R}\right)\,,~~~\text{with}~~~ R=\sqrt{\frac{3 K_6 \rho_0}{8 \pi G}}\xi_1 = \sqrt{\frac{\lambda_6\rho_0}{384 \pi Gm^6}}\xi_1\,,
\label{eq:3bodyprofile} 
\end{equation}
where we have used Eq.~\eqref{eos 2 cases} to substitute for~$K_6$, and where~$\xi_1\simeq 2.75$ is the first zero of the exact solution. In contrast with the previous case, the size of the soliton depends on the central density.

All of these results are valid as long as~$\zeta\gg 1$, and break down whenever quantum pressure starts contributing significantly to the dynamics of the system. Although this regime can be studied, no exact solutions that interpolate between the two limits exist. We refer the reader to Refs.~\cite{Chavanis:2011zi, Chavanis:2011zm} for a semi-analytic analysis of the full equation \eqref{eq:MadelungEqGr}.

However, let us stress that early studies paid little attention to how the superfluid phase (and its associated solitonic cores) actually forms within DM halos. It was often assumed that once a halo reached equilibrium, it would settle into a simple configuration consisting of a single central soliton surrounded by an isothermal envelope~\cite{Goodman:2000tg}. In the next Section, we show that the reality is far more intricate. By following the evolution of a halo from an out-of-equilibrium state into the superfluid phase, we find that the interplay between DM thermalization and condensation leads to a much richer and more dynamic phenomenology than originally envisioned.

\subsection{Formation of the superfluid phase}
\label{sec:FormationSF}
In order for solitons to form, a superfluid phase has to be established in DM halos. 
Self-interacting bosons can generate a superfluid state through Bose-Einstein condensation, which
is expected to occur in a degenerate system of bosons. However, for DM superfluidity to be effective in generating macroscopic ($\gtrsim$~kpc scales) structures, we also need the system to be in thermal equilibrium. Although these two properties will be discussed in detail in the rest of this Section, it is important to
clarify their role in the formation of the DM superfluid phase.

Degeneracy occurs whenever the de Broglie volume of the system is highly occupied, and it can
be satisfied {\it independently of whether the system is in equilibrium or not}. When this condition is satisfied, the system can be treated as a highly coherent background on scales smaller than the de Broglie volume. Examples are models of fuzzy DM or cosmological axions, in which self-interactions are never sufficiently efficient to establish
equilibrium, but which are degenerate at typical astrophysical densities. In general, it is possible to define the coherence scale~$\lambda_{\rm c}$ of the system, which sets the scale
below which the ensemble of DM particles can be considered as an effective condensate. It is given by
the smallest of the scale of gravitational stability and the de Broglie wavelength~\cite{Guth:2014hsa}:
\begin{equation}
    \lambda_{\rm c}= \min\left(\lambda_{\rm J},\lambda_{\rm dB}\right)\,,~~~\text{with}~~~\lambda_{\rm dB}=\frac{2\pi}{m v}\,.
\end{equation}
Let us investigate what properties DM should have
to determine a coherence length of order kpc. Virialized and out-of-equilibrium DM particles have a typical velocity dispersion in galaxies of order~$v\simeq 10^{-3}$. Extremely light bosonic masses~($m \simeq 10^{-21}$ eV)  have a de Broglie wavelength comparable to the Jeans
length, and a coherent, kpc-size, degenerate soliton can be formed.
On the other hand, if one focuses on heavier particles, it would be the de Broglie wavelength that
sets the scale of coherence, with sub-kpc extent. 
This explains why thermalization is crucial to obtain a superfluid
state which is coherent over kpc length, as it provides a mechanism to decrease the velocity dispersion of the system, allowing to set the coherence length to the Jeans scale.

Before discussing the two criteria in more detail, let us introduce the initial conditions that are employed in understanding the requirements for the SDM phase to form. According to numerical simulations of self-interacting DM~\cite{Kaplinghat:2015aga}, self-interactions play a negligible role during the initial stages of halo formation, which is primarily driven by gravity. Therefore, in this initial stage, the density distribution of DM particles is similar to the one of collisionless DM. 
For purely gravitationally interacting models,
N-body simulations reveal the more or less universal NFW density profile:
\begin{equation}
\label{NFW}
\rho(r)=\frac{\rho_\text{\tiny nfw}}{\frac{r}{r_s}\left(1+\frac{r}{r_s}\right)^2}\,,
\end{equation}
where~$\rho_\text{\tiny nfw}$ is the characteristic density, and~$r_s$ is the scale radius, both of which vary from halo to halo. The size of the
halo itself is conventionally defined by the virial radius~$R_\text{\tiny V}$, defined as the radius within which the average
density of the halo is about 200 times the critical density. For this analysis, we start with the NFW profile and investigate the conditions under which a substantial modification of the density profile occurs due to self-interactions.

From these initial conditions, the system progressively thermalizes, approaching a maximum-entropy state in which a superfluid phase can develop. While the detailed evolution of these processes would need to be followed numerically, it is instructive to draw an analogy with the behavior of self-interacting DM.
In this scenario, self-interactions drive the formation of approximately isothermal central regions, with the central cusp expanding and becoming shallower. However, this stage is only transient: as energy continues to flow outward, the core eventually undergoes gravothermal collapse, a runaway instability that would result in an even denser central region than the starting configuration.

This instability originates from the surprising thermodynamic fact that, for a self-gravitating system confined in a perfectly reflecting sphere, there is no global maximum of the entropy, but only local ones~\cite{Antonov_1985}. Moreover, if the confining sphere is allowed to expand beyond a critical radius, these local entropy maxima become inflection points, and the system becomes unstable.
This process can be understood from the fact that virialized self-gravitating systems have negative heat capacity~\cite{Lynden-Bell.1968}. Hence, when a flow of heat is allowed from inner to outer regions of halos, such as in the presence of DM self-interactions, as the core loses energy, it shrinks and becomes hotter. If there is no way for this inner core to equilibrate with outer regions (that is, if the outskirts are quite insensitive to the heat transfer),  the process continues until the system recollapses in a denser and compact core. The earlier phase of core expansion thus represents a transient stage during which the system attempts to reach thermal equilibrium. However, as the central distribution approaches a Maxwellian form and the system moves toward its local entropy maximum, the gravothermal instability sets in, driving the subsequent collapse.

If the system is assumed to maintain hydrostatic equilibrium and spherical symmetry\footnote{See also Ref.~\cite{Bautista:2025lxk} for a recent analysis on non‑spherical self-interacting DM halo shapes.} throughout its evolution, the dynamics of the core collapse are governed by the gravothermal fluid equations, a coupled set of four differential equations~\cite{1980MNRAS.191..483L, Essig:2018pzq, Balberg:2001qg, Koda:2011yb}.
The first two equations enforce the hydrostatic equilibrium and the total mass conservation during all the stages of the collapse, and are provided by
\begin{equation}
\label{gravothermal mass}
    \frac{\partial M(r)}{\partial r}= 4\pi r^2 \rho(r)\,,\qquad \frac{\partial}{\partial r}\left[\rho (r)\langle v^2\rangle\right]=-\frac{G M(r)\rho (r)}{r^2}\,.
\end{equation}
Here, $\langle v^2\rangle$ is the local average squared velocity of particles in the halo which is determined in terms of a Maxwellian distribution, $M(r)$ is the mass enclosed in a radius $r$ from the center of the halo, and $\rho(r)$ is the local density.
The second set of equations is provided by a heat and entropy equation 
\begin{equation}
\label{gravothermal Heat}
    \frac{L(r)}{4 \pi r^2}=-\kappa\frac{\partial T(r)}{\partial r}\,,\qquad \frac{1}{{4 \pi r^2}}\frac{\partial L(r)}{\partial r}=-\frac{\rho(r)\langle v^2\rangle}{\gamma-1}\left(\frac{\partial}{\partial t}\right)_M \log\left(\frac{\langle v^2\rangle}{\rho(r)^{\gamma-1}}\right)\,.
\end{equation}
Here, $T(r)$ is the local temperature,  $L(r)$ is the energy flux, $\gamma$ is the adiabatic index of the DM fluid, and $\kappa$ is the conductivity coefficient. The Lagrangian time derivative $(\partial/ \partial t)_M$ denotes differentiation at constant enclosed mass.

The conductivity coefficient $\kappa$ characterizes the efficiency of the system in transporting heat. It also plays a central role in connecting the microphysics of the fluid to the macroscopic transport behavior, since its value is determined by the way in which particles interact with one another.
In models without self-interactions, the thermal conductivity is determined solely by gravitational interactions. While this contribution is negligible for DM halos, other systems, such as globular clusters, can experience significant heat transfer through gravitational encounters~\cite{Lynden-Bell.1968, Chernoff:1989na, Makino:1996vi, Zhong:2025epi}. When  self-interactions dominate over gravitational effects in heat conduction, the thermal conductivity   is approximately determined by~\cite{Essig:2018pzq,Yang:2022zkd}
\begin{equation}
    \kappa\propto   \frac{n \lambda^2_\text{\tiny mfp}}{t_\text{\tiny rel}}\,.
\end{equation}
Here, $n$ is the local DM density, $\lambda_\text{\tiny mfp}$ is the mean free path, which is connected to the underlying scattering cross section $\sigma$ by the relation $\lambda_\text{\tiny mfp} \sim 1/(\sigma n)$, while the relaxation time reads $t_\text{\tiny rel}\sim \lambda_\text{\tiny mfp}/ v_\text{\tiny rel}$ in terms of the average relative velocity $v_\text{\tiny rel}$ of DM particles. 
Once the  conductivity $\kappa$ is provided, one can solve Eqs.~\eqref{gravothermal mass} and~\eqref{gravothermal Heat} and completely determine the dynamics of the gravothermal collapse for self-interacting DM halos.

Why does SDM behave differently in this respect? Unlike self-interacting DM, where the pressure from interaction is negligible on astrophysical scales, in SDM the high-density contracting core generates a pressure from particle interactions that counteracts the usual gravothermal contraction. From a thermodynamic perspective, this effectively reverses the sign of the heat capacity of the core. Consequently, the core cools, gradually re-equilibrates with the outskirts, and condenses. Collapse into an extremely dense configuration is avoided, and the system settles into the cored structures described in the previous section. As noted above, however, the formation of the superfluid phase has not yet been numerically validated in the literature.

\subsubsection{Degeneracy}

 In order for condensation to occur, DM has to be \textit{degenerate} at typical galactic densities. A system is said to be degenerate if there are many particles per de Broglie volume. That is, 
 \begin{equation}
     \mathcal{N}\equiv n \lambda^3_\text{dB} = \frac{\rho}{m}\left(\frac{2\pi}{m v}\right)^3 \gg 1  \,,
\label{cal N def}
 \end{equation}
where~$n$,~$\rho = mn$, and~$v$ are respectively the number density,  mass density, and velocity dispersion of the system, with the orbital velocity approximating the latter. As a result, this condition is parametrically equivalent to the effective temperature of a weakly-interacting gas well below the critical temperature, $T \ll T_c$. We also defined the degeneracy (or Bose-enhancement) factor~$\mathcal{N}$, which counts the number of particles in a given energy level, and should be understood as the mean occupation number of the system. We can define a degeneracy radius~$R_\text{\tiny deg}$ as the scale where the degeneracy factor becomes unity: 
\be
\label{degcond}
\mathcal{N} (R_\text{\tiny deg}) = \left. \frac{\rho}{m}\left(\frac{2\pi}{m v}\right)^3\right\vert_{r = R_\text{\tiny deg}} = 1\,.
\ee
We henceforth refer to the region~$r < R_\text{\tiny deg}$ within which degeneracy is achieved as the {\it degeneracy region}.
 
Notice that this definition is valid as long as the velocity dispersion is comparable to the mean velocity in the system. While this assumption breaks down once a superfluid phase forms, the DM velocity distribution in the early stages of halo formation is expected to resemble that of a virialized gas, with orbital velocity~$v=\sqrt{{G M(r)}/r}$, which matches the velocity dispersion at different radii. Here,~$M(r)$ is the mass enclosed within an orbit of radius $r$. Thus, $\mathcal{N}$ accurately describes the system before self-interactions impact the DM phase space distribution.

While it is generally not possible to derive an analytic expression for~$R_\text{\tiny deg}$, one can obtain an approximate result for a NFW profile~\eqref{NFW}
in the limit where~$R_\text{\tiny deg}$ is either much larger or much smaller than the scale radius~$r_s$: 
\begin{equation}
R_\text{\tiny deg}= \begin{cases} \left( \frac{ 8\pi^3}{G^3 m^8  \rho_\text{\tiny nfw} r_s} \right)^{1/5} & \mbox{for } R_\text{\tiny deg}  \ll r_s 
\\
\frac{\pi}{G m^{8/3} \rho_\text{\tiny nfw}^{1/3} r_s} & \mbox{for } R_\text{\tiny deg}  \gg r_s \,.
\end{cases}
\end{equation}
The condition of degeneracy therefore occurs all over the halo if DM particles satisfy
\begin{equation}
 R_\text{\tiny deg} \gtrsim R_\text{\tiny V} \quad  \Longrightarrow   \quad m\lesssim \left(\frac{M_\text{\tiny halo}}{10^{12}M_\odot}\right)^{-1/4} \text{eV}\, ,
\end{equation}
where~$M_\text{\tiny halo}$ is the halo mass. In the last step we have substituted the typical values~$\rho_\text{\tiny nfw} \simeq 10^{-25} \text{g}/\text{cm}^3, R_\text{\tiny V} =200$ kpc and concentration parameter~$c=r_s/R_\text{\tiny V}=6$ for a Milky-Way-like halo~\cite{Berezhiani:2015bqa, Berezhiani:2015pia}.\footnote{In fact, the aforementioned relation is derived by evaluating the degeneracy factor in terms of the virial density~$\rho_\text{\tiny V}\simeq 180 \,\rho_\DM^{(0)}\simeq 5.4 \times 10^{-28}$ g/cm$^3$, where~$\rho_\DM^{(0)}\simeq 3 \times 10^{-30}$ g/cm$^3$ is the present DM background density, and virial velocity~$v_\text{\tiny V}\simeq 110 \left({M_\text{\tiny halo}}/{10^{12}M_\odot}\right)^{1/3}$~\cite{Peacock:1999ye}.}
In other words, the DM mass has to be sub-eV for the condition of degeneracy to be satisfied. Note that one can obtain a weaker bound by requiring degeneracy only in the central region of galaxies, where the density is higher. For instance, for a Milky Way-like galaxy, demanding degeneracy only within the inner~10~kpc from the galactic center requires mass~$m\simeq 50$~eV, in contrast to the~$m\simeq 1$ eV bound required to have the same halo in a degenerate state up to the virial radius~$R_\text{\tiny V}$~\cite{Berezhiani:2021rjs}.

\subsubsection{Thermalization} 
\label{sec:therm}

Thermalization in halos is established when the bosonic particles reach their maximal entropy state via self-interactions.
According to simulations of self-interacting DM~\cite{Kaplinghat:2015aga}, the density profile of halos starts deviating from the collisionless case whenever particles scatter at least once during the age~$t_g$ of the halo. This condition defines another important length scale of the problem besides the Jeans and degeneracy scales, which is the thermal radius~$R_\text{\tiny T}$: 
\begin{equation}
    \Gamma(R_\text{\tiny T}) t_g=1\,.
    \label{eq:defTR}
\end{equation}
Here,~$\Gamma(R_\text{\tiny T})$ is the relaxation rate of the system, approximated by the scattering rate according to the discussion above, and~$t_g\simeq 13 \text{ Gyrs}$ is approximated as the age of the universe. We henceforth refer to the region~$r<R_\text{\tiny T}$,
within which thermal equilibrium is established, as the {\it thermal region}.

The size of the thermal region can vary depending on the nature of DM self-interactions.
For DM particles interacting mainly through two-body contact interactions, we find
\begin{equation}
\Gamma_\text{\tiny 2-body}=\left(1+\mathcal{N}\right) \frac{\sigma_2}{m} \rho v\,;  \qquad \sigma_2= \frac{\lambda_4^2}{128\pi^2 m^2}\,,
\label{eq:2bodyrate}
\end{equation}
where~$\sigma_2$ is the~$2\rightarrow 2$ scattering cross section for the scalar field theory~\eqref{eq:LagrangianDM} with quartic potential. 
Notice that the interaction rate is enhanced by the degeneracy factor~$\mathcal{N}$ (Eq.~\eqref{cal N def}), due to the highly occupied phase space. 
Specifically, this factor accounts for the multiplicity of particles in the final states in two-body scatterings~\cite{Sikivie:2009qn,Erken:2011dz}. 

If the gas interacts mainly through three-body interactions, one may still use the scattering rate to estimate the relaxation rate. 
In this case, an approximate version of the three-body scattering rate can be derived using the quasi-particle description~\cite{Berezhiani:2022buv} for a scalar field theory with hexic potential,\footnote{In Ref.~\cite{Berezhiani:2022buv}, the three-body relaxation rate is derived for a complex scalar field with a hexic potential of the form~$\lambda_6|\Phi|^6/3$. To normalize it similarly to the theory in Eq.~\eqref{eq:LagrangianDM}, we multiply by the factor~$(1/12)^2$, being the ratio of the squared amplitudes in the two theories, which accounts for their different multiplicities.} giving
\begin{equation}
    \Gamma_\text{\tiny 3-body}=\left(1+\mathcal{N}^{\,2}\right)\rho^2 \frac{\sigma_3}{m^5} v^4\,; \qquad \sigma_3=\frac{\lambda_6^2 m^2}{16 (4\pi)^4}\,.
    \label{eq:3bodyrate}
\end{equation}
Since we are considering three-body self-interactions, the density dependence is stronger compared to the two-body case. The same applies to the~$\mathcal{N}$ dependence of the rate, since there are now three particles involved in the scattering. 

The rates~\eqref{eq:2bodyrate} and~\eqref{eq:3bodyrate} can be substituted into Eq.~\eqref{eq:defTR} to determine~$R_\text{\tiny T}$ as a function of theory parameters
and density profile. As before, one can obtain an approximate expression for the thermal radius for a NFW profile~\eqref{NFW} in the limit that~$R_\text{\tiny T}$ is either much larger
or much smaller than~$r_s$, assuming~$\mathcal{N} \gg 1$. In these two limits, one finds~\cite{Berezhiani:2022buv} 
\begin{equation}
R^{(2)}_\text{\tiny T} \simeq { }r_s \left( \frac{2 \pi^2 \rho_\text{\tiny nfw}}{Gr_s^2 m^4} \frac{\sigma_2}{m}t_{\rm g} \right)^{\gamma}\,;\quad {\rm with} \quad \gamma = \begin{cases} \frac{1}{3} & \mbox{for } R_\text{\tiny T}\ll r_s  \\ \frac{1}{5} & \mbox{for } R_\text{\tiny T}\gg r_s \end{cases}
\label{RT2}
\end{equation}
for the two-body interaction case, and
\begin{equation}
R^{(3)}_\text{\tiny T} \simeq r_s\left(\frac{9 (4 \pi)^4\rho_\text{\tiny nfw}^3}{2 Gr_s^2 m^{12}}\frac{\sigma_3}{m} t_g\right)^\gamma\,,  \qquad \text{with}~~~~\gamma = \begin{cases} \frac{1}{5} & \mbox{for } R_\text{\tiny T}\ll r_s  \\ \frac{1}{11} & \mbox{for } R_\text{\tiny T}\gg r_s  \end{cases}
\label{RT3}
\end{equation}
for the three-body case. To provide some quantitative estimates, we can substitute~$\rho_\text{\tiny nfw} \simeq 10^{-25} \text{g}/\text{cm}^3$ for Milky Way-like halos and~$t_g\simeq 13 \text{ Gyrs}$ to obtain (assuming~$R_\text{\tiny T} \gg r_s$ in both cases)
\begin{align}
R^{(2)}_\text{\tiny T} & \simeq 10^6 \, {\rm kpc} \left(\frac{m}{\rm \mu eV}\right)^{-4/5}\left(\frac{\sigma_2/m}{10^{-8} \, {\rm cm^2/g}} \right)^{1/5} \,, \nonumber \\
R^{(3)}_\text{\tiny T} & \simeq 10^8 \, {\rm kpc} \left(\frac{m}{\rm \mu eV}\right)^{-12/11}\left(\frac{\sigma_3/m}{10^{-8} \, {\rm cm^2/g}} \right)^{1/11}\,.
\label{RT23}
\end{align}
Thus very large thermal radii are obtained for those choices of parameters.

\subsubsection{Degeneracy vs thermalization} 
\label{sec:degtherm}

At this point, it is natural to inquire about the relative size of the degenerate and thermal regions. There are two possibilities. The first possibility is when thermalization is achieved on scales larger than the degeneracy radius,~$R_\text{\tiny T}> R_\text{\tiny deg}$, which means that~$\mathcal{N} < 1$ at the thermal radius~$R_\text{\tiny T}$.
The second possibility is when~$R_\text{\tiny T} < R_\text{\tiny deg}$, corresponding to a degeneracy region that encompasses the thermal region. By studying these two regimes in turn, we will see that only the second case ($R_\text{\tiny T} < R_\text{\tiny deg}$) results in~$\gtrsim$~kpc-size superfluid regions, and is therefore of phenomenological interest for galactic dynamics~\cite{Berezhiani:2021rjs}. 
Notice that, when neither conditions for superfluidity are satisfied, one expects a non-degenerate classical gas of weakly interacting particles.

\begin{itemize}

\item~$R_\text{\tiny T}> R_\text{\tiny deg}$: At distances within the thermal region but outside the degenerate core,~$R_\text{\tiny deg} < r < R_\text{\tiny T}$, particles have experienced interactions and achieved thermal equilibrium. Since they are non-degenerate, however, we expect the profile to resemble that obtained in  models of self-interacting DM. Finally, at smaller distances~$r < R_\text{\tiny deg} < R_\text{\tiny T}$, both conditions for superfluidity are satisfied, and we get a superfluid state.

As shown in~\cite{Berezhiani:2021rjs}, the regime~$R_\text{\tiny T}> R_\text{\tiny deg}$
selects, for Milky Way-like halos, DM particles significantly heavier than eV. The resulting soliton size is~\cite{Berezhiani:2021rjs} 
\begin{equation}
\lambda_{\rm J} \lesssim 10^{-2} {\rm pc} \left( \frac{m}{\rm 20 \, eV} \right)^{-5/4}  \left( \frac{\sigma/m}{\rm cm^2/g} \right)^{1/4}\,,
\end{equation}
which is evidently much smaller than galactic scales of interest. Let us stress that, in this case, finite temperature effects in the superfluid description are important since they determine the transition between the normal and superfluid phases, at~$r=R_\text{\tiny deg}$.  

\item~$R_\text{\tiny deg} > R_\text{\tiny T}$: In this case there is an intermediate region~$R_\text{\tiny T} < r < R_\text{\tiny deg}$ wherein self-interactions are negligible, resulting in a density profile that should mimic the NFW profile. Since particles are out of equilibrium yet degenerate in this intermediate region, one expects Bose-Einstein condensates with typical coherence length of order de Broglie wavelength, with~$\mathcal{N} \gtrsim 1$.
Meanwhile, at distances~$r <R_\text{\tiny T} < R_\text{\tiny deg}$, both conditions for superfluidity are satisfied.
For a Milky-Way halo mass of~$10^{12}M_\odot$, one obtains a superfluid region of size~\cite{Berezhiani:2021rjs} 
\begin{equation}
\lambda_{\rm J} \simeq 30 \, {\rm kpc} \left( \frac{m}{\mu {\rm eV}} \right)^{-5/4}  \left( \frac{\sigma/m}{\rm 10^{-8} \, cm^2/g} \right)^{1/4}\,.
\label{Jeans MW}
\end{equation}
This is~$\gtrsim$~kpc for a wide range of parameters. Notice that, in this case, the transition between degenerate and superfluid phase is an out-of-equilibrium phenomenon. By choosing masses that are extremely sub-eV, finite temperature effects can be neglected for typical galactic density.

\end{itemize}

The two cases are summarized in Table~\ref{tab:res}. The natural feasibility of satisfying both superfluid conditions---degeneracy and thermal equilibrium---while also achieving astrophysical-size cores in the regime~$R_\text{\tiny deg} > R_\text{\tiny T}$, compels us to focus on this scenario in the remainder of the review to study the properties of SDM halos. This condition demands to focus on DM masses below the eV scale. The interested reader can find additional detail in Ref.~\cite{Berezhiani:2021rjs}.

Let us also stress that, aside from these two scales, one should also consider the virial radius~$R_\text{\tiny V}$. For sub-eV masses, the degeneracy radius always exceeds any astrophysical virial radius, implying that DM remains degenerate on all scales. 
Therefore, a thermal radius larger than the virial radius implies that the entire halo undergoes a superfluid phase transition. In contrast, if~$R_\text{\tiny T} < R_\text{\tiny V}$, then a residual outskirt of out-of-equilibrium particles is present in the region~$R_\text{\tiny T} < r < R_\text{\tiny V}$. As we will show in the next sections, both cases allow the formation of astrophysical-size SDM cores.

{%
\begin{table}[t]
\centering
    \begin{tabular}{c lll}
        Cases  & Degeneracy &  Thermalization & DM \\
        \midrule
        \midrule
         $r <  R_\text{\tiny deg} < R_\text{\tiny T} $ & \,\,\,\, \ding{51} & \,\,\,\, \ding{51} & SDM with $\lambda_{\rm J} \ll {\rm kpc}$ \\
        $ R_\text{\tiny deg} < r < R_\text{\tiny T} $ & \,\,\,\, \ding{55} & \,\,\,\, \ding{51} & $\sim$ self-interacting DM   \\
        $ R_\text{\tiny deg} <  R_\text{\tiny T} < r $ & \,\,\,\, \ding{55} & \,\,\,\, \ding{55} & $\sim$ CDM \\
        \midrule
        $r < R_\text{\tiny T} < R_\text{\tiny deg} $ & \,\,\,\, \ding{51} & \,\,\,\, \ding{51} & SDM with $ \lambda_{\rm J} \gtrsim {\rm kpc}$ \\
        $ R_\text{\tiny T} < r < R_\text{\tiny deg} $ & \,\,\,\, \ding{51} & \,\,\,\, \ding{55} & $\sim$ Bose-Einstein condensate \\
        $ R_\text{\tiny T} <  R_\text{\tiny deg} < r $ & \,\,\,\, \ding{55} & \,\,\,\, \ding{55} & $\sim$ CDM \\
        \midrule
        \bottomrule
    \end{tabular}
    \caption{Summary of the cases discussed in Sec.~\ref{sec:degtherm}.}
\label{tab:res}
\end{table}
}%

\subsection{Fragmentation of the thermal core from Jeans instability}
\label{sec:fragmentation}

The upshot of the above discussion is that the phenomenologically-relevant regime is~$R_\text{\tiny T} <  R_\text{\tiny deg}$, in which case the thermal radius~$R_\text{\tiny T}$ sets the size of the region within which DM undergoes a superfluid phase transition. Meanwhile, the Jeans scale~$\lambda_{\rm J}$ sets the size of a stable superfluid soliton. Depending on the relative hierarchy between these two scales, the final structure of the halo after the superfluid phase transition can vary significantly.

If~$R_\text{\tiny T}<\lambda_{\rm J}$, the halo is expected to contain a single superfluid core of size~$R_\text{\tiny T}$. In contrast, if~$R_\text{\tiny T} >\lambda_{\rm J}$, then the thermal region will fragment into a collection of self-gravitating solitons (see also~\cite{Delgado:2022vnt}). A priori, it seems that both possibilities are allowed, since~$R_\text{\tiny T}$ and~$\lambda_{\rm J}$ are 
controlled by different combinations of the mass and coupling of the theory. 

However, we must bear in mind that we are by assumption working in the interaction pressure regime~($\zeta \gg 1$), where the restoring force against gravitational collapse is provided by self-interactions. As shown in~\cite{Berezhiani:2021rjs}, the phenomenologically viable hierarchy in this regime is  
\begin{equation}
    R_\text{\tiny T}\gtrsim \lambda_{\rm J}\,.
\end{equation}
This implies that the thermal core is unstable, and is prone to fragmenting into a collection of superfluid droplets of size~$\lambda_{\rm J}$. 
This result is obtained by performing a numerical comparison of the Jeans scale~\eqref{eq:Js} and the thermal radius derived from the scattering rates~\eqref{eq:2bodyrate} and~\eqref{eq:3bodyrate}. To evaluate these, we assume typical NFW parameters for Milky Way-like galaxies, such as~$\rho\simeq 10^{-25} \text{g/cm}^{3}$ for the average
density of the thermal core before fragmentation. For the Jeans scale~$\lambda_{\rm J}$, we use the full formula \eqref{eq:Js} to capture both the degeneracy and interaction pressure cases,  substituting the sound speed~$c_s^2 = {\rm d}P/{\rm d}\rho$ for the two- and three-body equations of state in Eq.~\eqref{eos 2 cases}, expressed in terms of the respective cross sections given in Eqs.~\eqref{eq:2bodyrate} and~\eqref{eq:3bodyrate}.   

\begin{figure} 
	\centering
	\includegraphics[scale=0.45]{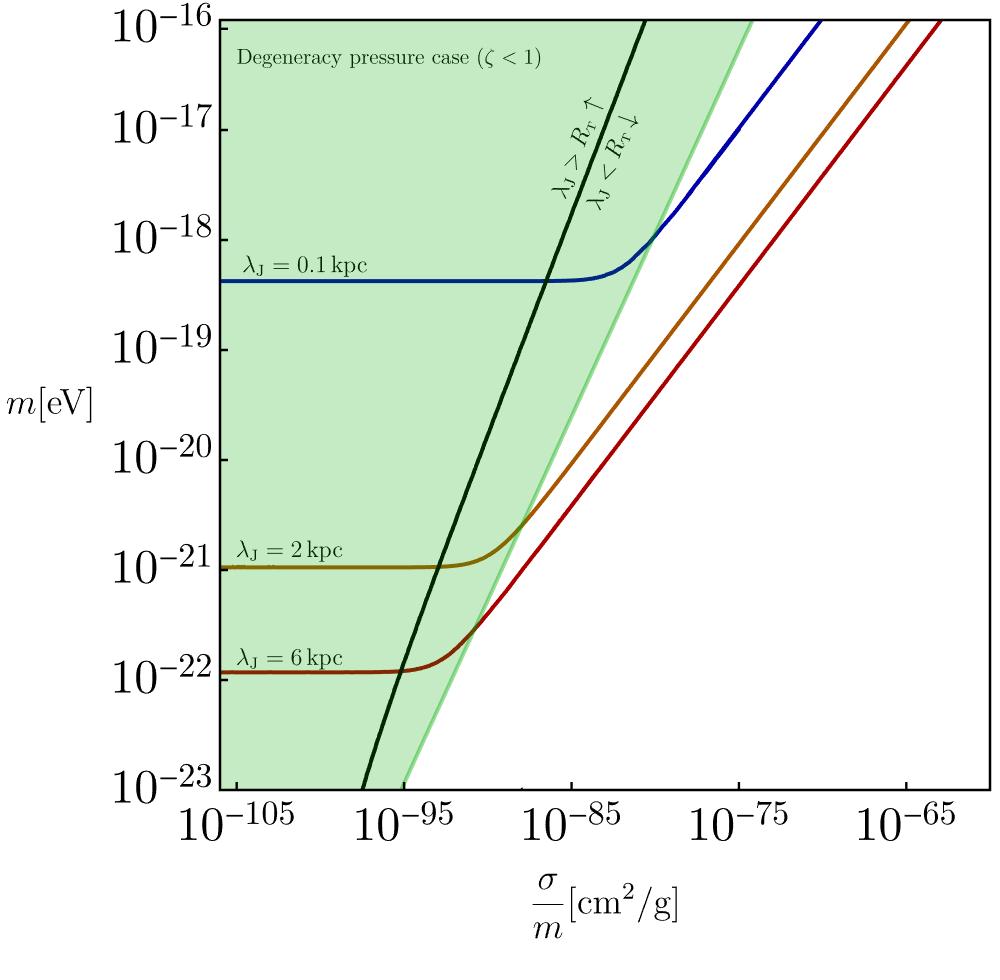} 
\caption{Comparison between the thermal radius~$R_\text{\tiny T}$ and Jeans scale~$\lambda_{\rm J}$ for superfluids with two-body interactions. 
Blue/orange/red solid lines indicate Jeans scales of size~$\lambda_{\rm J} = 0.1,2,6$ kpc in degenerate regions of the Milky Way DM halo. The green region corresponds to degeneracy pressure case ($\zeta\lesssim 1$). 
The black solid curve indicates the parameter space that generates a thermal radius~$R_\text{\tiny T} = \lambda_{\rm J}$. On its right, the Milky Way DM halo is in global thermal equilibrium. We see that it is impossible to have an interaction pressure-dominated core with~$\lambda_{\rm J} \gtrsim R_\text{\tiny T}$. 
 Figure inspired from Ref.~\cite{Berezhiani:2021rjs}. }
	\label{fig:JS}
\end{figure}

In Fig.~\ref{fig:JS} we compare different values of~$\lambda_{\rm J}$ and~$R_{\rm T}$ in the two-body interacting case, as a function of the DM particle mass~$m$ and interaction cross section. The degeneracy pressure regime~($\zeta \lesssim 1$) is indicated by the green region. Notice that~$\lambda_{\rm J}$ becomes independent of~$\sigma/m$ in the degeneracy pressure regime. In all cases, we see that~$R_\text{\tiny T}\gtrsim \lambda_{\rm J}$ in the interaction pressure regime (white region), as claimed.\footnote{We should stress that the line~$\zeta = 1$, delineating the transition from degeneracy to  interaction pressure, depends on density.}  
 Let us stress that, although $\lambda_{\rm J}$ is density-independent in the interaction pressure case, this is not the case for the regime~$\zeta\leq 1$. 
 Hence, although the sloped part of the solid curves remains the same across all astrophysical structures, the same cannot be said for the turning point and the flat segment.

Based on these findings, we can provide insights into the conclusions drawn in Ref.~\cite{Slepian:2011ev}. It was argued by these authors that the original model of Goodman~\cite{Goodman:2000tg} was ruled out, because it failed to simultaneously reproduce physically-acceptable rotation curves and satisfy the Bullet Cluster bound~\cite{Markevitch:2003at}. A key assumption of the paper, however, is that the collapse of the halo is spherically symmetric, adiabatic, and capable of maintaining equilibrium in the outskirts. Only in this way, the final SDM halo can be modeled as a single superfluid core surrounded by an isothermal envelope, which is the case for which the above analysis applies.
As we have seen in this Section, however, the spherical symmetry of the halo seems hard to justify,\footnote{When relaxing the assumption of spherical symmetry, one expects also the emergence of disk-like dark substructures in SDM models, featuring only an axial symmetry associated to rotations about a fixed axis~\cite{Alexander:2019qsh}. These could exist as isolated substructures and, in principle, even seed satellite galaxies, and may be probed through astrometry or strong gravitational lensing. Furthermore, similar to spherically symmetric halos, they may experience tidal disruption and result in streams~\cite{Alexander:2019qsh}.} and a more general class of destabilizing fluctuations should be considered, leading to a thermal region that is prone to fragmentation into a collection of superfluid droplets. The resulting DM density profile is therefore
expected to be more complex, given that it is possible to relax both assumptions of global thermal equilibrium and single core configuration compared to the original proposal of Ref.~\cite{Goodman:2000tg}.\footnote{There are also certain shortcomings in the finite temperature equation of state introduced in Ref.~\cite{Slepian:2011ev}, which are discussed in more details in Refs.~\cite{Sharma:2018ydn, Sharma:2022jio}.}

We conclude by briefly commenting on three-body interacting superfluids. 
By performing the same comparison using the three-body scattering rate of Eq.~\eqref{eq:3bodyrate}, no significant deviations are found from the two-body case. In other words, in the interaction pressure case, the thermal region is expected to fragment into smaller cores also for three-body interactions. The only notable difference is that, for a hexic potential, the Jeans scale exhibits a weak dependence on density also in the interaction pressure regime. 

\subsection{Tidal disruption of solitons}
\label{tidal subsec}

The upshot of the above analysis is that the thermal region is expected to fragment into smaller superfluid droplets of size~$\sim \lambda_{\rm J}$.
The density profile within each droplet should be well-approximated by~\eqref{eq:2bodyprofile} and~\eqref{eq:3bodyprofile}, with~$R$ set by~$\lambda_{\rm J}$.
This assumes, however, that the solitons can be treated as isolated systems, influenced only by self-gravity, and not affected by external potentials.\footnote{For instance, underlying the derivation of density profiles~\eqref{eq:2bodyprofile} and~\eqref{eq:3bodyprofile} is the assumption of spherical symmetry and the choice of the considered equations of state, which would be affected by the presence of an external potential.} While this may be justified for solitons near the center of the halo, superfluid droplets far enough in the outskirts will experience the gravitational potential sourced by other solitons within their orbit. In particular, we will see that such solitons tend to be tidally disrupted by the collective gravitational field of inner solitons.

Let us briefly review the basic concepts of tidal disruption~\cite{2008gady.book.....B}. Consider an extended body of mass~$M_\text{\tiny body}$ orbiting at a distance~$r$ in the gravitational potential of an enclosed mass~$M(r)$. Due to the spatial inhomogeneity of the gravitational potential, different parts of the body are pulled differently depending on their position in the potential itself. The net results are tidal forces, which cause an elongation of the orbiting body along the force gradient. 
In extreme cases, if tidal forces are strong enough, the orbiting body may undergo tidal disruption and shatter.
The strength of tidal effects can be estimated by introducing the tidal radius~$r_\text{\tiny tid}$, defined as~\cite{2008gady.book.....B, Kesden:2006vz}
\begin{equation}
r_\text{\tiny tid}=r\left(\frac{M_\text{\tiny body}}{M(r)}\right)^{1/3}\,.
\label{ch3:tidal}
\end{equation}
The tidal radius sets an upper limit on the size~$\ell$ of the body, that is stable under tidal deformations. When~$\ell > r_\text{\tiny tid}$, the body is susceptible to tidal disruption and can fragment into a collection of debris.\footnote{While tidal disruption may not occur immediately when~$\ell \lesssim r_\text{\tiny tid}$, it is important to note that tidal deformations can still take place. Indeed, in cases of significant deformation, after several cycles, the orbiting probe can become stretched to the extent that its size along the stretched direction approaches that of the tidal radius. This can ultimately result in the shattering of the body. This effect has been studied in simulations of fuzzy DM solitons orbiting in a NFW profile~\cite{Glennon:2022huu}, which showed that repulsive self-interactions accelerate disruption, and that there is a degeneracy between soliton mass and self-coupling strength in determining disruption timescales.}

In the case of interest, focusing on a two-body interacting superfluid for concreteness, the soliton size is approximated by the Jeans  length:~$\ell_\text{\tiny soliton} = \lambda_{\rm J}$. Its mass~$M_\text{\tiny body} = M_\text{\tiny soliton}$ is found by integrating the density profile~\eqref{eq:2bodyprofile} up to~$\lambda_{\rm J}$, with the result~\cite{Berezhiani:2022buv} 
\begin{equation}
\label{mass-soliton}
M_\text{\tiny soliton}=\frac{\rho_0\lambda_{\rm J}^3}{2\pi}\,.
\end{equation}
We keep the soliton central density~$\rho_0$ general for now, since it is determined by the fluctuations that collapsed in forming it. 

To evaluate the enclosed mass~$M(r)$, we approximate the density distribution within the soliton's orbit to be given by a coarse-grained version of NFW profile~\eqref{NFW}. 
This is supported by the following considerations. While we have described the formation of the superfluid solitons as a sequence of three events---thermalization, phase transition, and fragmentation---these are expected to happen simultaneously in realistic scenarios.\footnote{The concurrency of these events is justified by the corresponding separation of timescales: the formation of solitons is expected to happen on a timescale determined by the dynamical time, which is much shorter than the Hubble scale involved in the process of thermalization.} Small regions of high density within the halo are likely to undergo thermalization first, and transition into a superfluid state, eventually giving rise to solitons of characteristic size~$\lambda_{\rm J}$. Then, thermalization continues to impact regions with lower densities, leading to the formation of additional solitons with comparable size. Therefore, we expect that the average density of each soliton should be set by the local density of the halo in which they formed. Because the NFW profile was our starting input, the density distribution of droplets within a radius~$r$ is expected to follow an NFW profile as well.
 
With these assumptions, the ratio of the tidal radius over the size of a soliton is given by~\cite{Berezhiani:2022buv}
\begin{equation}
\frac{r_\text{\tiny tid}}{\lambda_{\rm J}}\simeq\left(\frac{\rho_0}{40\rho(r)}\right)^{1/3}\,.
\end{equation}
This would imply that stable solitons must have a central density at least 40 times the mean density of the environment in which they form.
However, superfluid solitons do not typically remain in circular orbits; for instance, those on eccentric orbits will traverse the central region, where the local density is high.
A more conservative stability requirement, which allows for general orbital trajectories,
is obtained by replacing~$\rho(r)$ with the NFW characteristic density~$\rho_\text{\tiny nfw}$:
\begin{equation}
\rho_0>40\rho_\text{\tiny nfw}\,.
\label{eq:boundrho}
\end{equation}
Solitons that do not satisfy this condition experience tidal disruption, ultimately giving rise to streams of superfluid remnants. 

It is instructive to translate the condition \eqref{eq:boundrho} into a lower bound on the size of the halo's initial density fluctuation that would lead to the formation of a stable soliton. This is obtained by equating the mass~$M_\text{\tiny fluc}\simeq \rho_\text{\tiny nfw} R^3$  
enclosed in a homogeneous density fluctuation of size~$R$, to the mass of a stable soliton $M_\text{\tiny soliton}$ given in Eq.~\eqref{mass-soliton}. Stable solitons are then obtained by fluctuations satisfying 
 \begin{equation}
     R=\left(\frac{1}{2\pi}\frac{\rho_0}{\rho_\text{\tiny nfw}}\right)^{1/3} \lambda_{\rm J} \gtrsim 2 \lambda_{\rm J}\,.
 \end{equation}
In other words, it appears that fluctuations sufficiently larger than the Jeans scale can collapse into stable solitons. Nevertheless, several approximations have been adopted. First, the analysis assumes that the initial density seeds can be treated as homogeneous, which may not hold since the relevant scales involved in the collapse are of order~$\lambda_{\rm J}\sim \text{kpc}$. More importantly, tidal disruption is expected to play a key role even before the overdensity can fully collapse into a soliton: in particular, tidal effects may prevent soliton formation in the first place, as both Jeans collapse and tidal disruption occur on a similar timescale, roughly given by the dynamical time $t_\text{\tiny dyn}\sim1/\sqrt{G\rho}$. As a result, only a small fraction---if any---of the solitons formed through fragmentation are expected to survive tidal disruption, with the central core being the primary exception to tidal interactions because of the underlying spherical symmetry of the halo.

\begin{figure}[t!]
	\centering
	\includegraphics[width=0.5\textwidth]{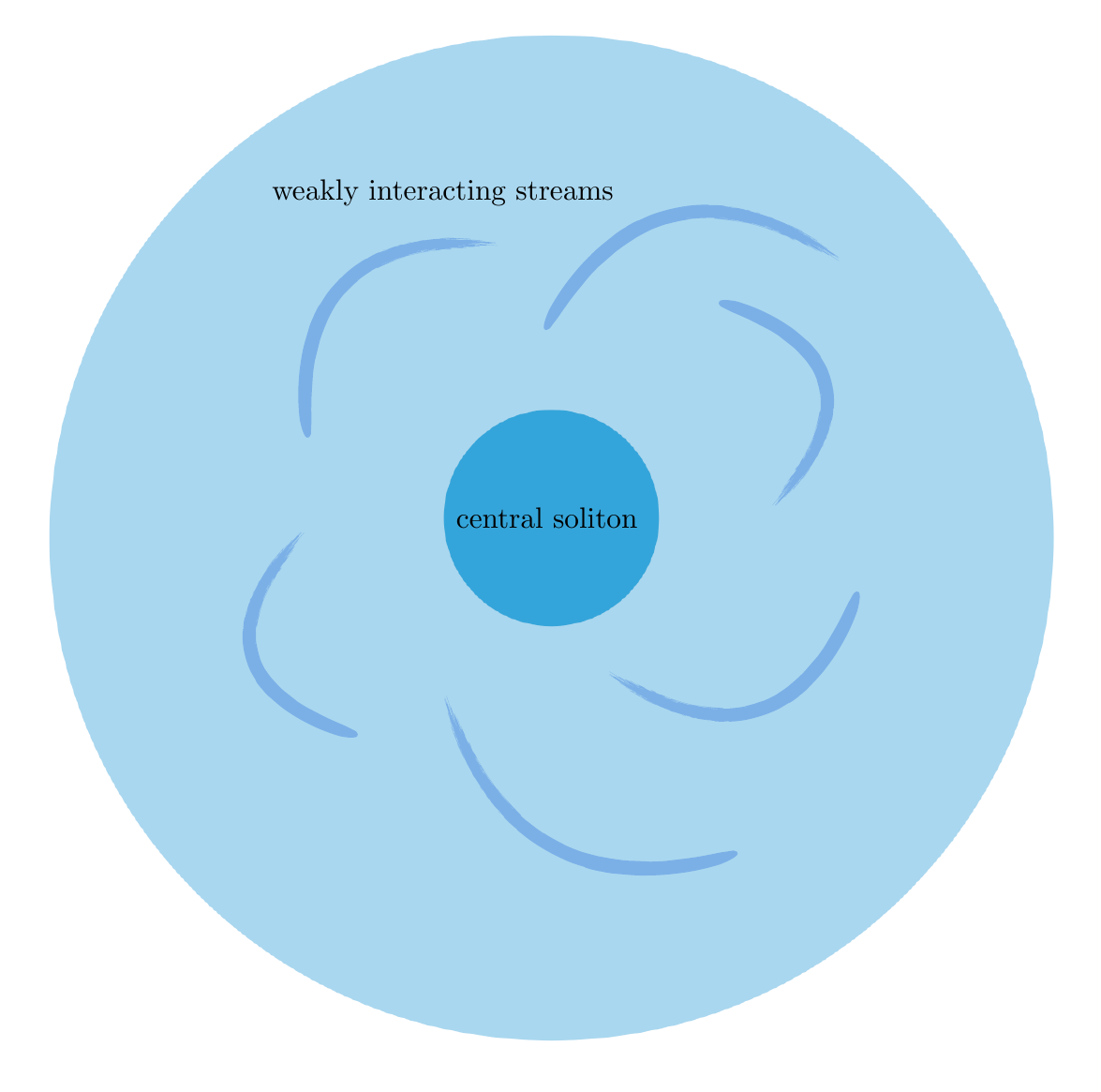}
	\caption{Visual depiction of the central superfluid area, approximately of size~$R_\text{\tiny T}$. This zone features a central superfluid soliton of size $\lambda_{\rm J}$, encircled by a series of lightly interacting streams of superfluid remnants revolving around it. These remnants stem from the tidal disintegration of superfluid droplets that originated in the periphery of the thermal core. Whether the halo is in a state of thermal equilibrium or not determines the presence of an outer region consisting of out-of-equilibrium degenerate particles, extending up to the virial radius of the halo. Figure taken from Ref.~\cite{Berezhiani:2023vlo}.}
	\label{streams}
\end{figure}

The expected final picture is that the halo density distribution features a central superfluid core surrounded by superfluid debris left behind by the tidal disruption of outer solitons.  These debris, when coarse-grained over distances greater than~kpc, should approximately follow a NFW profile. Figure~\ref{streams} provides a schematic representation of the thermal region of size $R_\text{\tiny T}$ following these events.
However, since the above analysis relies on analytical estimates, numerical studies are clearly necessary to validate this conceptual framework or reveal its limitations. This would allow a deeper understanding of the complex non-linear dynamics at play in SDM structure formation. Recent works along this direction have already made some progress to understand the formation of SDM solitons in galaxy halos. For instance, through a 3D Schr\"odinger-Poisson simulation of a virialized halo plus stochastic fluctuations in repulsive scalar‐field DM models, Ref.~\cite{Garcia:2023abs} demonstrated that halos with sizes much larger than the healing length are likely to harbor the formation of a central soliton, with halo assembly history potentially leading to a large scatter in the soliton masses in cosmological halos.

\subsection{Summary}

Let us briefly summarize the results of this Section, pertaining to the formation and evolution of the superfluid phase under its own self-interactions and gravity.
We have seen that there are three important length scales at play: 

\begin{itemize}

\item The degeneracy radius~$R_\text{\tiny deg}$, within which DM particles have a high occupancy per de Broglie volume. 

\item The thermal radius~$R_\text{\tiny T}$, within which DM particles achieve thermal equilibrium.

\item The Jeans scale~$\lambda_{\rm J}$, which sets the size of gravitationally-stable superfluid solitons. 

\end{itemize}

Since superfluidity requires both degeneracy and thermal equilibrium, clearly a superfluid region can be at most of size~$\sim {\rm min}(R_\text{\tiny T},R_\text{\tiny deg})$.
We have seen that the hierarchy~$R_\text{\tiny T} \lesssim R_\text{\tiny deg}$ is necessary to obtain superfluid solitons of size~$\lambda_{\rm J} \gtrsim$~kpc, provided that DM particles are sufficiently light~($m \lesssim \mu {\rm eV}$). If the thermal radius is smaller than the full extent of the DM halo, an intermediate region emerges,~$R_\text{\tiny T} < r < R_\text{\tiny deg}$, where DM particles form a Bose-Einstein condensate but do not achieve a superfluid state, with a typical coherence length on the order of the de Broglie wavelength. Deeper inside, in the region~$r < R_\text{\tiny T}  < R_\text{\tiny deg}$, superfluidity can be achieved.

Furthermore, we have also established the inequality~$\lambda_{\rm J}\lesssim R_\text{\tiny T}$, in the interacting pressure case of interest. This implies that the thermal core is unstable, and is prone to fragmenting into a collection of superfluid droplets of size~$\lambda_{\rm J}$. To summarize, the phenomenologically-viable hierarchy of scales for DM superfluidity is
\begin{equation}
\lambda_{\rm J}\lesssim R_\text{\tiny T} \lesssim R_\text{\tiny deg}\,.
\end{equation}
Dynamically, we expect halo formation to proceed as follows. In the first stages of its formation, the DM halo evolves in a similar way as standard collisionless DM. This continues until self-interactions start affecting the density profile, generating a region of the halo of size~$R_\text{\tiny T}$ through thermalization, within which superfluidity is established. 

In the last part of the discussion (Sec.~\ref{tidal subsec}), we have seen that superfluid droplets that form in the outskirts of the halo, and whose central density is therefore lower, are prone to tidal disruption from the gravitational field of inner solitons. An estimate of the soliton stability bound is given in Eq.~\eqref{eq:boundrho}. The end result is a collection of streams of superfluid remnants, orbiting around a central, surviving core, as depicted in Fig.~\ref{streams}.

The final density distribution is expected to be characterized by the following distinct phases:

\begin{itemize}
    \item The most inner region~$r\lesssim \lambda_{\rm J}$ of the halo hosts a central, quasi-homogeneous superfluid soliton. Assuming a central density of~$\rho_0\sim 10^{-25}\text{g}/{\text{cm}^3}$, the mass of the central soliton is estimated to be, from~\eqref{mass-soliton}
    \begin{equation}
        M_\text{\tiny soliton}\simeq 2\times 10^5\left(\frac{\lambda_{\rm J}}{\text{kpc}}\right)^3M_\odot\,.
    \end{equation}
    \item The region~$\lambda_{\rm J}\lesssim r<R_\text{\tiny T}$ of the halo is comprised of streams of superfluid debris originating from the tidal disruption of outer solitons. These remnants behave as non-interacting objects, since their interactions are not expected to be Bose-enhanced due to the low velocity dispersion in the debris environment (for a related discussion see also Sec.~\ref{sec:Bullet}). Unless the process of fragmentation and tidal disruption is capable of generating a non-coherent and out-of-equilibrium fraction of DM, these debris should be interpreted as small chunks of superfluid phase, interacting only through gravity. In principle, this region could also host solitons that are dense enough to not be tidally disrupted, according to the relation~\eqref{eq:boundrho}, which could be as massive as~\cite{Berezhiani:2022buv}
    \begin{equation}
    M_\text{\tiny soliton}^\text{\tiny debris}\gtrsim  10^7\left(\frac{\lambda_{\rm J}}{\text{kpc}}\right)^3M_\odot\,.
    \end{equation}
    When compared with the results of CDM simulations~\cite{Robles:2019mfq}, the similarity between this lower bound (for $\lambda_{\rm J} \sim \mathcal{O}(\rm kpc)$) and the heaviest resolved subhalos in simulations (of the order of $10^7 M_\odot$) suggests that none of the subhalos generated through hierarchical structure formation would survive tidal effects in our context. 
    
    \item In the external region~$R_\text{\tiny T} \lesssim r \lesssim R_\text{\tiny V}$, DM behaves as collisionless particles, since self-interactions are not efficient enough. Their velocity distribution is similar to that of CDM, albeit with a highly degenerate phase space. As we discuss in Sec.~\ref{sec:Bullet} on merging galaxies, this region is extremely important in constraining this class of models. In particular, during the collision of DM halos, this region experiences larger interactions because of Bose enhancement, and it is thus responsible for bounds on DM self-interactions, extrapolated from cluster collisions. Nevertheless, if galaxy clusters are in global thermal equilibrium ($R_\text{\tiny T}>R_\text{\tiny V}$), this region is absent and this bound relaxes to its nondegenerate counterpart.
    
\end{itemize}

\begin{figure}
\centering
	\includegraphics[scale=0.72]{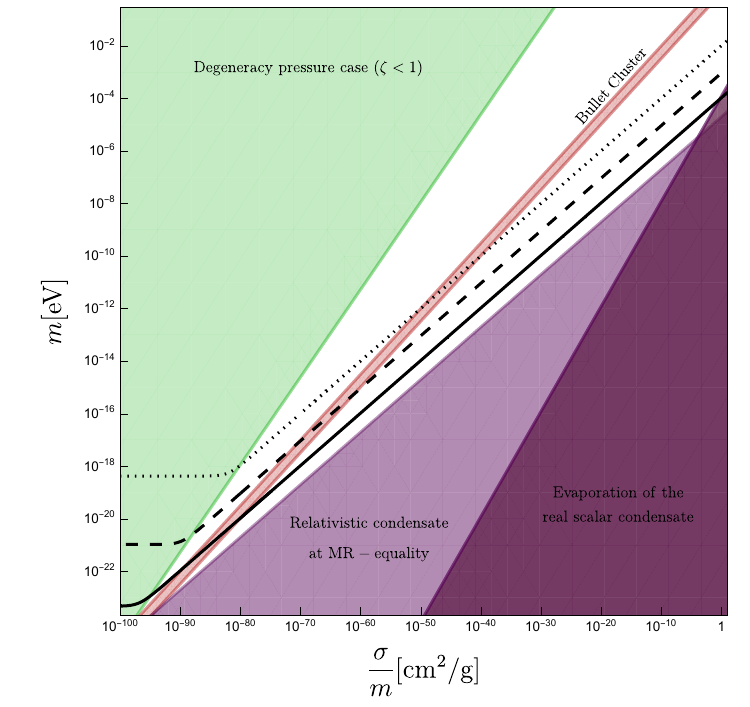} 
\caption{Parameter space for the model of  SDM with two-body interactions. The quartic coupling has been rescaled in terms of the scattering cross section per unit of DM mass. The colored regions are constrained by various observational bounds, while the white ones are allowed.
The dotted, dashed and solid black lines denote solitons of size~$\lambda_{\rm J}=0.1,2$~and~$30$ kpc, respectively.
See the main text for details.}
\label{ch3:general2}
\end{figure}

Figure~\ref{ch3:general2} shows a comprehensive plot of the parameter space of two-body interacting superfluids, for various values of the DM mass and cross section. 
The dotted, dashed and solid black lines correspond to solitons of size~$\lambda_{\rm J}=0.1,2$~and~$30$ kpc, respectively. 
The green region indicates the parameter space in which solitons are mainly sustained by quantum pressure, with density profiles resembling those of fuzzy DM models.
The light purple region is excluded by the requirement that the cosmological pressure provided by the condensate is negligible at matter-radiation equality, which we will discuss in more details in Sec.~\ref{sec: earlyuniv}. If we impose that condensates must be non-relativistic already at~$T=\text{keV}$, then the excluded region is extended up to in between the dashed and solid black lines. The dark purple region corresponds to condensates that would have evaporated into dark radiation by the present time through number-changing processes, assuming that DM is comprised of real scalar bosons. In particular, within this range of parameters, one would find that the typical depletion time-scale of the condensate \eqref{eq:depletiontime} would be shorter than a Hubble time.
Finally, the red stripe corresponds to the Bullet Cluster bound for degenerate DM particles, which will be discussed in detail in  Sec.~\ref{sec:Bullet}.

We identify two allowed sub-regions in the plot.  The white region on the left of the red stripe corresponds to halos in which only central regions present a superfluid phase. In this case,~$R_\text{\tiny T}<R_\text{\tiny V}$, allowing only solitons up to~$\lambda_{\rm J}\sim 6$ kpc. Conversely, the white region on the right side corresponds to halos that are entirely in thermal equilibrium ($R_\text{\tiny T}>R_\text{\tiny V}$). In this case, Jeans scales up to~$\mathcal{O}(100 \text{ kpc})$ are allowed, provided that the Bullet Cluster bound can be relaxed for halos in global thermal equilibrium. 
However, as we will show in the next Section (see also the light purple region), these values are marginally consistent with having a non-relativistic equation of state at matter-radiation equality.

\section{From cosmological production to virialization}
\label{sec: cosmo}

This Section provides a brief overview about the cosmological evolution in the context of the SDM scenario, and its implications for galactic-scale dynamics. Specifically, we will discuss the arguments for possible phase-space rearrangement through virialization, which leads to the requirement of thermalization, as examined in Sec.~\ref{sec:therm}.

\subsection{Early universe dynamics and bounds}
\label{sec: earlyuniv}

DM particles in the sub-eV mass range could be produced as non-thermal relics,\footnote{Sub-eV mass bosons can also be produced thermally in the early universe through interactions with SM particles. However, such a component would contribute as dark radiation, thus being constrained by measurements of the effective number of relativistic species through CMB and BBN data~\cite{Cadamuro:2010cz,Cadamuro:2011fd, Millea:2015qra, Green:2021hjh}, or suppress structure formation through free streaming effects~\cite{Xu:2021rwg}.} {\it e.g.}, via the vacuum misalignment mechanism~\cite{Preskill:1982cy, Abbott:1982af, Dine:1982ah}, to make up a significant fraction of the DM in the universe. This mechanism, when applied to an ultra-light scalar field model, assumes that at some time in the early universe, the scalar field was initially configured homogeneously across the horizon scale. This configuration could have arisen from a prior phase transition, after which the scalar field acquired different values at different spatial locations, later homogenized by the Hubble flow. 

How homogenization proceeds depends on the production time relative to cosmic inflation. If the phase transition occurs before inflation, the subsequent inflationary expansion naturally smooths out the field configuration. Conversely, if the transition takes place after inflation, the field takes random values in causally disconnected regions, leading to large inhomogeneities and the formation of topological defects~\cite{Kibble:1976sj, Battye:1994au}. In this post-inflationary case, the spontaneous breaking of the global $U$(1) symmetry triggers, via the Kibble–Zurek mechanism, the formation of a vortex network with an initial coherence length set by the Hubble radius. As the horizon grows, regions containing vortices come into causal contact, temporarily increasing the apparent inhomogeneity. However, analytical and numerical studies of global-string networks show that their non-linear evolution rapidly drives the system toward a scaling regime with roughly one string per Hubble patch~\cite{Hiramatsu:2012gg,Kawasaki:2014sqa, Vaquero:2018tib, Buschmann:2019icd,OHare:2021zrq, OHare:2024nmr}. In this attractor state, vortices intercommute, self-intersect, and fragment into loops that efficiently radiate the Goldstone mode, converting the network’s energy into phonons on Hubble timescales. Provided the transition occurs well before matter-radiation equality, the decay of loops and cosmic expansion wash out large-scale gradients, yielding an effectively homogeneous superfluid background on Hubble scales.

For the purposes of our discussion, we simply assume the field is homogeneous. To illustrate the implications of this assumption, it is useful to first consider the example of a non-self-interacting massive scalar field in an expanding universe,
\beq
\ddot{\varphi}+3H\dot{\varphi}+m^2\varphi=0\,.
\label{freephi}
\eeq
As is well known (see, {\it e.g.},~\cite{Preskill:1982cy, Abbott:1982af, Dine:1982ah}), the nearly frozen value of the scalar field acts as an attractor in the high-curvature regime ($H\gg m$) due to the dominance of the Hubble friction term. In this limit, the scalar field has the equation of state of dark energy (note that DM constitutes a subdominant component in the energy budget in the early universe). Once the initial kinetic energy has been redshifted away in this high-curvature regime, the equation governing the scalar field \eqref{freephi} simplifies to
\beq
3H\dot{\varphi}+m^2\varphi\simeq 0\,.
\eeq
This equation can be readily integrated in a radiation-dominated universe~($H=\frac{1}{2t}$), yielding
\beq
\varphi=\varphi_0 {\rm e}^{-\frac{m^2}{12}\big(H^{-2}-H_0^{-2}\big)}\,,
\eeq
where~$H_0$ and~$\varphi_0$ denote the initial values.

As the universe expands and cools during the radiation-dominated era, the curvature scale eventually becomes comparable to~$m$. As a result, the scalar field begins to oscillate and behaves like an ordinary condensate of collisionless massive particles. As long as~$H\ll m$, its equation of state is that of dust, with number density redshifting inversely proportional to the volume.

At the onset of oscillations~($H\simeq m$), the energy density of the scalar field can be estimated as its potential energy, which in turn is set by the amplitude of the scalar field. Notice that, at this point, we are no longer in the slow-roll regime, but it nevertheless serves as a good estimate of the initial density for our purposes. As the universe expands further and reaches matter-radiation equality, the energy density of the scalar field should match the required DM density,~$\rho_{\rm eq}\simeq 0.4~{\rm eV^4}$. Using the Friedmann equation~$3\mpl^2 H_{\rm eq}^2=\rho_{\rm eq}$, one readily obtains the required scalar field amplitude~$\varphi_m$ at the moment when~$H\simeq m$ to be
\beq
\varphi_m=\sqrt{3}\mpl \left(\frac{H_{\rm eq}}{m} \right)^{1/4}\simeq 10^{-7}\mpl\left(\frac{m}{\rm eV} \right)^{-1/4}\,,
\eeq
where we have used~$H_{\rm eq}\simeq 10^{-28}~{\rm eV}$. The energy density of the scalar field at~$\varphi_m$ is thus
\beq
\rho_m\simeq \big(20~{\rm GeV}\big)^4 \left(\frac{m}{\rm eV}\right)^{3/2}\,.
\eeq
In a scenario where such a scalar field is introduced solely to account for DM without an underlying particle physics motivation, the initial conditions must be finely tuned to achieve the appropriate DM abundance. Given that the variation of the scalar field is negligible at densities~$\rho >\rho_m$,
the initial conditions can be set at curvature scales much larger than~$m$. 

In axion-like scalar DM models, self-interactions are considered to be extremely weak, to the extent that they cannot generate substantial pressure throughout the regime in which axions act as DM. DM superfluidity, however, requires significant self-interactions in order to reshape the galactic dynamics. This requirement leads to the modification of the equation of state of the scalar field. In other words, the previous discussion must be revised to account for the self-interaction potential, which may be strong enough to generate significant interaction pressure and ultimately pose challenges for the scalar DM model.

Notably, there exists a one-to-one correspondence between the Jeans scale~$\lambda_{\rm J}$ ({\it i.e.}, the size of self-sustained superfluid solitons) and the equation of state at matter-radiation equality~\cite{Berezhiani:2022buv}
\beq
\frac{P}{\rho}\Bigg|_{\rm equality}\simeq 10^{-5}\left( \frac{\lambda_{\rm J}}{\rm kpc} \right)^2\,.
\label{eq:MR}
\eeq
Taking into account that~$P/\rho\propto \rho$ for a quartic theory, we can compute the radiation temperature~$T_*$ above which the equation of state of our DM candidate becomes relativistic
\beq
T_*\simeq 46~{\rm eV} \left( \frac{\lambda_{\rm J}}{\rm kpc} \right)^{-2/3}\,.
\eeq
This can be somewhat problematic if the candidate makes up most of the DM, as it becomes relativistic too close to equality.
Similarly, $\lambda_{\rm J} \gtrsim 10 \, {\rm kpc}$ re-enter the cosmological horizon around the same time,
and consequently fail to adequately seed structure formation (which would require an earlier horizon crossing)~\cite{Dodelson:2020bqr}.
These results have also been confirmed through effective hydrodynamical simulations in Refs.~\cite{Dawoodbhoy:2021beb, Shapiro:2021hjp, Hartman:2022cfh}, which showed that self-interacting scalar field DM models with core radii of order $1 \, {\rm kpc}$ significantly suppress the formation of halos with masses below~$10^{14} M_\odot$. Notably, this constraint is more stringent than the one estimated above.\footnote{Similarly, Ref.~\cite{Hartman:2021upg} showed that the model could also be constrained using data from the cosmic microwave background, baryonic acoustic oscillations, growth factor measurements, and type Ia supernovae distances, resulting in a comparable bound on the size of the core.}

One way to resolve this conflict, while still achieving superfluid solitons with kpc size, is to modify the scalar field effective potential at high densities thorough the addition of higher dimensional operators. Flattening the potential in this way can reduce the effective pressure at matter-radiation equality. Further work along this direction is needed to clarify the viability of the model against early-universe bounds.\footnote{For models of fuzzy DM, there have been recent works discussing free streaming and interference effects when the ultralight bosons are warm, either from the moment they are produced or subsequently due to interactions~\cite{Ling:2024qfv, Liu:2024pjg, Liu:2025lts, Amin:2025dtd, Capanelli:2025nrj}. For SDM, because of the heavier particle masses, we expect these effects to be smaller. However, given the aforementioned uncertainties in the superfluid production mechanism and early universe model, we cannot make a firm prediction, and leave a better understanding of these issues to future work.}

\subsection{Caveats of structure formation}
\label{sec:Caveat}
Assuming that the aforementioned caveat regarding the early universe can be addressed, the next question pertains to the subsequent evolution of the scalar field configuration. According to the vacuum misalignment argument, the scalar field exists in a highly homogeneous and coherent state across the entire universe at the moment it begins to oscillate.

As long as this state is of high occupancy, which holds true for the case of interest, the DM distribution can be adequately described by the classical scalar field,~$\phi_{\rm cl}(x,t)$. Fundamentally, this function corresponds to the expectation value of the field operator~$\hat{\phi}$ in the \textit{coherent state}
\beq
\label{clinit}
\la \phi_{\rm cl} |\hat{\phi}(x,t)| \phi_{\rm cl} \ra=\phi_{\rm cl}(x,t)\,.
\eeq
As long as coherence is maintained, the evolution of this one-point function follows from the classical equation of motion of the gravitating scalar field. In the non-relativistic approximation, the evolution is governed by the Schr\"odinger-Poisson system of equations, that readily follows from~\eqref{eq:GPgravity}. With some exceptions, these equations are equivalent to hydrodynamic equations. In other words, as long as the assumption of coherence remains valid, the evolution of the DM field is governed by the corresponding classical dynamics. 

If this assumption remained valid throughout the entire virialization process, one would no longer need to invoke the thermalization condition of Sec.~\ref{sec:therm}, as the superfluid phase of our DM particles would be ensured from the outset. In fact, this assumption would imply the validity of the quasi-hydrodynamic description, meaning that the DM phase space would not be gaseous, with a velocity distribution similar to that of CDM, thus requiring DM particles to maintain coherence throughout the non-linear regime of structure formation. The approach we have adopted is more conservative, allowing for the possibility that coherence---and the corresponding description in terms of a single classical one-point function---is lost during virialization. Furthermore, we have argued that superfluidity will be re-established subsequently, if the DM scattering cross-section is sufficiently large to achieve thermalization. 

That being said, it is nevertheless instructive to assess the departure from coherence and the exchange of particles from an ordered to a disordered phase. There has been longstanding interest in the depletion of the one-point function in a variety of contexts; see, \textit{e.g.}, \cite{Baacke:1996se} and references therein. More broadly, the concept of loss of coherence has been extensively studied in \cite{Dvali:2013vxa,Dvali:2013eja,Berezhiani:2016grw,Dvali:2017eba,Dvali:2017ruz,Berezhiani:2021gph,Dvali:2022vzz}, with particular emphasis on its relevance to black holes and cosmological backgrounds. Fundamentally, the quantum state of the system is a distribution function. The classicality ({\it i.e.}, the applicability of the description merely by the one-point function) emerges when higher moments of the distribution function are suppressed relative to the expectation value. For instance, the initial coherent state can be constructed as
\beq
|\phi_0\ra={\rm e}^{-{\rm i}\int {\rm d}^3 x \, \phi_0\hat{\Pi}}|\Omega\ra\,, \qquad \text{with}\qquad \phi_0(x)\equiv \phi_{\rm cl}(x,0)\,,
\label{cohstate}
\eeq
where~$\hat{\Pi}$ is the conjugate momentum of~$\hat{\phi}$, and~$|\Omega\ra$ satisfies~$\la\Omega|\hat{\phi}|\Omega\ra=0$, which can be taken to be the vacuum of the theory.\footnote{Although the perturbative consistency and renormalizability of one-loop UV divergences within interacting field theories requires~$|\Omega\ra$ to be a background-dependent squeezed vacuum, higher-loop consistency requires even non-Gaussian modifications~\cite{Berezhiani:2023uwt} (see also~\cite{Berezhiani:2021gph,Berezhiani:2020pbv}). For the sake of our discussion, we can ignore these peculiarities.} Such condition ensures that most of the scalar field energy budget is encoded in the one-point function, as it implies that the variance of the quantum state is suppressed in comparison. In principle, if instead of a vacuum state one were interested in populating the state with particles with a certain momentum distribution, then one could make the replacement
\beq
|\Omega\ra\rightarrow {\rm e}^{\int \frac{{\rm d}^3k}{(2\pi)^3}f(k) \left(\hat{a}_k\hat{a}_{-k}-\hat{a}^\dagger_k \hat{a}^\dagger_{-k}\right)}|\Omega\ra\,,
\label{squeezedvac}
\eeq
where~$\{\hat{a}_k^\dagger,\hat{a}_k\}$ denote ladder operators, and~$f(k)$ parametrizes the momentum distribution of particles outside the one-point function. Notice that the introduction of this exponential operator, commonly referred to as the squeezing of the state (see, {\it e.g.}, Ref.~\cite{Berezhiani:2023uwt}),  does not alter \eqref{clinit}.  This makes it clear that, in the Schr\"odinger picture, the question of losing coherence boils down to the possibility of an initial state given by~\eqref{cohstate} evolving into the strongly squeezed one with~\eqref{squeezedvac}. Quantitative analysis can be performed by studying the evolution of perturbations over the classical background that track structure formation throughout the non-linear regime.

Alternatively, instead of a coherent state, DM can be in the form of an \textit{ultra-cold gas of massive particles}. While these two possibilities are indistinguishable at the background level, the evolution of perturbations might be different. In this scenario, since we are dealing with a collection of weakly interacting particles, the DM particles are expected to acquire a velocity dispersion determined by the virial theorem. In reality, without a concrete UV completion of the scalar field model (unlike axions) and an explicit production mechanism, the initial state should be considered a mixture of both coherent and gaseous components.

\begin{figure} 
\centering
\includegraphics[scale=0.26]{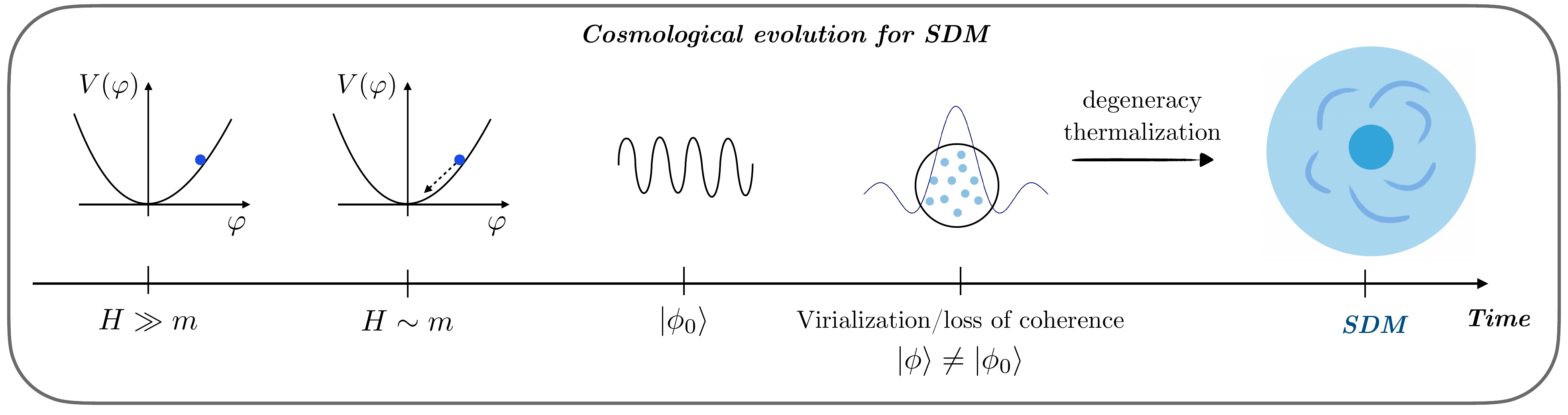}
\caption{
Pictorial representation of the cosmological evolution for SDM.}
	\label{figcosmology}
\end{figure}

A detailed analysis of these effects is beyond the scope of this review and is left for future work.  For the purpose of our discussion, we have simply assumed the worst case scenario for galactic scale superfluidity, namely that virialization entails a significant loss of coherence, and asked what it would take for superfluidity to be re-established within virialized DM halos. For a pictorial illustration of the expected cosmological evolution of the SDM candidate, see Fig.~\ref{figcosmology}.

To be more quantitative, let us briefly outline the systematic procedure for assessing the breakdown of the one-point function description. More fundamentally, the Schr\"odinger equation for the DM field must be promoted to the operator equation
\begin{equation}
\label{opeq}
     {\rm i} \frac{\partial}{\partial t} \hat{\Psi}+\frac{\Delta}{2 m } \hat{\Psi} - \frac{\lambda_4}{4 m^2}|\hat{\Psi}|^2 \hat{\Psi} - m \Phi_\text{\tiny N} \hat{\Psi} = 0\,,
\end{equation}
where we have kept the Newtonian potential~$\Phi_\text{\tiny N}$ as classical, which is a reasonable approximation to assess the leading order departure from coherence. Thus, for our purposes,~$\Phi_\text{\tiny N}$ is sourced by the classical part of the density operator 
\begin{equation}
\label{poissonpert}
    \Delta \Phi_\text{\tiny N}=4\pi Gm |\la \hat{\Psi} \ra |^2\,.
\end{equation}
By computing the expectation value of \eqref{opeq}, it is straightforward to recognize that~$\la \hat{\Psi} \ra$ obeys the classical Schr\"odinger equation supplemented with quantum corrections that depend on higher-order correlation functions evaluated at coincidence.\footnote{Here, the contributions that are sensitive to UV cutoff can be dropped, as they must be subtracted off by the renormalization procedure, which we will not go into in this work.} 

At leading order, the departure from coherence is captured by substituting the decomposition~$\hat{\Psi}\equiv\la \hat{\Psi} \ra+\hat{\psi}$, and keeping terms linear in the perturbation~$\hat{\psi}$. The result is
\beq
{\rm i} \frac{\partial}{\partial t} \hat{\psi}+\frac{\Delta}{2 m } \hat{\psi} - \frac{\lambda_4}{4 m^2}\left(2 | \la \hat{\Psi} \ra |^2 \hat{\psi}+\la \hat{\Psi} \ra ^2 \hat{\psi}^\dagger\right) - m \Phi_\text{\tiny N} \hat{\psi} = 0\,.
\label{pert}
\eeq
It must be stressed that our primary interest here is not the classical growth of perturbations but rather the deviation from coherence as it occurs. Therefore, the c-number functions~$\la \hat{\Psi} \ra$ and~$\Phi_\text{\tiny N}$ in Eq.~\eqref{pert} are assumed to be solutions of the classical equations of motion through the non-linear process of structure formation/virialization.\footnote{
Depending on the scales of interest, Eq.~\eqref{pert} can be simplified accordingly. For instance, noting that~$| \la \hat{\Psi} \ra |^2\sim \rho/m$ and~$\Phi_\text{\tiny N}\sim G\rho k^{-2}$, it is straightforward to see that the last term in Eq.~\eqref{pert} can be neglected on scales significantly shorter than the Jeans scale. This is indeed a valid approximation, since we are interested in the plausibility of the production of particles with de Broglie wavelength set by the characteristic CDM velocity, and thus possibly smaller than the Jeans scale.} Solving for the mode functions of~$\hat{\psi}$ in the background of~$\Phi_\text{\tiny N}$ and~$\la \hat{\Psi} \ra$ would allow us to calculate the tree-level correlation functions, which in turn can be used to calculate the depleting backreaction on the one-point function itself.

To compute such backreaction one can use the corresponding equation for the one-point function itself, that can be easily fished out of Eq.~\eqref{opeq} to be of the form
\begin{equation}
\label{Psibackreaction}
     {\rm i} \frac{\partial}{\partial t} \langle\hat{\Psi}\rangle+\frac{\Delta}{2 m } \langle\hat{\Psi}\rangle - \frac{\lambda_4}{4 m^2} |\langle\hat{\Psi}\rangle|^2  \langle \hat{\Psi} \rangle - m \Phi_\text{\tiny N} \langle\hat{\Psi}\rangle =  \frac{\lambda_4}{4 m^2}\left(2\langle |\hat{\psi}|^2 \rangle  \langle \hat{\Psi}\ra+\langle \hat{\psi}^2   \rangle  \la \hat{\Psi} \ra^* \right)\,,
\end{equation}
where we have used $ \langle \hat{\psi} \rangle = 0$, and collected on the right-hand side  the leading terms sourcing the loss of coherence. Notice that, in principle,  the Newtonian potential operator is determined in terms of~$|\hat{\Psi}|^2$  through the Poisson equation and correspondingly contributes to depletion. While this would be the only source for loss of coherence for fuzzy DM, it might be less relevant for our scenario and we have suppressed it for illustrative purposes.
Following this procedure, {\it i.e.}, solving for the backreaction effect from~\eqref{Psibackreaction}, would allow one to assess the transfer of energy from the coherent component to the gaseous one.

In the context of fuzzy DM, quantum corrections to the coherent classical evolution have been numerically studied in Refs.~\cite{Eberhardt:2022exf, Eberhardt:2023axk} in the non-relativistic approximation. There, the authors studied the spread of the wavefunction around its peak value, and found that there is an exponential growth in configurations undergoing a classical non-linear evolution. In contrast, the growth is only quadratic in time for stable, virialized systems.~\footnote{This agrees with the general statement of Ref.~\cite{Dvali:2013vxa} which shows that, for systems exhibiting classical (or Lyapunov) instabilities, the timescale over which classical evolution remains valid scales logarithmically with the number of constituent particles of the system---a large number for coherent configurations. In contrast, for stable systems, this timescale follows a power-law scaling. As a result, the breakdown of classicality occurs much more rapidly in the presence of classical instabilities~\cite{Dvali:2013vxa,Dvali:2013eja}.} These numerical studies conclude that the loss of coherence is always a subleading effect in fuzzy DM. In the context of SDM, on the other hand, the different mass range and the role of self-interactions may play a more important role in the way the loss of coherence is realized.
Therefore, further investigations are essential to carry out this computation in details, ensuring the validity of this framework and refining the understanding of the cosmological evolution of the SDM candidate. 

We conclude this Section by mentioning DM decoherence. While the loss of coherence discussed above refers to the tendency of a quantum system to deviate from its classical evolution due to internal interactions, decoherence denotes, in contrast, the process by which a quantum subsystem becomes classical due to interactions with its surrounding environment. This phenomenon has been studied in  Ref.~\cite{Allali:2020ttz}, showing that Bose-Einstein condensates of sub-eV bosons can  undergo decoherence in a Hubble time as a result of gravitational interactions with their environment.

\section{Phenomenological implications}
\label{sec: pheno}

The realization of a superfluid state in galaxies leads to important effects and various phenomenological implications for the dynamics of cosmic structures. 
In this Section we are going to review several aspects of SDM phenomenology in galaxies, in particular existing observational constraints and unique predictions of this model.\footnote{For instance, the interplay between a SDM state and cosmic strings was recently investigated in Refs.~\cite{Bernardo:2023ehz, Favero:2024xwh}.}

\subsection{Merging galaxies: Bullet Cluster}
\label{sec:Bullet}

One of the most stringent bounds on DM self-interactions comes from the Bullet Cluster~\cite{Markevitch:2003at, Randall:2008ppe, Kahlhoefer:2013dca}. 
This is a system of two colliding galaxy clusters, in which the less massive cluster---the {\it bullet}---has undergone an almost head-on collision with the more massive {\it target}. One of its most striking features is the spatial offset between the baryonic gas, observed in X-ray, and the DM distribution, inferred from gravitational lensing. The baryonic gas lags behind due to ram pressure experienced during the collision, while the lensing peaks align with the distribution of the visible galaxies, which are not significantly slowed down and pass through the interaction region largely unaffected. This observed offset is consistent with DM having negligible self-interactions during the collision.

With this in mind, we focus on understanding the Bullet Cluster constraint in the context of ultralight DM particles, following Refs.~\cite{Berezhiani:2021rjs,Berezhiani:2022buv}.  The DM components of the two clusters behave as a collisionless system if, on average, a particle from the bullet experiences fewer than one scattering event while passing through the target.  This translates into the condition
\begin{equation}
    \langle n_\text{\tiny sc}\rangle<1\,, 
    \label{eq:Nsc1}
\end{equation}
where the number of scattering~$n_\text{\tiny sc}$ is averaged over the DM particles in the collision. 
This can be estimated using the scattering rate~$\Gamma$, according to 
\begin{equation}
     \langle n_\text{\tiny sc}\rangle=\Gamma\frac{2R_\text{\tiny V}^\text{\tiny target}} {v}\,,
     \label{eq:Nsc2}
\end{equation}
where~$v \simeq 10^{-2}$ is the infall velocity of the bullet, and~$R_\text{\tiny V}^\text{\tiny target} = 2$ Mpc is the virial radius of the target cluster~\cite{Clowe:2003tk}.
Furthermore, for the estimates that follow, the characteristic density of the target cluster halo is assumed to be~$\rho_\text{\tiny target}\sim 10^{-25}$ g/cm$^3$, obtained by averaging over a region of 500~kpc from its center~\cite{Clowe:2003tk}.

The upshot of Eqs.~\eqref{eq:Nsc1} and~\eqref{eq:Nsc2} is that the Bullet Cluster system effectively places a constraint on the interaction rate~$\Gamma$. The resulting constraint on the model parameters depends on the specific DM theory under consideration, but also on the phase space distribution of DM within halos.

Let us consider a DM theory in which non-degenerate particles interact mainly through 2-body interactions. In this case, the constraint on the scattering rate implies the familiar upper bound on the scattering cross section per unit mass:
\begin{equation}
    \frac{\sigma}{m} \lesssim 1 \frac{\text{cm}^2}{\text{g}}\,,\qquad \text{if}~~ \Gamma =\rho_\text{\tiny target} \frac{\sigma}{m} v\,.
\end{equation}  
However, in the context of sub-eV DM particles, degeneracy at typical galactic density implies that the interaction rate is enhanced by the Bose-factor~$\mathcal{N}$ in the same way as discussed for thermalization in Sec.~\ref{sec:therm}. Taking this into account results in a tighter bound:
\begin{equation}
    \frac{\sigma}{m} \lesssim \frac{1}{\mathcal{N}} \frac{\text{cm}^2}{\text{g}}\simeq 10^{-2} \left(\frac{m}{\text{eV}}\right)^4 \frac{\text{cm}^2}{\text{g}}\,,\qquad\text{if} ~~\Gamma= \mathcal{N} \rho_\text{\tiny target} \frac{\sigma}{m} v\,.
    \label{eq:deg2b}
\end{equation} 
In the context of the Bullet Cluster system, the enhancement only applies to particles in the degenerate outskirts, outside of the thermal radius, since particles comprising the superfluid phase are mostly in the single ground-state level.\footnote{Here, we assume finite temperature effects to be negligible. This assumption holds since the virial temperature is several orders of magnitude lower than the critical temperature, for the particle masses under consideration. If this is not the case, the fraction of the fluid at finite temperature  could contribute in a non-negligible way during the collision of halos.} Therefore, the assumption underlying Eq.~\eqref{eq:deg2b} is that the out-of-equilibrium component in the outskirts is non-negligible compared to the superfluid component.

This assumption is violated if the interaction rate is sufficiently high that the entire halo has reached global thermal equilibrium. In that case, the bound must be relaxed, since the collision of two clusters would just correspond to the interaction of two highly coherent configurations. Therefore, we expect that the bound \eqref{eq:deg2b} relaxes to its non-degenerate version if~$R_\text{\tiny T}>R_\text{\tiny V}^\text{\tiny target}$, which is achieved for cross sections satisfying the condition~$\frac{\sigma}{m}>3 \left(\frac{m}{\text{eV}}\right)^4 \frac{\text{cm}^2}{\text{g}}$~\cite{Berezhiani:2021rjs,Berezhiani:2022buv}.

We can therefore summarize the Bullet Cluster bound for degenerate SDM in the following three points:
\begin{itemize}
    \item Cross sections satisfying 
    \begin{equation}
         \frac{\sigma}{m} \lesssim  10^{-2} \left(\frac{m}{\text{eV}}\right)^4 \frac{\text{cm}^2}{\text{g}}
    \end{equation}
    are allowed. For these combinations of masses and cross-section, the DM halos of clusters contain a region in the superfluid phase and a degenerate outskirt. Although the interaction between outskirts is Bose-enhanced, this would still result in negligible DM self-interactions during the cluster collision.
    \item Cross section satisfying
    \begin{equation}
           10^{-2} \left(\frac{m}{\text{eV}}\right)^4 \frac{\text{cm}^2}{\text{g}} \lesssim  \frac{\sigma}{m} \lesssim 3 \left(\frac{m}{\text{eV}}\right)^4 \frac{\text{cm}^2}{\text{g}}
    \end{equation}
    are not allowed. This case is similar to the previous one, but during the collision, the degenerate outskirts interact significantly.
    \item Lastly, cross-sections within the range 
    \begin{equation}
           3 \left(\frac{m}{\text{eV}}\right)^4 \frac{\text{cm}^2}{\text{g}} \lesssim  \frac{\sigma}{m} \lesssim  1 \frac{\text{cm}^2}{\text{g}}
    \end{equation}
    are allowed.
    In this case, clusters have reached global thermal equilibrium and the bound reverts to a non-degenerate case. This is because the collision of the two clusters would be similar to the scattering of two highly coherent beams of DM particles. 
\end{itemize}
The Bullet Cluster constraint translates to bounds on the allowed parameters for SDM, shown by the red band in Fig.~\ref{ch3:general2}.

Let us stress that the approximation underlying Eq.~\eqref{eq:Nsc2} is valid in the context of SDM provided that the infall velocity exceeds the sound speed of the superfluid. In this regime, the collision between the two superfluid halos can be approximated as a head-on collision of on-shell constituent particles, which is also the standard approximation typically adopted for CDM. If this condition is not met, the correct degrees of freedom describing the collision are phonons with an almost linear dispersion relation. In this case, the dynamics of the process can no longer be described as the scattering of individual constituents; rather, it must be analyzed in terms of interactions between the background fields, the resulting collective (and soft) phonon excitations, and potentially significant surface effects. However, in SDM, the sound speed is constrained to remain below $c_s\sim10^{-3}$ in order to satisfy the bound from matter–radiation equality~\eqref{eq:MR}, so it is at least one order of magnitude smaller than the infall velocity. This justifies the approximation employed in this Section of considering the 2-body scattering of constituents as the leading interaction channel between the solitons.
A detailed discussion of why the scattering between superfluids with relative velocities much greater than the sound speed reduces to the standard constituent scattering is presented in Appendix~\ref{sec:Scattering}.

We conclude this Section by mentioning that it is possible to derive an analog bound on the strength of DM self-interactions for three-body interacting superfluids. To do so, we should use the interaction rate of Eq.~\eqref{eq:3bodyrate}. In this case, one finds that the following combination of parameters is excluded
\begin{equation}
   \left( \frac{m}{\text{eV}} \right)^{11/2} \text{keV}^{-2}\lesssim \lambda_6 \lesssim 10^3 \left(\frac{m}{\text{eV}}\right)^{11/2} \text{keV}^{-2},\qquad \text{if} \qquad m\ll \text{eV}\,.
\end{equation}
For these values of mass and coupling, DM in the degenerate and out-of-equilibrium outskirts is interacting. 
In particular, the right inequality is saturated for clusters that are in global thermal equilibrium, under 3-body interactions.
If DM is either non-degenerate or has achieved a thermal configuration in clusters, then the excluded parameter space reverts to the non-degenerate version: 
\begin{equation}
    \lambda_6 \gtrsim 30 \left(\frac{m}{\text{eV}}\right)^{3/2} \text{keV}^{-2}\,.
\end{equation}
The dimensionful nature of the coupling constant indicates the characteristic energy-momentum transfer at which scattering amplitudes of DM particles begin to violate unitarity, and thus require UV completion. However, due to the fact that the DM candidate in question is not a thermal relic, the smallness of the cut-off scale has more pressing repercussions for the equation of state. Indeed, as shown in~\cite{Berezhiani:2022buv}, in order for the three-body case to exhibit a non-relativistic equation of state at matter-radiation equality, while providing superfluid cores of galactic size, the self-interaction potential needs to be suitably modified at high density (see Sec.~\ref{sec: earlyuniv} for related discussion of the two-body case).

\subsection{Rotation curves}
\label{sec:rotationcurves}
The shape of galaxy rotation curves provides one of the most compelling evidence for the existence of DM at galactic scales. These curves describe the orbital velocity of stars and gas moving in the galactic environment, as a function of their distance from the galactic center.
Historically, DM was proposed to account for the discrepancy between the observed and expected rotation curves. In the absence of additional mass, Newtonian dynamics predicts a Keplerian fall-off in the rotation curve beyond the baryonic component. Instead, observations show that rotation curves remain approximately flat at large radii, implying the presence of a dominant, non-luminous mass component extending well beyond the visible disk. See Ref.~\cite{Sofue:2000jx, Bertone:2016nfn} for some historical reviews.

In this Section, we discuss rotation curves in the context of SDM. First, it is important to recall that the original proposal of SDM with quartic interactions of Ref.~\cite{Goodman:2000tg} was deemed to be ruled out in Ref.~\cite{Slepian:2011ev} due to the inability of the model to reproduce realistic galactic rotation curves.
The analysis was based on the key assumption of global thermal equilibrium of the halo, taken to have a central superfluid core in equilibrium with an outer region of regular DM (with the entire system described by an equation of state that interpolates between the two distinct phases). The main shortcoming of the model is that, for cross sections compatible with the Bullet Cluster bound, the central core is too dense compared to the thermal outskirt, and rotation curves have a shape that is not compatible with observations.

We revisit the analysis above following up on the discussion of Sec.~\ref{sec:fragmentation}. First, we note that it is possible to find a viable region of parameter space where a superfluid phase forms in the inner regions of the halo, while the outer regions remain out of equilibrium, breaking global thermal equilibrium. Additionally, accommodating a single superfluid core surrounded by a normal-phase envelope is difficult, as multiple solitons are more likely to form, with most prone to tidal disruption. This leads to superfluid debris outside the core, which is unlikely to have a simple equation of state or remain in equilibrium with the core. However, as we have pointed out in Sec.~\ref{tidal subsec}, they are expected to be in the form of weakly-interacting virialized streams with coarse-grained density distribution resembling a NFW profile. As such, the rotation curves of large, Milky Way-like galaxies should be similar to CDM. 
Deriving the detailed halo density distribution is challenging and will likely require numerical simulations.

Nonetheless, there are still cases where we can make some qualitative predictions about the halo density distribution. In particular, while the dynamics of larger galactic halos, such as that of the Milky Way, are expected to be dominated by the out-of-equilibrium and debris components, this is not the case for {\it dwarf galaxies}.  By adjusting the parameters of the quartic superfluid theory, it is possible to have a central core extending approximately up to~$10-15$ kpc from the galactic center, thereby engulfing the majority of the DM fraction of dwarf galaxies, according to the relation 
\begin{equation}
\lambda_{\rm J}=2\left(\frac{\sigma/m}{\text{cm}^2/\text{g}}\right)^{1/4}\left(\frac{m}{\text{eV}}\right)^{-5/4} \text{kpc}\,,
\end{equation}
describing the diameter of the central soliton. Therefore, we expect the DM content for small halos to be characterized by the density profile in \eqref{eq:2bodyprofile}, assuming the baryonic density to be negligible throughout the halo. This density distribution would lead to rotation curves that exhibit a slow rise, increasing up to a radial distance~$\simeq \lambda_{\rm J}/2$. Furthermore, the typical plateau associated with larger galaxies should not be present, as there would be no outskirts in these types of galaxies. This is the typical shape expected in low surface brightness galaxies, such as IC2574. Given that the size of the central core is density-independent for quartic superfluids, this shape is expected to be universal for all  halos whose central dynamics is  DM dominated.

However, this is not always the case, as the central soliton may get distorted by the presence of baryons. In fact, it is desirable for this distortion to occur, considering that not all dwarf galaxies exhibit the same shape in their rotation curves~\cite{Oman:2015xda}.
To quantify the effect of baryons, let us introduce in this discussion the following two-parameter “spherical exponential” profile~\cite{Berezhiani:2017tth}
\begin{equation}
    \rho_\text{\tiny b}(r)=\frac{\gamma_\text{\tiny b}}{8\pi L^3} e^{-r/L}\,,
\end{equation}
which we use as a toy density distribution for baryons.
The scale parameter~$L$ controls the spread of the baryon density distribution, while~$\gamma_\text{\tiny b}$ sets the central baryon density. By keeping the baryon density fixed, we can adjust the DM density distribution in such a way that the two components are in hydrostatic equilibrium, thus obtaining a stationary configuration of the halo. This is done under the assumption that the DM equation of state is a polytropic equation of index $n=1$, which is a reasonable assumption as long as we focus on radii $r<\lambda_{\rm J}/2$, {\it i.e.}, well within the superfluid core.

\begin{figure}
\centering
	\includegraphics[scale=0.5]{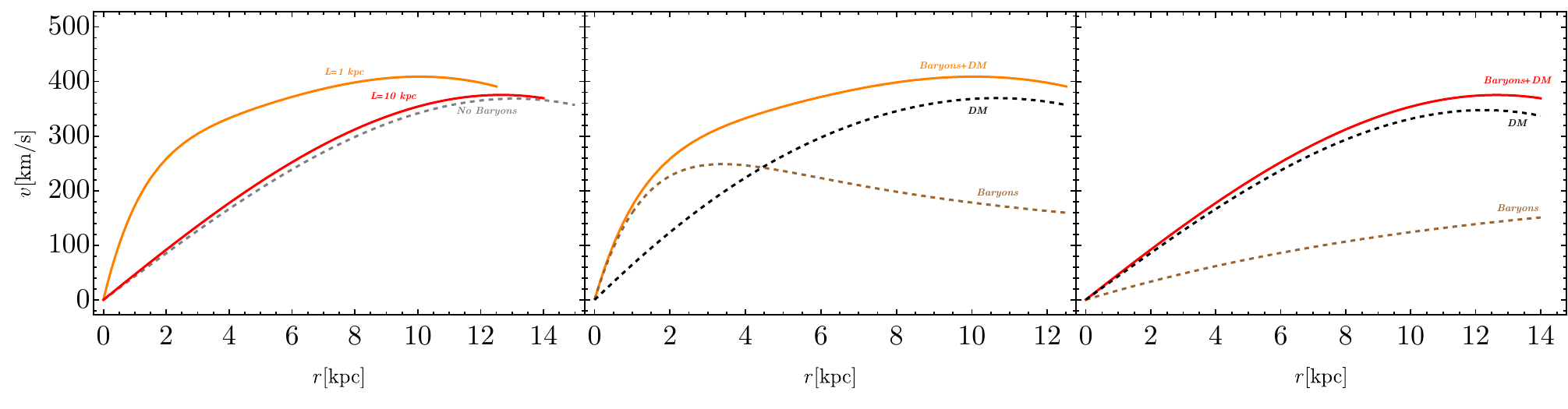} 
\caption{The left panel displays the rotation curves for a quartic superfluid in the presence of baryons, with the gray, red, and orange lines corresponding to the cases of no baryons, a homogeneous baryon distribution, and a cuspy baryon distribution, respectively. In the absence of baryons, the total DM mass is rescaled by a factor of 1.2 to account for the missing mass. The central and right panels illustrate the partial contributions to the rotation curves from baryons (brown lines) and DM (black lines), for the cuspy and homogeneous baryon distributions, respectively. Figure inspired from Ref.~\cite{Berezhiani:2022buv}. }
\label{rotationcurves}
\end{figure}

The rotation curves derived from these density distributions are shown in the left panel of Fig.~\ref{rotationcurves}, taken from Ref.~\cite{Berezhiani:2022buv}. These are derived assuming a fixed DM and baryon mass of $M_\text{\tiny DM}=10^{10} M_{\odot}$ and $M_\text{\tiny b}=2\times 10^9 M_{\odot}$, respectively. The main difference is provided by the parameter $L$, which takes the value $L=1$~kpc (cuspy baryon distribution, orange line) and $L=10$~kpc (homogeneous baryon distribution, red line). The factor $\gamma_\text{\tiny b}$ is fixed accordingly to the total baryon mass and the scale parameter, while the theory parameters are fixed in order to have a central soliton of size $\lambda_{\rm J} = 15$ kpc in the absence of baryons. The plots show that, if the baryon distribution is negligible all over the halo (see right panel), the homogeneous core observed in the DM-only scenario is maintained, resulting in a gradual rise of the rotation curve. In contrast, when baryonic matter dominates at the center (see central panel), the superfluid soliton shrinks and develops a cuspy profile, resulting in a sharper rise in the rotation curve.
The shrinking of the soliton is evident by checking that the DM contribution to the rotation curve starts to decay at smaller radii for $L=1$ kpc compared to $L=10$ kpc.\footnote{This is similar to the way the diversity of rotation curves is explained in the context of self-interacting DM, see Ref.~\cite{Kamada:2016euw}.}

In conclusion, further analyses incorporating physically motivated baryonic models are essential to robustly assess the viability of the SDM parameter space in light of constraints from galaxy rotation curves.

\subsection{Structure and dynamics of superfluid dark matter halos}
In this Section, we present an overview of the qualitative expectations for the
structure and dynamics of DM halos within the SDM framework, based on the
analytic considerations developed thus far. Firm, quantitative predictions will
ultimately require dedicated numerical simulations—which we leave for future
work—especially to capture the intricate dynamics of tidal disruption within
halos and the subsequent evolution of the superfluid debris. Nevertheless, the
existing analytic treatment already points toward several generic features of
the model. In particular, cores are naturally formed in SDM due to the formation
of central solitons, providing a potential resolution to the cusp-core problem
observed in galaxies. We outline the expected scaling behaviors, discuss the
tendency toward cored versus cuspy density profiles, and comment on how SDM
may account for the observed diversity of galactic rotation curves. These
considerations sketch the main phenomenological implications of SDM on galactic
scales and help guide the direction of future, simulation-based studies.

Following the detailed discussion in Sec.~\ref{Sec: SDMgalaxies}, the main
expectation for the central regions of DM halos in the SDM framework is
the formation of a central soliton, supported by the balance between repulsive
self-interactions and gravity. The size and mass of this core depend sensitively
on the microscopic theory---in particular on the equation of state and the
density-dependence of the sound speed $c_s(\rho)$, which can be nontrivial in
models beyond the simple quartic case (where $c_s$ depends linearly on the density). Surrounding the
soliton, the superfluid phase is generically prone to gravitational
fragmentation, governed by the Jeans scale of the condensate. For a superfluid
with density $\rho$, the Jeans scale reads from Eq.~\eqref{IPJS}, 
\(\lambda_{\rm J}\simeq \sqrt{\frac{\pi c_s^2}{G \rho}}\), 
which also sets a characteristic minimum mass for the halo, 
$M_\text{\tiny DM halo} \gtrsim \frac{4\pi}{3}\,\rho\,\lambda_{\rm J}^{3}$.
This implies that halos capable of hosting a central soliton cannot be lighter
than this quantity, which depends sensitively on the microphysical parameters of the theory. Most of the resulting
droplets from fragmentation are expected to experience strong tidal disruption,
dissolving into extended streams of superfluid debris rather than surviving as
long-lived bound clumps. Capturing this nonlinear, multi-scale dynamics---
especially the shredding, mixing, and re-condensation of tidally disrupted
fragments---requires numerical simulations.

We also emphasize that outliers may exist: in particular, regions near
the outskirts of halos may fail to form a coherent soliton, and satellite halos
may instead originate from the condensation of tidal superfluid streams far from
the central core, with masses smaller than $\frac{4\pi}{3}\,\rho\,\lambda_{\rm J}^{3}$. Although these analytic considerations outline a qualitative
picture---a central soliton, a fragmented superfluid, and debris-dominated
outskirts---the detailed structure of halos in SDM ultimately awaits numerical
simulations capable of resolving tidal disruption, phase transitions, and the
evolution of superfluid streams, both within the central and far regions of DM halos. This preliminary understanding, however, allows
us to begin addressing a few of the common puzzles and features that
characterize CDM, as well as other DM candidates such as fuzzy DM, in terms of
halo structure, core formation, and diversity of galactic rotation curves.

Let us start by discussing the \emph{diversity problem} in galactic rotation
curves. In standard CDM, galaxies with similar total halo masses are predicted
to host relatively self-similar, cuspy density profiles
\cite{Oman:2015xda, Oman:2017vkl, Santos-Santos:2019vrw, Roper:2022umd, Read:2016xbf},
which implies a narrow range of circular velocities at fixed outer halo
velocity. Observations, however, show a much wider spread in inner rotation
curves, indicating that CDM alone may not fully capture the processes shaping
the inner halo structure. Baryonic feedback can partially modify central
densities, but it is likely insufficient to explain the full observed diversity,
particularly in low-mass or DM-dominated systems
\cite{Santos-Santos:2019vrw}. In models of ultralight DM, such as fuzzy DM, one expects the presence of central solitonic cores embedded in a larger halo. In this theory,
simulations show that the core-halo relation exhibits substantial scatter
\cite{Chan:2021bja, Schive:2014hza}, primarily due to the ambiguous transition
between the solitonic core and the outer NFW-like halo, as well as due to the halo's
merger history, tidal interactions, and relaxation state. Despite this, halos
always host a central core, and including baryonic effects may further modify
the core-halo relation. This highlights that the inner halo structure in fuzzy DM is
highly sensitive to both microphysics and dynamical history, offering a useful
point of comparison for SDM, where similar effects are likewise expected to play a role. As we discussed in Sec.~\ref{sec:rotationcurves} on rotation curves, the presence of baryons in
halos could similarly modify the core properties in SDM, as illustrated in Fig.~\ref{rotationcurves}. Just as in fuzzy DM, where the core-halo structure is sensitive to merger
history, tidal interactions, and relaxation state, the inner structure of SDM
halos can be affected by both the superfluid microphysics and baryonic processes.
This analogy suggests that the diversity of inner halo properties, including
core size and density, may naturally arise in SDM, providing a framework to
interpret the range of observed galactic rotation curves.

A further aspect worth considering is the {\it scaling relation} between core density
and core radius in galaxies. Not only should these models predict the existence
of galactic cores, but they must also consistently account for the cores observed
across many---if not all---galaxies. While model parameters can be adjusted to
fit the core of a single galaxy, a robust model must accurately reproduce the
observed scaling relation across a broader population of galaxies. In the case
of the quartic superfluid theory, the prediction for the core would appear as
roughly a vertical line in a plot of core density versus core radius. However,
in general it is not straightforward to clearly identify the core, due to the
presence of superfluid debris and fragmented structures surrounding the central
region.
The general problem of reproducing the observed core
density–radius relation has been investigated both in the context of fuzzy and
superfluid DM models~\cite{Deng:2018jjz, Burkert:2020laq, Pozo:2020ukk, Benito:2025xuh,
Hartman:2022cfh, Blum:2025aaa}. For the latter, recent hydrodynamical simulations indicate that the scaling
relations of simulated SDM halos fail to match those of Milky Way dwarf
spheroidals and nearby galaxies---even for kpc-scale core radii---thereby
falling short of resolving the cusp-core and too-big-to-fail problems
simultaneously~\cite{Hartman:2022cfh}. A thorough understanding of the SDM
scaling relations remains an open and active area of research. In particular,
incorporating baryonic physics is necessary for a more definitive comparison
with data. Improvements along these directions are left to future work.

As a final point, we comment on {\it ultra‑faint dwarf galaxies}, the smallest
and most DM‑dominated galaxies known. Their stellar half‑light radii
are only of order tens of parsecs (often $\lesssim 50$\,pc), and their velocity
dispersions are extremely small, implying mass‑to-light ratios among the highest
observed. Because of their low stellar content and high DM densities,
ultra‑faint dwarf galaxies provide a nearly “clean” laboratory for testing wave‑like (ultra‑light) candidates with minimal baryonic complications. They have been used to place informative constraints on the particle mass of
fuzzy DM, mainly due to dynamical heating from wave interference in the
DM, which is mostly set by the de-Broglie wavelength of the particles~\cite{Dalal:2022rmp, May:2025ppj, Yang:2025bae, Eberhardt:2025lbx}.
Additional effects, such as tidal forces from the Milky Way, stochastic interference,
soliton fluctuations, and tidal stripping, can modify the expected heating of stars
and must be carefully considered when interpreting constraints from these systems~\cite{Dalal:2022rmp, May:2025ppj, Yang:2025bae, Eberhardt:2025lbx}.

Let us now discuss the formation of these galaxies in the context of SDM. As shown above,
the central soliton typically has a size of order kpc for certain choices of the
microscopic parameters, such as the particle mass and self-interaction strength.
One possible mechanism to form bounded, core‑like satellite structures is through
the superfluid debris surrounding the central soliton. For such debris, the
coherence scale is not simply given by the Jeans scale of Eq.~\eqref{IPJS}, since
the assumption of a self-gravitating, spherically symmetric structure is broken
({\it e.g.}, due to tidal stripping from the host halo). In general, the relevant
coherence scale of the debris is expected to be smaller, likely set by the tidal
radius of the forming satellite, giving rise to potentially small structures, $\lambda_{\rm J}^\text{\tiny satellites} \ll \lambda_{\rm J}$. If the mini halo forms instead in the field, rather than as a satellite of a
larger system, we expect a genuine central core to develop. However, its size may
be relatively small because the central density in such isolated mini halos is
typically low, resulting in correspondingly small core masses. Once such satellite or field structures form, it is unclear whether tidal heating
occurs in the same way as in fuzzy DM. In SDM, the effective de-Broglie
wavelength is much shorter because the particle mass is significantly larger than
in fuzzy DM, and typical virial velocities are similar, thus potentially alleviating the effects on stars. Moreover, self‑interactions
can further modify the dynamics. Consequently, more detailed studies and
dedicated simulations are needed to assess whether bounds analogous to those
derived from ultra‑faint dwarf galaxies in fuzzy DM can be established for SDM, and if so, how they
differ in magnitude and scaling.

Further work is needed to fully address these
issues, particularly through numerical simulations capable of resolving tidal
disruption, debris evolution, and core formation in diverse halo environments.

\subsection{Portals to Standard Model}
\label{sec: axions}
In the foregoing Sections, we have presented a model in which DM is constituted by a self-interacting scalar field, whose dynamics and phenomenology are determined entirely within the dark sector. Importantly, in its simplest incarnation, the model is minimal in the sense that:
\begin{itemize}
    \item There are no a priori couplings linking the dark sector to the SM. In particular, we do not assume any portal interactions ({\it e.g.}, Higgs, vector, or axion-like portals);
    \item The dark sector evolution proceeds purely through its self-interactions and gravitational coupling to the SM.
\end{itemize}
Because of this minimal construction, all conventional bounds that rely on energy transfer to SM particles are automatically evaded. In this sense, the dark fluid can be viewed as a self-contained and gravitationally coupled component of the universe.

However, the absence of SM couplings is a choice of minimality, not a fundamental requirement. If one were to introduce portal interactions, the model would naturally fall into the broader class of scalar or axion-like DM scenarios, though it differs in key aspects---most notably by exhibiting stronger self-interactions. In that case, the model can be subjected to associated astrophysical and laboratory constraints.

To draw an analogy with alternative DM model in the same ultralight mass range, one can consider the QCD axion~\cite{Weinberg:1977ma, Wilczek:1977pj}, as briefly discussed in Sec.~\ref{Intro-models}. The QCD axion is the pseudo Nambu–Goldstone boson $a$ of a spontaneously broken Peccei–Quinn (PQ) symmetry~\cite{Peccei:1977hh, Peccei:1977ur}, with potential arising from non-perturbative effects of the form~\cite{Vafa:1983tf}
\begin{equation}
V(a) = m_a^2 f_a^2 \left[1 - \cos \left(\frac{a}{f_a} \right) \right]\,,
\end{equation}
in terms of the axion mass $m_a$ and PQ symmetry-breaking scale $f_a$, related by the condition~\cite{Gorghetto:2018ocs}
\begin{equation}
    m_a \simeq 5.7~\mu\mathrm{eV} \left( \frac{10^{12}\,\mathrm{GeV}}{f_a} \right)\,.
\end{equation}
By expanding the potential for  small field amplitudes, one gets
\begin{align}
V(a) = \frac{1}{2} m_a^2 a^2 - \frac{1}{4!}\frac{m_a^2}{f_a^2} a^4 + \mathcal{O}(a^6)\,,
\end{align}
which corresponds to an attractive quartic self-interaction
\begin{equation}
\label{lambda-axions}
\lambda_4 = - \frac{m_a^2}{6f_a^2}\,.
\end{equation}
Equation~\eqref{lambda-axions} highlights a striking contrast between superfluid and axion DM: while the stability of the former relies on strong repulsive self-interactions that counteract gravity, the latter features much weaker ($m_a \ll f_a$) attractive self-interactions that instead enhance gravitational collapse.

The QCD axion can couple to the SM fields (including photons, electrons and neutrons) through higher-dimensional operators of the form (see~\cite{DiLuzio:2020wdo} for a recent review)
\begin{equation}
    \mathcal{L}_{\text{\tiny int}} \supset 
    \frac{g_{a\gamma\gamma}}{4}\, a F_{\mu\nu} \tilde{F}^{\mu\nu}
    + {\rm i} g_{aee}\, \partial_\mu a \, \bar{e}\gamma^\mu\gamma_5 e
    + {\rm i} g_{aNN}\, \partial_\mu a \, \bar{N}\gamma^\mu\gamma_5 N + \cdots\,,
\end{equation}
with typical magnitudes (in terms of $\mathcal{O}(1)$ coefficients)
\begin{equation}
    g_{a\gamma\gamma} \sim \frac{\alpha}{2\pi f_a}\,, \qquad 
    g_{aee} \sim C_e \frac{m_e}{f_a}\,, \qquad 
    g_{aNN} \sim C_N \frac{m_N}{f_a}\,.
\end{equation}
These couplings are extremely small, typically in the range 
$g_{a\gamma\gamma} \sim (10^{-13} \div 10^{-10})\,\mathrm{GeV}^{-1}$, 
but potentially not vanishing, allowing the axion to interact weakly with SM particles while remaining a viable DM candidate~\cite{Adams:2022pbo}. Let us emphasize that an analogous discussion applies to ALPs, which share many features with QCD axions, including similar EFTs and a potential role as DM. However, unlike QCD axions, ALPs do not couple to gluons in the same way and, since their masses are not generated by non-perturbative QCD effects, their mass $m_a$
  and decay constant $f_a$	
  are independent parameters, potentially allowing stronger self-interactions.

If similar interactions were introduced between our scalar and SM fields---for instance, through a dimension-five operator of the form 
$ g_{\phi \gamma \gamma }\phi F_{\mu\nu}\tilde{F}^{\mu\nu}$ 
or via  terms 
$g_{\phi ee}\partial_\mu \phi \, \bar{e} \gamma^\mu \gamma_5 e$---then similar astrophysical and cosmological limits derived for axions and ALPs with sub-eV masses could become relevant to this framework, even though the presence of strong repulsive self-interactions for SDM may change their effective applicability.
The most significant constraints  on axion DM in this mass range include (see Ref.~\cite{Adams:2022pbo} for a more comprehensive review):
\begin{itemize}
     \item {\it Stellar cooling:} Observations of horizontal-branch stars, red-giants in the globular
cluster $\omega$Centauri, and white dwarves constrain the electron coupling to $g_{aee} \lesssim 10^{-13}$ for $m_a \lesssim 10^3~\mathrm{eV}$~\cite{Straniero:2020iyi,Capozzi:2020cbu}.
    
    \item {\it Helioscope searches:} For $m_a \lesssim 0.1~\mathrm{eV}$, the non-observation of solar axions (reconverted into X-ray photons via strong laboratory magnetic fields) implies limits onto the photon coupling of $g_{a\gamma\gamma} \lesssim 0.6\times10^{-10}\,\mathrm{GeV}^{-1}$ (see, {\it e.g.}, the CAST and IAXO experiments~\cite{CAST:2015qbl, Irastorza:2011gs, IAXO:2019mpb}).
    
    \item {\it Cosmological hot DM bounds:} Thermalized axions with $m_a \gtrsim 0.7~\mathrm{eV}$ would contribute excessively to the relativistic energy density, violating CMB and large-scale structure constraints~\cite{Cadamuro:2010cz,Cadamuro:2011fd, Millea:2015qra, Green:2021hjh,Xu:2021rwg}.
    
    \item {\it Laboratory experiments:} Additional laboratory searches (including abracadabra-10 cm~\cite{Ouellet:2018beu,Salemi:2021gck,Ouellet:2019tlz}, SHAFT~\cite{Gramolin:2020ict}, and ADMX haloscopes~\cite{ADMX:2009iij,ADMX:2018gho,ADMX:2019uok,ADMX:2021nhd}) exclude complementary regions in the $(m_a, g_{a\gamma\gamma})$ parameter space.

\item {\it Gravitational waves:}  Couplings between ultralight DM and SM particles can induce oscillations in the fine-structure constant and electron mass, causing tiny periodic changes in interferometer components that modulate arm lengths and the measured gravitational-wave strain, allowing LIGO-Virgo-KAGRA to constrain such couplings~\cite{LIGOScientific:2025ttj}.
    
\end{itemize}

In contrast, in the minimal version of the SDM model, these couplings are set to zero, leaving only gravitational interactions, thus implying that the model remains unconstrained by conventional searches for light, weakly coupled particles (but see discussion in Sec.~\ref{sec:Forces} when considering condensate-induced long-range forces among baryons).

\subsection{Vortices}

One of the hallmark features of superfluids is their ability to form quantized vortices. In the context of DM halos, vortex formation is in fact the only mechanism by which the superfluid can retain the angular momentum initially present prior to the phase transition. This arises from the nature of the condensed phase, which is described by a highly occupied single-particle wavefunction. Consequently, the fluid supports only irrotational, or potential, flow, with the velocity field determined by the gradient of the phase of the wavefunction~$\Psi \propto{\rm e}^{{\rm i}\theta}$ as
\begin{equation}
\vec{v} = \frac{1}{m} \vec{\nabla} \theta \,.
\end{equation}
Famously, this implies vanishing vorticity and velocity circulation everywhere where $\vec{v}$ is smooth. Therefore, the superfluid component cannot rotate uniformly as a normal fluid, instead it can retain the angular momentum by having vortex lines permeating throughout, satisfying the condition
\begin{equation}
\vec{\nabla} \times \vec{v} = \frac{1}{m}  2\pi \delta^{(3)} \big(\vec{r} - \vec{r}_0\big) \hat{z}\,,
\end{equation}
where the singularity describes the core of a vortex, centered at~$\vec{r}_0$, and whose lines are located along a given direction~$\hat{z}$ associated to the rotation axis of the superfluid. In other words, the rotation of a superfluid is inhomogeneous through the formation of quantized vortices~\cite{tsubota2010quantized, Tsubota_2013, Barenghi2001QuantizedVD}.

Let us outline some of the properties of superfluid vortices with the concrete example of a self-interacting complex scalar field as a prototype model, governed by~\eqref{eq:LagrPhi4}. (We could just as well consider the non-relativistic Gross-Pitaevskii approximation that would also apply to the real scalar field, albeit only in the non-relativistic limit.) Vortices are axisymmetric topological defects that require vacuum manifold with nontrivial first homotopy group. They are configurations that map points on the vacuum manifold to the axisymmetric spatial boundary. In the case of interest, this is provided by the spontaneously broken~$U(1)$ symmetry. For a complex scalar field, this symmetry is self-evident. For a real scalar field, on the other hand, it corresponds to the approximate~$U(1)$ symmetry of the non-relativistic approximation responsible for particle number conservation.

The configuration of the superfluid, distorted by the presence of the vortex line in the center, can be obtained by modifying the homogeneous configuration \eqref{sect2:fieldconf} and introducing nontrivial angular winding of the phase. This in turn requires a non-trivial radial dependence of the modulus of the field to avoid a singularity at the center. The resulting ansatz is
\begin{equation}
\Phi_0(t)=vf(r)\,{\rm e}^{{\rm i}\left(\mu t+n\varphi\right)}\,; \qquad \mu=\sqrt{m^2+\lambda_4 v^2}\,,
\label{vortex_ansatz}
\end{equation}
where~$r$ is the distance from the vortex,~$\varphi$ is the polar angle, and~$n$ is the integer winding number.\footnote{The winding number should not be mistaken for the number density of Eq.~\eqref{numberdensity}.} The function~$f(r)$ determines the profile of the configuration that interpolates from the vortex line to the asymptotically unperturbed superfluid ({\it i.e.},~$f(r\rightarrow\infty)=1$). Substituting the above ansatz into~\eqref{eq:eomPhi} gives the equation for~$f$:
\begin{equation}
\frac{1}{r}\frac{{\rm d}}{{\rm d} r}\left( r \frac{{\rm d} f}{{\rm d} r}\right)-\frac{n^2}{r^2}f+2m^2 c_s^2 f\left(1-f^2\right)=0\,.
\end{equation}
Although the analytic solution is unknown, it is straightforward to find approximate expressions in certain limits. Far from the defect, the configuration approaches the unperturbed superfluid exponentially fast, with~$f(r\rightarrow \infty)=1-{\rm e}^{-2mc_s r}$, with the exponential fall-off dictated by the healing length~$\xi = (2 m c_s)^{-1}$. Deep in the core of the vortex, on the other hand, $f(r\rightarrow 0)\propto r^n$, which reproduces the approximate behavior~$\rho \propto r^2$ for the simplest vortex of winding~$n =1$~\cite{Rindler-Daller:2011afd,Hui:2020hbq}.

As we can see from the asymptotic configuration, the size of the vortex is set by the healing length~$\xi$ of the condensate~\cite{Rindler-Daller:2011afd, Kain:2010rb, Berezhiani:2015bqa}\footnote{The healing length should not be confused with the dimensionless radial variable~$\xi$ in the Lane-Emden equation~\eqref{eq:LaneEmden}.}
\begin{equation}
\label{Rvortex}
R_\text{\tiny vortex} \simeq \xi = \frac{1}{2 m c_s}   \simeq  \lp \frac{m}{{\rm \mu eV}} \rp^{1/4} \lp \frac{\sigma/m}{{\rm g/cm^2}} \rp^{-1/4} \lp \frac{\rho}{10^{-25} {\rm g/cm^3}} \rp^{-1/2} {\rm m}\,.
\end{equation}
The angular momentum of superfluid vortices can be readily obtained from the corresponding Noether charge
\begin{equation}
L_i=-\epsilon_{ijk}\int {\rm d}^3 x \big(x_j T_{0k}-x_k T_{0j}\big)\,,
\end{equation}
where $T\mn$ is the energy-momentum tensor for the scalar field theory in question. Evaluating this on the vortex configuration \eqref{vortex_ansatz} is straightforward. In fact, for the size of the superfluid core $R\gg (mc_s)^{-1}$, the integral is dominated by the asymptotic configuration. The throat of the vortex, meanwhile, gives a subleading contribution, as can be seen by comparing the two:
\begin{equation}
L_z=n\mu v^2\int {\rm d}^3 x f^2\simeq \pi n \frac{\rho}{m}R^3\cdot\begin{cases} 1 & 
~~~~\textit{asymptotics}\,;\\
(n+1)^{-1}(2 m c_sR)^{-2} & 
~~~~\textit{vortex throat}      
      \,,
    \end{cases} 
\end{equation}
As we can easily see, the purely winding contribution dominates when the superfluid core is larger than the healing length. This is precisely the case of our interest, as it is self-evident from \eqref{Rvortex} and from the fact that the size of the superfluid core is of order kpc. Therefore, we arrive at the well-known fact that the angular momentum stored in the superfluid component is quantized in units of $\pi \rho R^3/m$.

However, the formation of high winding vortices is energetically disfavored~\cite{1981phki.book.....L}, due to the fact that their energy scales as~$n^2$. Therefore, the lowest energy configuration with overall winding~$n$ corresponds to~$n$ widely separated solitons. This, on the other hand, does not affect the aforementioned angular momentum quantization for obvious reasons. Furthermore, taking into account that when the superfluid is spun, it may be energetically favored to store the angular momentum in excited particles, {\it i.e.}, in the normal component, the critical angular velocity required to create a vortex may be larger than the bare minimum dictated by angular momentum quantization. In fact, it is logarithmically larger than the latter~\cite{1981phki.book.....L,Rindler-Daller:2011afd}:
\begin{equation}
\label{critOmega}
\Omega_{\rm c} \simeq \frac{1}{m R^2} \log \lp \frac{R}{\xi} \rp\,.
\end{equation}
This naturally brings us to the point of vortex formation in our SDM halos,
which may have some nonzero angular momentum,\footnote{The presence of angular momentum in scalar DM models can also arise in an inhomogeneous density field, or if the patch has a nonspherical boundary, such as in Riemann S-ellipsoids~\cite{Rindler-Daller:2011afd}.} see Refs.~\cite{Silverman:2002qx,Yu:2002sz,Brook_2022, Sikivie:2009qn, Kain:2010rb} for related works. Indeed, halos tend to form with rotation angular frequency parametrized as~$\Omega \simeq \chi_{\rm s} \sqrt{G \rho}$, where the spin parameter~$\chi_{\rm s}$ takes values in the range~$0.01 \lesssim \chi_{\rm s}\lesssim 0.1$, as inferred from N-body simulations~\cite{1987ApJ...319..575B}.

\begin{figure}[t!]
	\centering
	\includegraphics[width=0.49\textwidth]{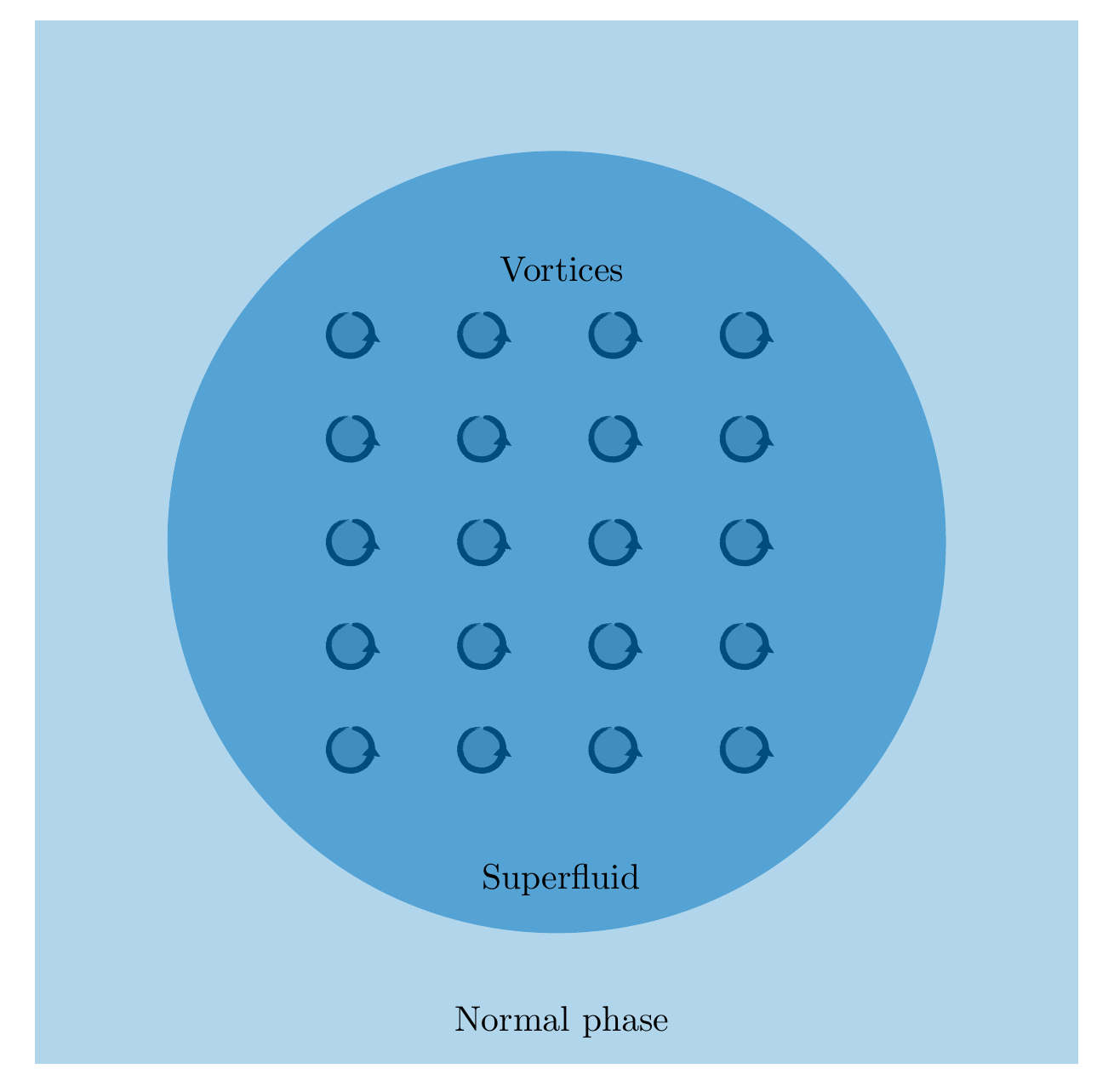}
	\caption{Pictorial representation of a central SDM soliton containing a lattice of ordered rotating vortices, along the line of Fig.~\ref{streams}. In this picture, we have assumed that the initial angular momentum stored in the DM halo before the superfluid phase transition is mostly transferred to the central superfluid core, while we have neglected the possible production of disordered vortices from stochastic interferences in the normal component.}
	\label{figvortices}
\end{figure}

Assuming a typical NFW density~$\rho \sim 10^{-25} {\rm g/cm^3}$, one obtains an angular velocity of order~$\Omega \sim 10^{-16} \chi_{\rm s} \, {\rm s}^{-1}$. On the other hand, for halos of characteristic size~$R \sim 30 \, {\rm kpc}$ and based on a DM candidate with mass~$m \simeq \mu {\rm eV}$, the critical angular velocity
is~$\Omega_{\rm c} \sim 10^{-35} \, {\rm s}^{-1}$ (dropping the logarithmic factor from Eq.~\eqref{critOmega}), which implies that vortex formation is unavoidable in SDM halos,~$\Omega \gg \Omega_{\rm c}$~\cite{Berezhiani:2015bqa}. To estimate their abundance, we assume that the angular velocity of the core is of the same order of magnitude as the average angular velocity of the entire halo, and that in the superfluid phase transition process from the rotating DM core, a significant fraction of its angular momentum is transferred to the superfluid component.
Then, the number of vortices per unit surface area (perpendicular to the rotation axis) can be estimated as~$m \Omega/\pi$~\cite{FEYNMAN195517}, implying the total number within a SDM core of size~$R$ to be\footnote{Notice that, in principle, in the degeneracy region outside of the soliton core, stochastic interferences may also generate a maze of disordered vortices (contrary to the ordered array in the superfluid core), that mimic a classical system of collisionless particles supported by its velocity dispersion, similar to models of fuzzy DM~\cite{Irsic:2017ixq, Hui:2020hbq, GalazoGarcia:2022nqv, Liu:2024pjg}.}
\begin{equation}
N_\text{\tiny vortex} \sim m \Omega R^2 \simeq \chi_{\rm s} \left(\frac{R}{\xi}\right) \simeq 10^{18}~ \chi_{\rm s} \left(\frac{m}{\mu {\rm eV}} \right) \left( \frac{R}{30 \, {\rm kpc}} \right)^2 \left( \frac{\rho}{10^{-25} {\rm g/cm^3}} \right)^{1/2}\,.
\end{equation}
This shows that in a SDM model we expect the production of a large number of vortices. This feature has also been recently investigated using numerical simulations in two and three dimensions~\cite{Brax:2025uaw, Brax:2025vdh}, where, starting from stochastic initial conditions for a spherical halo with nonzero angular momentum, a network of vortex lines  aligned with the total spin is formed. A pictorial representation is shown in Fig.~\ref{figvortices}. Vortices are separated on average by a distance of order~$\chi_{\rm s}^{-1}\xi$. This implies that, for sufficiently high~$\chi_{\rm s}$, vortices are expected to be numerous enough to significantly alter the superfluid density profile, potentially inducing a rotating and oblate shape. This, in turn, could have observational impact on rotation curves.

It would be interesting to study whether superfluid vortices can be detected observationally.\footnote{In the context of MONDian superfluids, the possible impact of vortices on the surrounding baryons was studied in Ref.~\cite{Mauland:2021twm}, showing that they are unlikely to significantly alter the galaxy rotation curves because of their small size.} A promising observational probe is gravitational lensing, based on the detection of flux anomalies between images induced by the presence of small-scale structures. To distinguish the effect of vortices from that induced by substructures on small scales, one can consider the feature that vortices are associated with underdense fluctuations, leading to a de-magnification of the flux compared to the smooth background~\cite{Hui:2020hbq}. However, the measurement of this effect could be complicated for dynamical systems of vortices, which would require studying time-dependent observables~\cite{Hui:2020hbq}.

The above discussion focused on vortices formed thanks to the rotation of halos. However, additional vortices could be formed purely statistically due to the enhancement of perturbations at the critical point of the phase transition via the Kibble-Zurek mechanism~\cite{Kibble:1976sj,Zurek:1985qw,Zurek:1993ek, Zurek:1996sj}. The essence of the mechanism is that, in the environment of amplified fluctuations, the probability of the field configuration to develop an accidental winding is not expected to be negligible in causally disconnected domains of the superfluid. However, configurations with opposite winding will inadvertently unwind each other in the process of relaxation, if they are within the coherence length of the system. Assuming a cosmological formation of topological defects, due to the vacuum structure of the theory, this coherence length is given by the Hubble radius, which sets the scale of causal contact. As usual, this implies on average one topological defect per Hubble volume during a phase transition. In our case, the corresponding coherence length is given by the Jeans scale. This can be understood by remembering that the relaxation time is given by the dynamical time $t_{\text {\tiny dyn}}^{-1}=\sqrt{G\rho}$ and that perturbations propagate with the sound speed of SDM medium, resulting in the coherence length~$\sim c_s/\sqrt{G\rho}$.
In this case, we would expect about one vortex per superfluid core within a DM halo, even though a more detailed study is needed to have a more complete picture. This should have a negligible impact on a core-wide vortex structure due to the overabundance of vortices induced by angular momentum conservation.

Before closing this subsection, let us finally comment on the stability of these structures within the halo condensate. A few papers have discussed how vortices within a soliton could be unstable unless sufficiently strong repulsive self-interactions are present~\cite{Dmitriev:2021utv, Schobesberger:2021ghi}. This is precisely the regime of interest for SDM, where rotating solitons are dynamically stable as long as the rotational energy is less than the self-interaction and gravitational energies. In the absence of self-interactions, it was found that vorticity and signs of quantum turbulence could be generated during structure formation only outside of the solitonic cores, within regions of much lower density, or in the chaotic halo that surrounds the soliton.
There is however an alternative scenario besides repulsive self-interactions that can stabilize vortices within soliton cores, provided by the presence of a background gravitational well from other matter fields~\cite{Glennon:2023oqa}. This could be given by supermassive BHs at the center of the soliton core, with a mass of order that of the soliton itself, thanks to which the vortex can have a long lifetime before decaying. While the complete interplay between DM vortices and BHs is still under investigation, we next turn to discuss the behavior of SDM around central BHs.

\subsection{Behavior around black holes}
\label{sec:BHSDM}

Galaxies are expected to host supermassive BHs at their centers. The gravitational influence of these BHs modifies the density profile of the central superfluid soliton. In particular, the DM density distribution should steepen in the vicinity of the BH,\footnote{The generation of DM spikes around supermassive BHs is expected to affect also the shadow images of BHs~\cite{Jusufi:2020cpn}.} deviating from its otherwise nearly uniform core density~$\rho_0$, thereby enhancing DM self-interactions. In this subsection, we provide a concise overview of the key findings of Ref.~\cite{DeLuca:2023laa} on the impact of supermassive BHs on the SDM density profile.

Consider a non-spinning central BH with a mass~$M_\BH$ and Schwarzschild radius given by
\begin{equation}
\label{Schwarzschild}
r_\BH = 2 G M_\BH \simeq 9.6 \, \cdot 10^{-11}\, {\rm kpc} \lp \frac{M_\BH}{10^6 M_\odot}\rp \,.
\end{equation}
We are interested in supermassive BHs that are potentially observable with future gravitational wave (GW) experiments like the Laser Interferometer Space Antenna (LISA)~\cite{LISA:2017pwj,LISA:2022kgy, LISA:2024hlh}. The BH gravitational well is expected to alter the DM distribution within a characteristic radius
\begin{equation}
\label{BHinfluence}
r_h = \lp \frac{3M_\BH}{4 \pi \rho_0} \rp^{1/3}
\simeq 0.25 \, {\rm kpc} \lp \frac{M_\BH}{10^{6} M_\odot}\rp^{1/3} \lp \frac{\rho_0}{10^{-24} {\rm g/cm^3}} \rp^{-1/3}\,,
\end{equation}
which represents the distance at which the enclosed DM mass reaches the BH mass, such that their gravitational potentials are equal. 
To simplify the analysis, we assume a spherically symmetric DM distribution, isotropic in velocity, and relaxed to a near-equilibrium state. Additionally, we presume that the BH mass is significantly smaller than the total integrated DM halo mass~$M_\DM$.

We delineate the density profile across three distinct regions. First, in the outer region~$r > r_h$ beyond the BH's sphere of influence, the BH's impact on SDM is negligible. In the intermediate region, encompassing radii smaller than~$r_h$, the BH causes modifications to the DM density profile owing to its gravitational well. These alterations persist until relativistic effects on the DM motion become significant enough to influence the superfluid properties and equation of state, heralding the onset of an inner region where further modifications in the DM profile occur. See Ref.~\cite{DeLuca:2023laa} for a more exhaustive discussion.

First, let us turn our attention to the intermediate region. At distances~$r \lesssim r_h$, the BH eventually dominates the gravitational potential. In this regime, 
DM particles become gravitationally bound to the BH, with characteristic velocity dispersion
\begin{equation}
v(r) = \sqrt{\frac{G M_\BH}{r}}\,.
\label{v interm}
\end{equation}
Under the assumption of energy equipartition, their temperature is
\begin{equation}
T (r) = \frac{1}{3} \frac{G m M_\BH}{r}\,.
\label{TBH}
\end{equation}
Despite this influence, we assume that, within this region, the BH's impact is not strong enough to alter the equation of state of the DM fluid. Consequently, the DM density profile can be determined by solving the hydrostatic equilibrium equation, expressed as
\begin{equation}
\frac{1}{\rho (r)} \frac{\d P(r)}{\d r} = 
- \frac{4 \pi G}{r^2} \int^r_0 \d r' r'^2\rho(r') - \frac{G M_\BH}{r^2} \simeq - \frac{G M_\BH}{r^2}.
\label{hydrostatic 1}
\end{equation}
The first term in the middle expression stems from the superfluid self-gravity, and is subdominant compared to the BH potential,~$\Phi_\BH = - G M_\BH/r$. 

At sufficiently smaller distances, comparable to the Schwarzschild radius, the Newtonian approximation becomes inadequate. Again working in the approximation that the BH dominates the gravitational potential, {\it i.e.}, ignoring the contribution of DM to the stress-energy tensor, the
spherically symmetric space-time is well-approximated by the Schwarzschild metric,
\begin{equation}
{\rm d}s^2 = - \left(1-\frac{r_\BH}{r}\right) {\rm d}t^2 + \frac{{\rm d} r^2}{1-\frac{r_\BH}{r}} + r^2 {\rm d}\Omega^2\,,
\end{equation}
The relativistic extension of Eq.~\eqref{hydrostatic 1} in this metric is given by~\cite{Shapiro:2014oha}:
\begin{equation}
\frac{\d P(r)}{\d r} = - \frac{\rho (r)+P(r)}{1-2 G M_\BH/r} \frac{G M_\BH}{r^2}\,.
\label{hydro rel}
\end{equation}
Solving this equation yields the density profile of SDM in the intermediate region.

Derived from Eq.~\eqref{TBH}, the superfluid's temperature correlates with the distance from the BH, introducing a specific scale~$r_\text{\tiny deg}$ at which the system's temperature may surpass the superfluid's critical temperature~$T_{\rm c} \propto (\rho / m)^{5/3}$~\cite{DeLuca:2023laa}. Consequently, the condition of degeneracy for Bose-Einstein condensation (see Eq.~\eqref{degcond}) breaks down, leading to a transition to an inner region where DM is no longer sufficiently cold to maintain a superfluid state. Instead, its equation of state approximates the ideal gas law due to numerous interactions around the BH~\cite{Shapiro:2014oha}:
\begin{equation}
P = n T = \frac{\rho v^2}{3}\,.
\label{ideal}
\end{equation}
In the case of two-body interactions,~$r_\text{\tiny deg}$ remains smaller than the BH horizon, preserving the superfluid state down to the BH horizon. Conversely, in the three-body interaction scenario,~$r_\text{\tiny deg}$ exceeds the BH horizon, altering the evolution of the DM profile accordingly. Nonetheless, it is crucial to note that thermalization remains valid as the BH is approached. 

Following methodologies outlined in~\cite{Shapiro:2014oha, DeLuca:2023laa}, one can determine the DM density profile within this regime by solving the hydrostatic equilibrium equation alongside a heat equation under the gravothermal fluid approximation~\cite{Balberg:2002ue, 1980MNRAS.191..483L, Shapiro:2014oha, Koda:2011yb}
\be
\frac{\d T}{\d r} = \frac{D}{\kappa r^2 (1-2 G M_\BH/r)^{3/2}} - \frac{T}{1-2 G M_\BH/r} \frac{G M_\BH}{r^2}\,,
\label{heat 2}
\ee
which has been previously introduced in Eq.~\eqref{gravothermal Heat} of Sec.~\ref{sec:FormationSF} for the case of self-interacting DM.
This equation relates the temperature gradients of a 
virialized gas with temperature $T = \frac{1}{3}m v^2$ and thermal conductivity~$\kappa$, at rest in a stationary and spherically-symmetric gravitational field,
with the total radiated heat flux,
assumed to be constant (and proportional to the constant $D$) for a steady-state solution~\cite{Bahcall:1976aa}.
The expression for the thermal conductivity $\kappa$ is obtained by matching the relativistic heat flux equation to the Newtonian result of kinetic theory~\cite{Shapiro:2014oha}, giving
\begin{align}
\kappa \simeq \frac{\sqrt{3}}{2} A C \frac{\sigma\rho^2 r^2v}{m^2} \,,
\end{align}
in the long mean free path regime $\frac{m}{\rho \sigma} \gg r$~\cite{DeLuca:2023laa},
where~$\sigma$ is the scattering cross section, $C \simeq \frac{58}{77}$ is a constant determined by N-body simulations~\cite{Koda:2011yb}, and $A = \sqrt{\frac{16}{\pi}}$ for hard-sphere interactions.

These equations are to be solved down to the radius~$r_\text{\tiny mb} = 4 G M_\BH$~\cite{Sadeghian:2013laa,Shapiro:2014oha}, corresponding to marginally-bound circular orbits within the Schwarzschild geometry. At that point, DM particles are entirely accreted into the BH, completely determining the heat constant $D$
and resulting in a sharp decline in the DM density profile.

\begin{figure}[t!]
\centering
\includegraphics[width=0.6\textwidth]{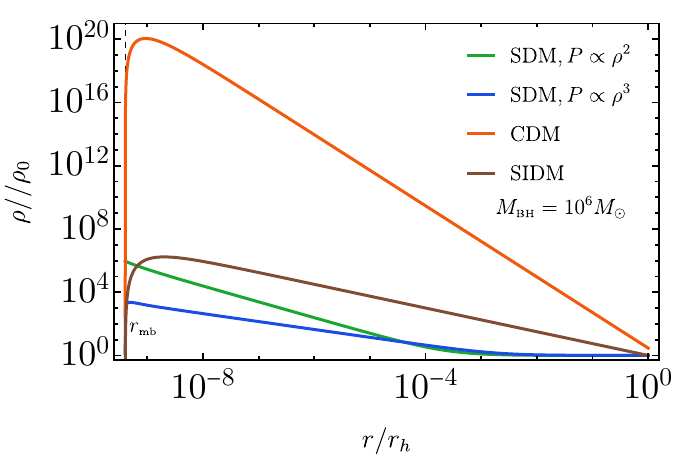}
\caption{Comparison of the DM profiles within the sphere of influence $r_h$ of a central BH with mass~$M_\BH = 10^6 M_\odot$, for collisionless (CDM, red), self-interacting (SIDM, brown), and superfluid (SDM, green and blue) models. Such profiles are all expected to drop at the accretion radius~$r_\text{\tiny mb} = 4 G M_\BH$. Figure taken from Ref.~\cite{Berezhiani:2023vlo}.}
\label{SFDMprofile}
\end{figure}

Figure~\ref{SFDMprofile} illustrates the density profile for the SDM  models discussed earlier, assuming a supermassive BH of mass~$M_\BH = 10^6 M_\odot$. The green and blue lines represent the profiles for the two- and three-body interacting cases, respectively. In both scenarios, the profile steepens significantly within the BH sphere of influence, increasing by orders of magnitude compared to the case without the BH, until it drops sharply at the accretion radius. Comparing the two interacting models reveals that the profile for~$P\propto \rho^3$ exhibits a milder density growth than in the~$P \propto \rho^2$ case, as expected since the equation of state is stiffer in the former case.

The growth in the density profile is significantly influenced by the presence of pressure or self-interactions in the fluid~\cite{DeLuca:2023laa,Aurrekoetxea:2024cqd}. This can be demonstrated by comparing the superfluid prediction, characterized by strong interactions, with the prediction by Gondolo and Silk for collisionless DM~\cite{Gondolo:1999ef}. In their work, they analyze the evolution of an initially isotropic distribution characterized by~$\rho = \rho_0 (r/r_h)^{-\gamma}$ and found that, as a central point mass accretes particles adiabatically, it results into the formation of a massive BH surrounded by a DM spike with density~\cite{Gondolo:1999ef}
\be
\label{GondoloSilkprofile}
\rho_\text{\tiny CDM} \simeq \rho_0 \lp \frac{M_\BH}{\rho_0 r_h^3} \rp^{\frac{\gamma-3}{\gamma-4}} \left(1 -\frac{r_\text{\tiny mb}}{r}\right)^3 \lp \frac{r}{r_h} \rp^{ -\frac{9-2\gamma}{4-\gamma}}\,.
\ee
In their work, the absence of self-interactions results in a much larger growth compared to SDM. Additionally, the importance of self-interactions is highlighted by studies on self-interacting DM, revealing a slope of~$\rho_\text{\tiny SIDM} \propto r^{-3/4}$ for velocity-independent cross sections~\cite{Shapiro:2014oha}. This underscores the trend depicted in Fig.~\ref{SFDMprofile}, where weaker interactions mildly enhance the profile compared to models with higher pressure, such as superfluids. 
The significant hierarchy among these diverse DM models will impact the strength of the dynamical friction force experienced by perturbers navigating through the fluids, as discussed in the next Section.

\section{Dynamical friction}
\label{sec:DynamicalFr}

One of the most interesting phenomena in Bose-Einstein condensates and superfluids is how they affect dynamical friction. The latter occurs when an object, known as the perturber, moves through either a discrete or continuous medium, causing the generation of density fluctuations, termed as wakes, within that medium. The gravitational force exerted by these wakes results in the gradual deceleration of the object's motion, similarly to the effect of radiation reaction in electrodynamics where the electromagnetic field of an accelerated charge slows it down~\cite{Landau:1975pou}.
The strength of the corresponding drag force depends on the trajectory of the body's motion, that is, whether the motion is linear, accelerated, or within binary systems. In linear trajectories, the generation of density fluctuations leads to a gradual slowdown of the object. Accelerated motion amplifies the density wakes and enhances the deceleration process. In binary systems, where two objects orbit each other, dynamical friction alters the orbital dynamics and leads to orbital decay over time. Hence, the nature of dynamical friction is inherently trajectory-dependent, playing a key role in shaping the long-term dynamics of celestial systems.

For a collisionless medium, the drag force on a linearly moving body has been computed in the pioneering study of Chandrasekhar~\cite{Chandrasekhar:1943ys}, while the linear motion in a gaseous medium has been discussed in Refs.~\cite{1971ApJ...165....1R, 1980ApJ...240...20R, Ostriker:1998fa}.
The case of circular binaries was investigated in Refs.~\cite{Kim:2008ab, 2009ApJ...703.1278K, Lee:2011px, Lee:2013wtc,2022ApJ...928...64D,Buehler:2022tmr}.
More diverse environments, such as fuzzy DM, self-interacting or superradiance-driven scalar clouds, and Bose-Einstein condensates, have also been explored recently in Refs.~\cite{Hui:2016ltb,Lancaster:2019mde, Annulli:2020lyc, Hartman:2020fbg, Wang:2021udl,Boudon:2022dxi,Boudon:2023qbu,Baumann:2021fkf,Tomaselli:2023ysb}. For superfluids---which exhibit frictionless flow---dynamical friction is expected to be significantly suppressed. This feature could lead to intriguing astrophysical implications and potentially resolve several challenges faced by CDM. For instance, the velocity of galactic bars within spiral galaxies is observed to remain nearly constant, contrary to the expected slowdown in CDM~\cite{Debattista:1997bi, Debattista:2000ey, Roshan:2021liy}. Similarly, the absence of mergers among the five globular clusters orbiting the Fornax dwarf galaxy poses a puzzle: in a cuspy CDM halo, these clusters should have long since migrated to the center due to dynamical friction~\cite{1976ApJ...203..345T}. Suppressed dynamical friction in a SDM framework could naturally account for their continued survival. See also Ref.~\cite{Lee:2025qvm} for a recent analysis in the context of self-interacting fuzzy DM.

Friction within superfluids does arise, however, from dissipative processes that occur when internal degrees of freedom are excited out of the condensate, resulting in a mixture of superfluid and normal components. In particular, the motion of an impurity through the fluid can trigger the emission of phonons, resulting in energy loss and effective dissipation. In what follows, we examine dynamical friction in SDM condensates, considering both the case of a single body moving linearly through the fluid and the orbital evolution of circular binaries. For a detailed treatment, we refer the reader to Refs.~\cite{Berezhiani:2019pzd, Berezhiani:2023vlo}.

\subsection{Linear motion}

The simplest superfluid model we consider is a degenerate state of weakly-interacting bosons in the zero-momentum state, where only longitudinal dynamical excitations are present at low energy. The latter are ultimately responsible for dissipation and dynamical friction. The corresponding force can be computed within the quasi-particle formalism, based on the low energy EFT of phonons gravitationally coupled to a probe particle, discussed in Sec.~\ref{sec:HigherOrder}. Energy dissipation corresponds to the gravity-mediated process of radiating phonons by a probe traveling through the superfluid.
As we will show, these results can be also reproduced using the hydrodynamical method described in Sec.~\ref{sec:Gross} in the context of the Gross-Pitaevskii equation. 

\subsubsection{Quasi-particle approach}

The simplest microscopic theory describing a probe moving through a superfluid is given by~\cite{Berezhiani:2019pzd}
\beq
\mathcal{L}=  \frac{1}{16\pi G}R-|\partial \Phi|^2-m^2 |\Phi|^2-\frac{\lambda_4}{2}|\Phi|^4 -\frac{1}{2}(\partial \chi)^2-\frac{1}{2}M^2\chi^2\,.
\label{quasiparticleaction}
\eeq
The massive self-interacting complex scalar field~$\Phi$ describes the fluid degree of freedom, while the massive real scalar field~$\chi$, minimally coupled to gravity, represents the perturber. According to the discussion of Sec.~\ref{sec:JeansQuasiP}, to describe a gravitating superfluid and track the gravitational degrees of freedom, we can  treat the fluid non-relativistically and focus on the Newtonian potential~$\Phi_\text{\tiny N}$. By perturbing the classical background as in Eq.~\eqref{perturbedSol}, linearizing gravity 
and integrating out the Newtonian potential, we obtain~\cite{Berezhiani:2019pzd}
\begin{align}
\mathcal{L}&=\frac{1}{2}\dot{\pi}^2+\frac{1}{2} \pi \left( m_g^2+c_s^2\Delta-\frac{\Delta^2}{4m^2} \right) \pi-4\pi GM^2\sqrt{\rho_0} \chi^2  \frac{1}{\sqrt{\Delta \left( m_g^2+c_s^2\Delta-\frac{\Delta^2}{4m^2} \right)}}\, \dot{\pi} \nonumber \\
& -~2\pi G M^4 \chi^2 \frac{c_s^2-\frac{\Delta}{4m^2}}{m_g^2+c_s^2\Delta-\frac{\Delta^2}{4m^2}} \chi^2 +\ldots
\label{highderL}
\end{align}
where we have assumed the perturber to be non-relativistic,~$\dot{\chi}\simeq M\chi$, and the non-relativistic chemical potential~$\mu_\text{\tiny NR}\ll m$ to ensure $c_s\ll 1$. 
Furthermore, we have introduced the tachyonic mass~$m_g^2\equiv 4\pi G\rho_0$ responsible for the Jeans instability, from which one easily reads off the dispersion relation for phonons in Eq.~\eqref{eq:phononGR}. The ellipses indicate the free part of the~$\chi$ action, as well as higher-order terms in~$\pi$ and~$\chi$. The second line describes non-local self-interaction terms for~$\chi$, which reduce to the Newtonian interaction between two~$\chi$-sources in the limit~$m_g\rightarrow 0$, deep inside the superfluid core. Furthermore, as one can appreciate, the information of the microscopic theory is captured by the sound speed~$c_s$, through its dependence on the couplings and density of the condensate.

Dissipative processes experienced by the probe~$\chi$ moving in the superfluid are described, to first approximation, by the interaction~$\chi \to \chi + \pi$ (see a similar discussion in Appendix~\ref{sec:Scattering}).
To describe this process, consider a~$\chi$ particle, moving freely with uniform velocity~$v$ and initial energy-momentum~$p_\chi^{{\rm in}}=(E_\chi^{{\rm in}},\vec{p}_{\chi}^{~{\rm in}})$, which radiates a phonon with~$p_\pi^f=\big(\omega_k,\vec{k}\,\big)$. 
Using standard scattering relations, the dynamical friction force then reads
\beq
|F_\text{\tiny DF}|\equiv
\int\omega_k\,{\rm d}\Gamma=\frac{4\pi G^2M^2\rho_0}{v^2} \int_{-1}^{1} {\rm d}{\rm cos}~\theta~ \int_{k_\text{\tiny min}}^{k_\text{\tiny max}}\frac{{\rm d}k}{k}\,\delta\left( {\rm cos }~\theta-\frac{\frac{k^2}{2M}+\omega_k}{kv}\right)\,,
\label{Fdelta}
\eeq
where~${\rm d}\Gamma$ is the standard textbook differential interaction rate. See Ref.~\cite{Berezhiani:2019pzd} for the complete derivation of Eq.~\eqref{Fdelta}.

To further simplify this expression, one must specify the limits of integration~$k_\text{\tiny min}$ and~$k_\text{\tiny max}$, which set the range of scales over which the superfluid effective description is valid. For the high-momentum cutoff, a first natural scale is the size of the extended perturber,~$k_\text{\tiny max} \approx R_\text{\tiny probe}^{-1}$, beyond which one should also consider its internal degrees of freedom. Furthermore, the non-vanishing of the angular integral~$\cos \theta < 1$ sets another scale, given by the solution of~$k_\star^2/2M +\omega_{k_\star} = k_\star v$. It follows that~$k_\text{\tiny max} = {\rm min} \big(2 \pi R_\text{\tiny probe}^{-1}, k_\star\big)$. Similarly, for the low-momentum cutoff, the assumption of a homogeneous superfluid breaks down at its spatial extent,~$k_\text{\tiny min} \approx R_\text{\tiny SF}^{-1}$. On the other hand, the longest possible wavelength for which the description can be trusted is given by the Jeans scale,~$k_{\rm J}$, given in Eq.~\eqref{eq:Js}, beyond which modes would exhibit the gravitational instability. Thus,~$k_\text{\tiny min} = {\rm max} \big(2 \pi R_\text{\tiny SF}^{-1}, k_{\rm J}\big)$. 

With these limits, the integration in Eq.~\eqref{Fdelta} can be carried out, resulting in the dynamical friction force~\cite{Berezhiani:2019pzd}
\beq
|F_\text{\tiny DF}| = \frac{4\pi G^2M^2\rho_0}{v^2}~{\rm log}\left[ \frac{ {\rm min} \left( 2\pi R_\text{\tiny probe}^{-1}, k_\star\right)}{ {\rm max} \left( 2 \pi R_\text{\tiny SF}^{-1}, k_{\rm J} \right)} \right]\,.
\label{Fkint}
\eeq
See also Refs.~\cite{Boudon:2022dxi,Boudon:2023qbu} for similar results for self-interacting DM.

\subsubsection{Hydrodynamical approach}

The dynamical friction experienced by a probe moving in a superfluid can also be computed within the hydrodynamical approach, based on the mass conservation and Euler equations~\cite{Ostriker:1998fa, Berezhiani:2019pzd}. As discussed in Sec.~\ref{sec:Gross}, the equation of motion for a density perturbation~$\rho=\rho_0\left(1+\alpha\right)$ around a static background of homogeneous density~$\rho_0$ reads, from Eq.~\eqref{eq:EOMphononGPGr},
\begin{equation}
\label{eq:EOMphononGPGrv2}
    \ddot{\alpha}-c_s^2 \Delta 
    \alpha-m_g^2 \,\alpha +\frac{\Delta^2}{4m^2} \alpha=4\pi G \rho_\text{\tiny ext}\,.
\end{equation}
Such density perturbation is sourced by an external probe $\rho_\text{\tiny ext} = M \, \delta^{(3)}\big(\vec{r}-\vec{r}_p(t)\big)$, moving in a general trajectory, with~$\vec{r}_p(t)$ denoting the position of the perturber at time~$t$. 

This equation can be solved in terms of the retarded Green's function, through a convolution with the source term~\cite{2022ApJ...928...64D,Buehler:2022tmr}. Dynamical friction is consequently derived from the gravitational attraction between the wake and the probe as
\begin{equation}
  \label{eq:df1}
\vec{F}_\text{\tiny DF}= \frac{M}{v} \dot \Phi_\alpha= (4\pi G M)^2 \rho_0 \int_{-\infty}^\infty {\rm d}t' \int \frac{{\rm d}\omega \,{\rm d}^3 k}{(2\pi)^4 k^2} \frac{{\rm i}\vec{k} }{(\omega-{\rm i}\epsilon)^2-c_s^2 k^2 - \frac{k^4}{4 m^2}} {\rm e}^{{\rm i}\omega (t-t')-{\rm i} \vec{k}\cdot\left(\vec{r}_p-\pvec{r}'_p \right)} \,,
\end{equation}
where~$\Phi_\alpha$ is the gravitational potential sourced by the wake.  Here, we specialized on \textit{steady state motion}, corresponding to a source turned on in the asymptotic past. 
The small parameter~$\epsilon >0$ enforces causality, and~$\pvec{r}'_p=\vec{r}_p(t')$ indicates the position of the probe evaluated at the retarded time~$t'$.

Let us now specialize to the case of linear motion. Suppose that the perturber is moving at constant velocity~$v$ along the~$\hat{z}$ direction, such that~$ \vec{r}_p (t) =\left(0, 0, v t\right)$. Substituting this in the general solution~\eqref{eq:df1} and integrating over~$\omega$ using the residue theorem, the friction force simplifies to~\cite{Berezhiani:2019pzd}\footnote{In Ref.~\cite{Berezhiani:2019pzd}, the derivation employs the causal Green's function instead of the retarded one, and the time integration is performed before the frequency integration. Despite these differences, the final result remains the same.}
\be
F_\text{\tiny DF} = -\frac{4 \pi G^2 M^2 \rho_0}{v^2} \int \frac{{\rm d}k}{k}\, {\rm d}\cos\theta\, {\rm e}^{{\rm i} \omega_k t - {\rm i} k vt \cos \theta} \delta \left( \frac{\omega_k}{k v} - \cos \theta \right) \,.
\label{DF-kcos}
\ee
As one can appreciate, one recovers the non-vanishing condition for the~$\delta$-function,~$\omega_{k_\star} = k_\star v$, in the limit of large mass~$M\rightarrow \infty$, as discussed above, ignoring the recoil of the perturber. Following our earlier discussion for~$k_\text{\tiny min}$ and~$k_\text{\tiny max}$, one can perform both angular and~$k$ integrals to finally obtain
\be
\left| F_\text{\tiny DF} \right| = \frac{4 \pi G^2 M^2 \rho_0}{v^2} \log \left[ \frac{ {\rm min} \big( 2 \pi R_\text{\tiny probe}^{-1} , k_\star \big) }{ {\rm max} \big( 2 \pi R_\text{\tiny SF}^{-1}, k_\mathrm{J}  \big) } \right]\,.
\label{ForceDF}
\ee
This matches Eq.~\eqref{Fkint}, obtained using the quasi-particle description. We further stress that this result holds true within the approximation of a superfluid background with homogeneous density~$\rho_0$. The generalization for a radially-dependent density profile will be discussed in the next Section.

\subsubsection{Explicit evaluation of the friction force}

Let us now evaluate the dynamical friction force computed above in different regimes, based on the high- and low-momentum cutoffs, $k_\text{\tiny max}$ and $k_\text{\tiny min}$, respectively. In particular, we will discuss the regimes where the superfluid core of size~$R_\text{\tiny SF}$ is either much smaller or comparable to the Jeans scale~$k_\mathrm{J}^{-1}$. 

\vspace{0.2cm}
\noindent
{\bf $\bullet$  Small superfluid core~$R_\text{\tiny SF} \ll k_\mathrm{J}^{-1}$:}
This corresponds to~$k_\text{\tiny min} = 2 \pi R_\text{\tiny SF}^{-1}$. Let us first consider the subsonic regime,~$v < c_s$. In this case, one can immediately realize that there is no real and positive solution to the~$\delta$-function condition in~\eqref{Fdelta}. Therefore, the force vanishes for subsonic motion
\be
|F_\text{\tiny DF}| = 0\,;\qquad v < c_s\,.
\label{F=0}
\ee
The situation is different in the supersonic regime~$v\gg c_s$, for which we will distinguish the case of localized or extended perturber. For a  sufficiently localized perturber, one can assume~$R_\text{\tiny probe}^{-1} \gg k_\star$ and the UV cutoff scale~$k_\star$ is determined by
\beq
k_\star=\frac{2Mm^2}{M^2-m^2}\left( - v+\sqrt{\frac{M^2}{m^2}(v^2-c_s^2)+ c_s^2}\, \right)\,;\qquad v > c_s\,.
\label{kstar simple 2}
\eeq
If the mass of the perturber $M$ is much heavier than the DM mass $m$, as expected in astrophysical contexts, the cutoff scale reduces to~$k_\star=2mc_s\sqrt{\mathcal{M}^2-1}$,
in terms of the Mach number~$\mathcal{M} = v/c_s$.
In this limit, the corresponding drag force reads~\cite{Berezhiani:2019pzd}
\beq
|F_\text{\tiny DF}|
=\left\{\begin{array}{cl}
\frac{4\pi G^2M^2\rho_0}{v^2}~{\rm log}\Big( \frac{mc_sR_\text{\tiny SF}}{\pi} \sqrt{\mathcal{M}^2-1} \Big)  \hspace{20pt}&\text{for}\hspace{10pt} \mathcal{M} \geq  \sqrt{1 + \frac{\pi^2}{m^2c_s^2R_\text{\tiny SF}^2}}\,; \\ \\
0 \hspace{20pt}&\text{for}\hspace{10pt} \mathcal{M} \leq  \sqrt{1 + \frac{\pi^2}{m^2c_s^2R_\text{\tiny SF}^2}} \,,
\end{array}\right.
\label{DF-qp}
\eeq
which smoothly vanishes as we approach the sonic limit. Notably, a complete derivation shows that the friction force vanishes at velocities slightly exceeding the speed of sound---we refer the reader to the reference~\cite{Berezhiani:2019pzd} for further technical details.  For an extended perturber satisfying~$R_\text{\tiny probe}^{-1} \ll k_\star$, the corresponding force would instead read~$|F_\text{\tiny DF}| \simeq \frac{4\pi G^2M^2\rho_0}{v^2}~{\rm log}(R_\text{\tiny SF}/R_\text{\tiny probe})$.

\vspace{0.2cm}
\noindent
{\bf $\bullet$ Critical superfluid core~$R_\text{\tiny SF} \simeq k_\mathrm{J}^{-1}$:} This corresponds to~$k_\text{\tiny min} = k_{\rm J}$, where there could be macroscopic superfluid excitations with vanishing energy cost. This is the physical regime we expect based on the cosmological evolution of superfluid halos discussed in Sec.~\ref{Sec: SDMgalaxies}.

To simplify the picture, consider the case of a sufficiently heavy,~$M \gg m$, and localized perturber, with~$R_\text{\tiny probe}^{-1} \gg k_\star$, such that the UV cutoff is set by~$k_\text{\tiny max} = k_\star$. In this case, one obtains the analytic solution:
\beq
k_\star^2\simeq 2m^2c_s^2\left( - 1 +  \mathcal{M}^2+\sqrt{\left(1- \mathcal{M}^2\right)^2+\frac{m_g^2}{m^2c_s^4}} \right)\,.
\label{kstarjeans}
\eeq
From a comparison with~\eqref{eq:Js}, one could inquire that~$k_\star > k_{\rm J}$ for arbitrary velocities~$v$, such that the friction force, non-vanishing even for a subsonic motion, would read~\cite{Berezhiani:2019pzd}
\be
|F_\text{\tiny DF}| = \frac{2\pi G^2M^2\rho_0}{v^2}~{\rm log}\left(\frac{- 1 +  \mathcal{M}^2 +\sqrt{\left(1- \mathcal{M}^2\right)^2+\frac{m_g^2}{m^2c_s^4}}}{-1+\sqrt{1+\frac{m_g^2}{m^2c_s^4}}}\right)\,.
\label{F heavy pert}
\ee
However,  one should be careful in properly tracking subleading corrections in~$1/M$ in the strong subsonic motion~$\mathcal{M} \ll 1$, which may lower the value of~$k_\star$ and make it smaller than~$k_{\rm J}$ for small but finite velocity~$v$. In particular, one can solve the~$\delta$-function condition perturbatively in~$1/M$ to obtain
\be
\frac{k_\star}{k_{\rm J}} \simeq 1 + \frac{v}{2c_s^2 \sqrt{1+\frac{m_g^2}{m^2c_s^4}}} \left(v - \frac{k_{\rm J}}{M} \right) + \mathcal{O}\left(\frac{1}{M^2}\right)\,.
\label{kstar pert}
\ee
From this expression, one can identify a critical velocity~$v_{\rm c}= k_{\rm J}/M$ at which~$k_\star$ equals~$k_{\rm J}$. The resulting force in the deep subsonic regime reads
\beq
|F_\text{\tiny DF}|\simeq \frac{2\pi G^2M^2\rho_0}{c_s^2\sqrt{1+\frac{m_g^2}{m^2c_s^4}}}\left(1 - \frac{v_{\rm c}}{v}\right)\,;\qquad v_{\rm c} \leq v \ll c_s\,.
\eeq
Thus, the friction force vanishes below the critical velocity, $|F_\text{\tiny DF}| = 0$ for $v \lesssim v_c$,  albeit with a threshold which is smaller than the sound speed of the condensate. This concludes the various cases of dynamical friction when the perturber moves linearly.

\subsubsection{Accelerated linear motion in the quasi-particle formalism}
Before closing this Section, let us briefly comment on the case of a perturber undergoing accelerated motion, as discussed in Ref.~\cite{Berezhiani:2019pzd}. To highlight the effect of acceleration, we can focus on subsonic velocity,~$v < c_s$, with the homogeneity scale of the superfluid being much smaller than the Jeans scale,~$R_\text{\tiny SF} \ll k_{\rm J}^{-1}$. (For uniform motion, the friction force vanishes in this regime, as shown in Eq.~\eqref{F=0}.) In this case, it is useful to draw a parallel between the phenomenon of electromagnetic {\it bremsstrahlung}, and the associated  radiation reaction, with phonon emission in superfluids. To do so, we will consider a scattering process involving two impurities, focusing on the simpler case of phonon bremsstrahlung from a body moving in the external field of another source. 

In the quasi-particle description, the effective Lagrangian in the limit~$m_g \to 0$ reads~\cite{Berezhiani:2019pzd}
\beq
\mathcal{L}=\frac{1}{2}\dot{\pi}^2+\frac{1}{2} c_s^2 \pi \Delta \pi
-4\pi G\sqrt{\rho_0}
M^2 \chi^2 \frac{1}{c_s\Delta}\dot{\pi}+\frac{1}{2}\dot{\chi}^2-\frac{1}{2}\big(\vec{\nabla}\chi\big)^2-\frac{1}{2}M^2\chi^2+ \mathcal{L}_\text{\tiny higher}\,,
\label{EFT2}
\eeq
where higher-order terms are encoded in 
\beq
\mathcal{L}_\text{\tiny higher} \equiv-\frac{c_s}{2\sqrt{\rho_0}}\big(\vec{\nabla} \pi\big)^2 \dot{\pi}+\frac{1}{8}\frac{c_s^2}{\rho_0}\big(\vec{\nabla} \pi\big)^4+2\pi GM^2\big(\vec{\nabla} \pi\big)^2 \frac{1}{\Delta} \chi^2-2\pi GM^4\chi^2 \frac{1}{\Delta} \chi^2\,.
\label{EFT2-high}
\eeq
The first two terms in~$\mathcal{L}_\text{\tiny higher}$ originate from contact interactions among superfluid constituents, whereas the last two interaction terms are of gravitational origin, hence their non-local structure.

To provide a simple result, let us consider only the rate of radiating phonons from external lines, induced by the last term in Eq.~\eqref{EFT2-high}, given by~$\sim \chi^2 \Delta^{-1} \chi^2$ . In principle, one could allow for 
phonon-mediated scattering or, more generally, for processes where the external phonon line can be attached at more locations besides the external~$\chi$-line. However, in the case of soft bremsstrahlung, describing the emission of low-energy (soft) particles from an accelerated or scattered perturber, the Lagrangian is dominated by the operator~$\chi^2 \Delta^{-1} \chi^2$, such that the total energy radiated in phonons, as the perturber's velocity changes from~$\vec v$ to~$\vec v{\,'}$ due to an external field, is~\cite{Berezhiani:2019pzd}
\be
E_\text{\tiny rad} = \int_{k_\text{\tiny min}}^{k_\text{\tiny max}} {\rm d}k \, \omega_k\frac{{\rm d}{\cal P}}{{\rm d}k}\,, \qquad\text{with}~~~ \frac{{\rm d}{\cal P}}{{\rm d}k}\propto \frac{G^2M^2\rho_0}{c_s^5k^3} \big( \vec{v}-\vec{v}{\,'}\big)^2\,,
\label{Erad quantum}
\ee
where we have taken the subsonic limit~$v\ll c_s$. By considering the on-shell dispersion relation~$\omega_k = c_sk$ in the sound regime, and performing the momentum integral within the range~$k_\text{\tiny min}  = 2\pi R_\text{\tiny SF}^{-1}$ to~$k_\text{\tiny max}=2mc_s$ (beyond which phonons no longer behave as waves), one finally gets
\beq
E_\text{\tiny rad} \sim \frac{G^2M^2\rho_0 R_\text{\tiny SF}}{c_s^4} \big( \vec{v}-\vec{v'}\big)^2\,.
\label{qm-scatt}
\eeq
This recovers the same dependence on~$G$,~$M$ and~$\rho_0$ as the one for a freely-moving perturber. Moreover, contrarily to the logarithmic dependence of the free case, the radiated energy presents an IR divergence~$\propto 1/k_\text{\tiny min} \sim R_\text{\tiny SF}$, and is characterized by a factor~$\big(\vec{v}-\vec{v}{\,'}\big)^2$, as expected for radiation reaction effects, proportional to the square of the acceleration.

\subsection{Circular motion}

While linear motion offers a reasonable approximation of dynamical friction across various trajectories, it becomes inadequate when the timescale over which friction significantly affects the dynamics of the massive object is comparable to the time of deviation from linear motion. Binary systems fall in this category, where dynamical friction can influence motion over multiple orbital cycles. In this subsection, we explore the friction force of bodies in circular motion within a homogeneous superfluid medium, using the hydrodynamic framework of Refs.~\cite{2022ApJ...928...64D,Buehler:2022tmr, Berezhiani:2023vlo}. We label the DM mass as~$m_\DM$, to avoid confusion with the multipole moment~$m$ in spherical harmonic decompositions.

\subsubsection{Formalism}

Consider once again an external probe of mass~$M$  moving in a superfluid medium of homogeneous density~$\rho_0$. The perturbation equation for the density inhomogeneity~$\alpha$ is still determined by Eq.~\eqref{eq:EOMphononGPGrv2}. In the following, we neglect the gravitational mass term~$m_g$, which is a reasonable assumption for the circular motion we are interested in as long as the orbit's radius is smaller than the Jeans scale. 

For a circular orbit of radius~$r_0$ and constant angular velocity~$\Omega$, the position vector can be parametrized in spherical coordinates as~$\vec{r}_p (t) =\left(r_0, \, \pi/2, \, \Omega t\right)$, assuming motion in the equatorial plane without loss of generality. For simplicity, we assume that the acceleration induced by the drag force is smaller than the orbital acceleration in order to have a self-consistent picture. Following the derivation of~\cite{Berezhiani:2023vlo}, the friction force is given by
\be
\vec{F}_\text{\tiny DF} =- \frac{4\pi G^2 M^2 \rho_0}{c_s^2}\vec{\mathcal{F}}_\text{\tiny DF}\,, 
\ee
where the dimensionless force~$\vec{\mathcal{F}}_\text{\tiny DF}$ is expressed as a sum over angular multipoles~$(\ell,m)$:
\be
\vec{\mathcal{F}}_\text{\tiny DF} \equiv c_s^2 \sum_{\ell=1}^{\ell_\text{\tiny max}}\sum_{m=-\ell}^{\ell-2}\gamma_{\ell m}\bigg\{\text{Re}\left(S_{\ell,\ell-1}^m-{S^{m+1}_{\ell,\ell-1}}^*\right)\hat{r}+\text{Im}\left(S_{\ell,\ell-1}^m-{S^{m+1}_{\ell,\ell-1}}^*\right)\hat{\varphi}\bigg\}\,.
\label{FDF1}
\ee
This expression gives the perpendicular~$(\hat{r})$ and tangential~$(\hat{\varphi})$ components of the drag force experienced by a probe moving on a circular orbit. 
Spherical symmetry dictates that the force along the direction perpendicular to the orbital plane vanishes.
The coefficients~$\gamma_{\ell m}$ are completely determined by the angular trajectory and given by
\begin{equation}
    \gamma_{\ell m}=  (-1)^m \frac{(\ell-m)!}{(\ell-m-2)!}\left\{{\Gamma\left(\frac{1-\ell-m}{2}\right)\Gamma\left(1+\frac{\ell-m}{2}\right)\Gamma\left(\frac{3-\ell+m}{2}\right)\Gamma\left(1+\frac{\ell+m}{2}\right)}\right\}^{-1}\,,
\end{equation}
in terms of the Gamma function~$\Gamma$. The coefficients~$S_{\ell,\ell-1}^m$ are the~$1+1$-dimensional analog of the starting integral in Eq.~\eqref{eq:df1} after integrating over the solid angle. For steady-state motion, they take the form
\begin{flalign}
\label{IntS}
S_{\ell,\ell-1}^m =\int  {\rm d}k\, k \frac{j_\ell(k r_0)j_{\ell-1}(k r_0)}{c_s^2 k^2+\frac{k^4}{4m_\DM^2}-(m \Omega +{\rm i}\epsilon)^2}\,.
\end{flalign}
Notice that the microphysics of the DM condensate enters through the denominator of the integrand. Using the residue theorem, the above expression evaluates to~\cite{Berezhiani:2023vlo}
\begin{align}
S_{\ell,\ell-1}^{m}&= \frac{\pi {\rm i}}{2c_s^2 \sqrt{1+\frac{m^2}{\ell_{\rm q}^2}} }\biggr[(-1)^{1+\theta(m)}j_\ell\big(\ell_{\rm q} \mathcal{M}  f^-_m \big)j_{\ell-1}\big(\ell_{\rm q} \mathcal{M}  f^-_m \big)
    \nonumber \\
    & 
    ~~~~~~~~~~~~~~~ +~{\rm i} j_\ell\big( \ell_{\rm q} \mathcal{M}  f^-_m \big)y_{\ell-1}\big(\ell_{\rm q} \mathcal{M}  f^-_m \big) 
   +\frac{2}{\pi}i_\ell\big( \ell_{\rm q} \mathcal{M}  f^+_m\big)k_{\ell-1}\big( \ell_{\rm q} \mathcal{M}  f^+_m\big)\biggr]\,,
\label{fullFr}
\end{align}
where~$i_\ell$ and~$k_\ell$ are the modified spherical Bessel functions of the first and second kind, respectively.
In this expression we have used the functions
\begin{align}
f^{\pm}_m & = \sqrt{2\pm 2\sqrt{1+\left(\frac{m}{\ell_{\rm q}}\right)^2}} \quad \Longrightarrow
\quad 
\begin{cases} f_m ^+\simeq 2\,,~~ f^-_m \simeq \frac{m}{\ell_{\rm q}}\quad  ~~~~~\text{for} \quad m \ll \ell_{\rm q}\,;\\
f^{\pm}_m\simeq \sqrt{\frac{2m}{\ell_{\rm q}}} \quad ~~~~~~~~~~~~~~\text{for} \quad m \gg \ell_{\rm q} \,,
\end{cases}
\label{fpm def}
\end{align}
in terms of the Mach number~$\mathcal{M} = v/c_s = r_0 \Omega/c_s$. The parameter
\begin{equation}
    \ell_{\rm q}= \frac{r_0 m_\DM c_s}{\mathcal{M}} = \frac{m_\DM c_s^2}{\Omega}
\end{equation}
describes the value of the azimuthal quantum number~$m$ above which the~$k^4/4m_\text{\tiny DM}^2$ term in the dispersion relation becomes important. Similarly to the case of linear motion, a supersonic probe is correlated with a lower value of $\ell_{\rm q}$, and thus with a stronger role of quantum pressure in the dynamical friction process. 

The final ingredient needed to calculate the drag force is the maximum multipole moment~$\ell_\text{\tiny max}$ up to which the sum over~$\ell$ in Eq.~\eqref{FDF1} is performed. It is determined by the probe size~$R_\text{\tiny probe}$ and the orbital radius~$r_0$ as
\begin{equation}
\ell_\text{\tiny max} =\frac{\pi r_0}{R_\text{\tiny probe}}\,.
\label{lmax}
\end{equation}
One can then distinguish two cases depending on the hierarchy between~$\ell_\text{\tiny max}$ and~$ \ell_{\rm q}$, for fixed Mach number. When~$\ell_\text{\tiny max} > \ell_{\rm q}$, the tangential component begins to converge only after reaching the multipole moment~$\ell_{\rm q}$.   Conversely, if~$\ell_\text{\tiny max} < \ell_{\rm q}$, the size of the probe acts as the effective cutoff. This is analogous to the case of linear motion, discussed earlier. 

\begin{figure}[t!]
\centering
	\includegraphics[width=0.328\textwidth]{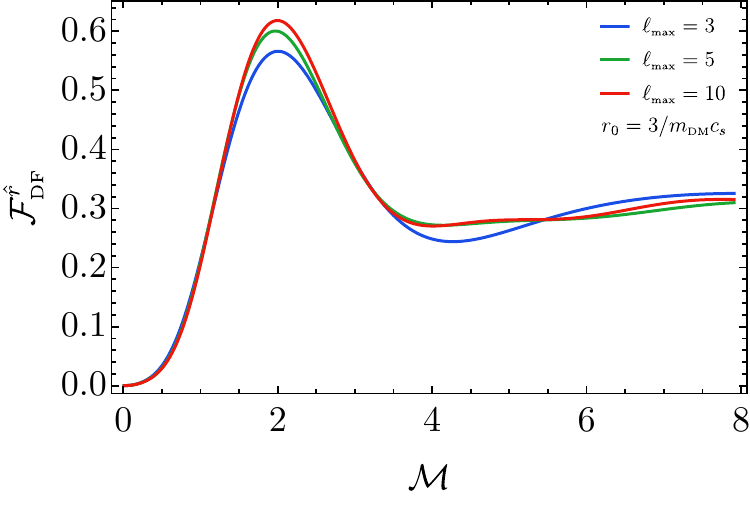}
 \includegraphics[width=0.328\textwidth]{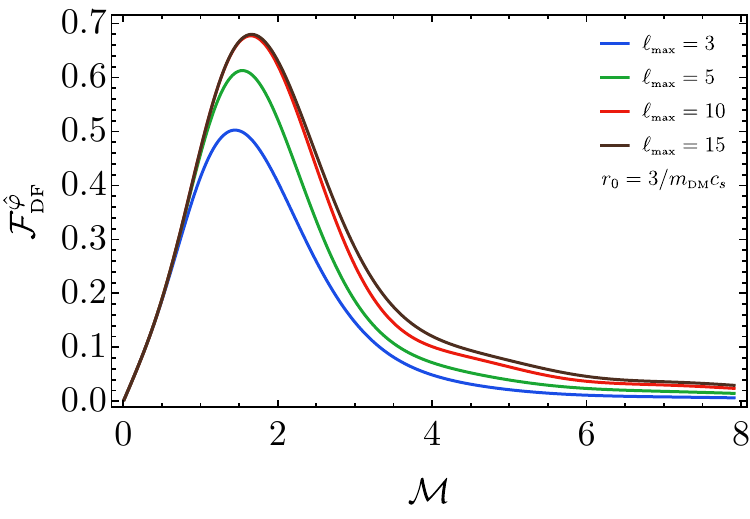}
 \includegraphics[width=0.328\textwidth]{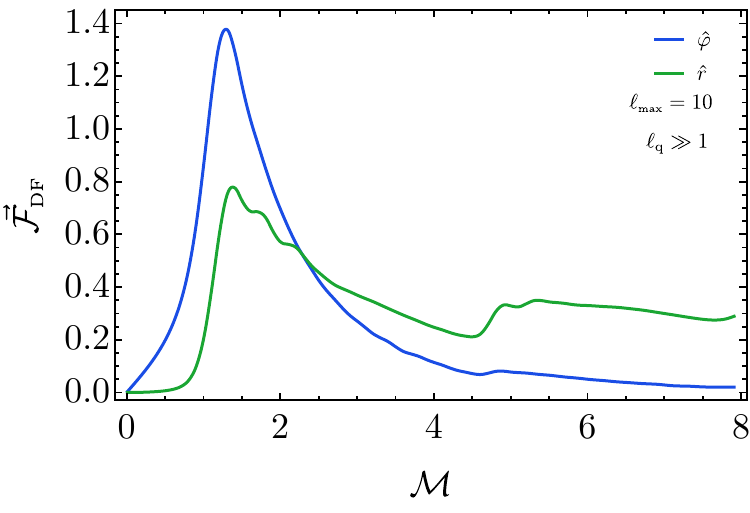}
	\caption{Illustration of the radial (left panel) and azimuthal (central panel) component of the dynamical friction force as a function of the Mach number~$\mathcal{M}$. The orbital radius is fixed at~$r_0 = 3/m_\text{\tiny DM}c_s$ to achieve an intermediate scenario ($\ell_{\rm q} \simeq \mathcal{O}(1)$) between the quantum pressure and sound regimes. Various lines represent different values of the maximum multipole~$\ell_\text{\tiny max}$. The right panel shows the drag force in the sound  regime, assuming~$\ell_\text{\tiny max} = 10$~\cite{Kim:2008ab} and~$r_0 = 100/m_\DM c_s$ to enforce~$\ell_{\rm q} \gg 1$.    
 Figure taken from Ref.~\cite{Berezhiani:2023vlo}.}
	\label{SFforce}
\end{figure}
The behavior of the friction force as a function of~$\mathcal{M}$ is shown in Fig.~\ref{SFforce}. The left and central panel focus on the intermediate regime $\ell_{\rm q} \sim \mathcal{O}(1)$, in which both quantum pressure and self-interactions are important. The radial component converges after summing up a few multipoles, and is suppressed (but non-vanishing) for subsonic velocities~$\mathcal{M}<1$. On the other hand, the tangential component of the friction (central panel) grows as more multipoles are included, reminiscent of the Coulomb logarithm encountered in the sound regime~\cite{2022ApJ...928...64D}. In this case, however, the divergence is regulated by the inclusion of quantum pressure in the dispersion relation~\cite{Buehler:2022tmr}. The persistence of nonzero friction in the subsonic regime is a consequence of the influence of quantum pressure at low~$\ell$.

For superfluids, the sound speed term plays a stronger role in the dispersion relation compared to quantum pressure,~$\omega_k \simeq c_s k$, as discussed in the previous sections (see, however, Refs.~\cite{Buehler:2022tmr, Berezhiani:2023vlo} for a dedicated discussion of the quantum pressure dominated regime). In this language, the sound regime is defined by the limit~$\ell_{\rm q} \to \infty$ (keeping the Mach number fixed), even though for subsonic motion the condition~$\ell_{\rm q}\gtrsim \mathcal{O}(1)$ provides already a reasonable approximation to compute the tangential component (contrary to the supersonic regime due to the Coulomb logarithm). The corresponding drag force is shown in the right panel of Fig.~\ref{SFforce}, with the tangential component being always non-vanishing in the subsonic regime and characterized by a large bump at~$\mathcal{M} \simeq 1.4$, analogous to the linear-trajectory case~\cite{Ostriker:1998fa}. On the other hand, the radial component vanishes for subsonic motion, while for supersonic velocities it exhibits a series of bumps caused by the overlap between the Mach cone and the sonic sphere. These bumps correspond to instances where the perturber intersects its own wake along its circular orbit~\cite{Kim:2008ab,2022ApJ...928...64D}. 

In the subsonic and sound regime,~$\mathcal{M} < 1$ and~$\ell_{\rm q}\rightarrow \infty$, the tangential component is always larger than its radial counterpart, and can be effectively summed up in the limit of infinite~$\ell_\text{\tiny max}$ to the form~\cite{Berezhiani:2023vlo}
\begin{equation}
\vec{F}_\text{\tiny DF} \cdot \hat{\varphi} =-\frac{4\pi G^2 M^2\rho_0}{v^2} \Big(\text{arctanh} \mathcal{M}-\mathcal{M}\Big)\,.
\label{eq:circularResummed}
\end{equation}
This expression is exact up to~$\mathcal{O}\left(m \Omega/m_\DM c_s\right)$ corrections, it breaks down only for~$\mathcal{M}\geq 1$, and reduces to the drag created by a perturber on a linear trajectory turned on at~$t = 0$~\cite{Ostriker:1998fa}. On the other hand, the radial component does not admit an analytical summed-up version and vanishes when one numerically sums over all multipoles for~$\mathcal{M} \ll 1$.

\subsubsection{Evolution of black hole binaries and gravitational wave dephasing}

The formalism developed above can be applied to study the dynamical friction force experienced by binary systems immersed within a superfluid soliton at the center of galaxies. 
As we discussed in Sec.~\ref{sec:BHSDM}, galaxies are expected to host supermassive BHs at their center, which are responsible for the generation of SDM spikes around them. Since these objects may assemble in binaries with lighter BHs, giving rise to extreme mass-ratio inspirals~\cite{Babak:2017tow}, it is natural to inquire about the strength of the drag force experienced by the light BH as it moves in a circular orbit around the heavier companion. Such binaries could be potentially measured by GW experiments like LISA~\cite{LISA:2017pwj,LISA:2022kgy, LISA:2024hlh} or pulsar timing arrays~\cite{Detweiler:1979wn, 1990ApJ...361..300F}, and therefore may provide information on the superfluid environment~\cite{Berezhiani:2023vlo}.

To simplify the setup, we assume that the spike is formed only around the central, non-spinning supermassive BH with mass~$M_\BH$, ignoring the spike around the lighter BH. Furthermore, we model the binary within BH perturbation theory, by studying the quasi-adiabatic orbital motion of a point-particle with mass~$m_\BH \ll M_\BH$, moving through the superfluid medium and subject to dynamical friction. 
As discussed in Refs.~\cite{DeLuca:2023laa, Berezhiani:2023vlo}, the presence of the DM spike crucially impacts the superfluid density profile, giving rise to an inhomogeneous background. 

To estimate the drag force, we can first appreciate that the light object follows a subsonic motion within the spike, reaching~$\mathcal{M} \approx 1$ only when the BH strongly modifies the DM density profile at~$r \lesssim 10^{-4} \, r_h$~\cite{Berezhiani:2023vlo}, where $r_h$ has been defined as the BH sphere of influence in Eq.~\eqref{BHinfluence}. Furthermore, the quantum-pressure multipole scale~$\ell_{\rm q}$ is found to be always much larger than the finite-size cutoff~$\ell_\text{\tiny max}$~\cite{Berezhiani:2023vlo}
\begin{equation}
 \frac{\ell_{\rm q}}{\ell_\text{\tiny max}}=  \frac{m_\DM v R_\text{\tiny probe}}{\pi\mathcal{M}^2}  \simeq  \frac{10^6}{\pi\mathcal{M}^2} \lp \frac{m_\DM}{\rm eV} \rp \sqrt{\frac{r_h}{r}} \gg 1\,,
\end{equation}
provided that we consider heavy enough DM particles~$m_\DM \gg 10^{-9}$ eV. (For the purpose of this estimate, we considered the conservative case where the Mach number is~$\mathcal{O}(1)$ at~$r \lesssim 10^{-4} r_h$.)
This allows us to ignore the influence of quantum pressure and identify a sound-dominated regime. In this case, as shown in Fig.~\ref{SFforce}, the radial drag component is suppressed when considering a homogeneous medium. 

However, the DM spike around the central BH can in principle invalidate the assumption of homogeneity underlying Fig.~\ref{SFforce}. To check this, one can estimate the radial growth of the density profile over the minimum length scale relevant to the drag force, set by the overdensity size~$G m_\BH/c_s^2$~\cite{Ostriker:1998fa}, to be
\begin{align}
 \frac{G m_\BH}{c_s^2} \bigg| \frac{{\rm d} \log \rho}{{\rm d} r} \bigg| \sim  \frac{G m_\BH}{c_s^2 r} \; \lesssim \; \lp \frac{m_\BH}{M_\BH} \rp \mathcal{M}^2 \ll 1\,.
\end{align}
Thus, we are justified in neglecting the radial component and focus only on the tangential one. The drag force is given by Eq.~\eqref{eq:circularResummed}, replacing the homogeneous density profile~$\rho_0$ with the radially-dependent profile~$\rho(r)$, computed in Sec.~\ref{sec:BHSDM} and shown in Fig.~\ref{SFDMprofile}. 

The prediction for superfluids can be immediately compared to the one for CDM~\cite{Chandrasekhar:1943ys}
\be
\vec{F}_\text{\tiny DF}^\text{\tiny CDM} \cdot \hat{\varphi} 
     = -\frac{4\pi G^2 m_\BH^2 \rho (r)}{v^2}\,,
\ee
where the density profile follows the prediction of Gondolo-Silk, given by Eq.~\eqref{GondoloSilkprofile}. The stronger enhancement of the density profile for collisionless DM compared to SDM, shown in Fig.~\ref{SFDMprofile}, indicates that the friction force for superfluids is orders of magnitude smaller than that for CDM~\cite{Berezhiani:2023vlo}.

\begin{figure}[t!]
	\centering
 \includegraphics[width=0.6\textwidth]{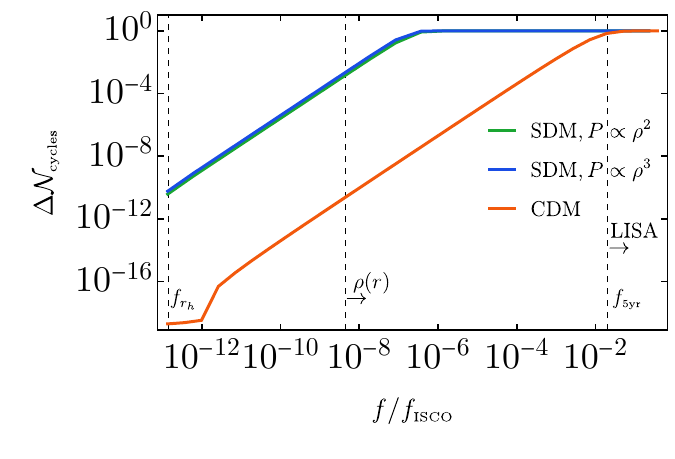}
	\caption{GW dephasing for an extreme mass-ratio inspiral, with BH masses~$M_\BH = 10^6 M_\odot$ and~$m_\BH = 10 \,M_\odot$, as a function of the GW frequency~$f$, for the superfluid models with two- (green) and three-body (blue) interactions, and for a collisionless/CDM medium (orange line). The vertical dashed lines indicate, from left to right, the frequency~$f_{r_h}$ at which~$r = r_h$, the frequency where the BH profile starts deviating from the core,~$\rho(r) \gtrsim \rho_0$, and the frequency~$f_\text{\tiny 5yr}$ corresponding to an observation time of five years for LISA. Figure taken from Ref.~\cite{Berezhiani:2023vlo}.}
	\label{Dephasing}
\end{figure}

This prediction is reflected on the impact of a SDM environment on the evolution of BH binaries. In particular, a non-vanishing drag force is responsible for torques that would change the trajectory of binaries compared to the standard GW emission~\cite{Eda:2014kra, Kavanagh:2020cfn, Coogan:2021uqv}. This can be quantified by computing the dephasing cycles, which account for changes in the GW phase induced by dynamical friction within the DM spike~\cite{Eda:2013gg,Eda:2014kra,Speeney:2022ryg}.\footnote{The interplay of eccentricity and dynamical friction, as well as their contribution to the GW waveform, have been discussed in Refs.~\cite{Yue:2019ozq, Becker:2021ivq}.}
Assuming the stationary phase approximation, the number of GW cycles is defined as
\begin{equation}
\label{eq:Ncycles_def}
\mathcal{N}_\text{\tiny cycles}(f) = \int_f^{f_\text{\tiny ISCO}}\frac{f'}{\dot{f}'}{\rm d}f'\,, 
\end{equation}
in terms of the frequency of the innermost stable circular orbit,~$f_\text{\tiny ISCO} \simeq 3 \cdot 10^{-3} {\rm Hz} \lp M_\BH/ 10^{6} M_\odot \rp^{-1}$, at which the phase of inspiral approximately ends~\cite{Favata:2010ic}, and the orbital frequency evolution~$\dot{f}$. The latter can be computed, in the adiabatic approximation, from the change in the orbital energy due to both GW emission and dynamical friction, respectively given by
\begin{align}
\dot{E}_{\text{\tiny GW}} = -\frac{32}{5} \frac{G^4 m_\BH^2 M_\BH^3}{r^5}\,; \qquad 
\dot{E}_{\text{\tiny DF}} = \vec{v} \cdot \vec{F}_{\text{\tiny DF}} \simeq -4\pi \frac{G^2 m_\BH^2 \rho(r)}{v} \frac{\mathcal{M}^3}{3}\,.
\end{align}
Here, the GW energy loss has been computed, at the lowest post-Newtonian order, from the quadrupole formula for circular orbits~\cite{Maggiore:2007ulw}, while relativistic effects for dynamical friction have been neglected in the orbital evolution~\cite{Barausse:2007ph,Traykova:2021dua,Vicente:2022ivh, Speeney:2024mas, Gliorio:2025cbh, Vicente:2025gsg}. Also, dynamical friction is evaluated using the formula \eqref{eq:circularResummed}, expanded for small Mach numbers.  By splitting the two contributions, the GW dephasing can then be computes as  
\begin{equation}
\Delta \mathcal{N}_\text{\tiny cycles} = \frac{\mathcal{N}_\text{\tiny cycles}|_\text{\tiny GW + DF}}{\mathcal{N}_\text{\tiny cycles}|_\text{\tiny GW}}\,.
\end{equation}
This is shown in Fig.~\ref{Dephasing} as a function of the GW frequency~$f$. Values smaller than unity,~$\Delta \mathcal{N}_\text{\tiny cycles} \lesssim 1$, indicate an efficient role of dynamical friction in driving the binary's inspiral. At small orbital frequencies, corresponding to wide binaries, GW emission is negligible, and dynamical friction provides the main channel through which the binary shrinks, giving rise to a large dephasing. At larger frequencies, on the other hand, GW emission starts to dominate the binary's inspiral, and the dephasing induced by friction is not large enough to be  probed by GW experiments like LISA. Furthermore, the more efficient role of dynamical friction for CDM  gives rise to a bigger dephasing in LISA, allowing to potentially distinguish these two DM candidates~\cite{Berezhiani:2023vlo, Aurrekoetxea:2024cqd}. 

Before closing this Section, let us briefly mention two other phenomena that may affect the binary evolution within a superfluid medium, and eventually impact on the GW dephasing. The first effect concerns the accretion of the light BH moving through the self-interacting scalar cloud~\cite{Yue:2017iwc, Brax:2019npi, Boudon:2022dxi,Boudon:2023qbu}, whose rate for a subsonic motion is given by~\cite{Brax:2019npi}
\begin{equation}
\dot{m}_\BH \simeq 12 \pi F_\star \frac{\rho_0 G^2 m_\BH^2}{c_s^2}\,,
\end{equation}
in terms of a numerical coefficient~$F_\star \simeq 0.66$. This result is lower than the standard Bondi prediction~$\sim \rho_0 G^2 m_\BH^2/v^3$~\cite{Bondi:1952ni}, because of the presence of repulsive self-interactions that can slow down the infall significantly. The second effect is the influence of DM feedback, which hinges on the coherent evolution of the DM distribution to accurately gauge its impact on the emitted GWs~\cite{Kavanagh:2020cfn}. This involves determining the immediate density of DM particles surrounding the secondary object in the case of dynamical friction~\cite{Kavanagh:2020cfn,Coogan:2021uqv}, and accounting for the depletion of particles from the distribution function due to their accretion by the secondary~\cite{Nichols:2023ufs,Boudon:2022dxi,Boudon:2023qbu, Karydas:2024fcn, Kavanagh:2024lgq}. Incorporating these effects would mark a significant stride toward thoroughly characterizing the evolution of binary systems within superfluids.

\section{Emergent phonon-mediated forces in superfluids} 
\label{sec:Forces}

While the~$\Lambda$CDM model has been remarkably successful at matching cosmological observations on large scales, its explanatory power on galactic scales has been the subject of ongoing debate~\cite{Bullock:2017xww}. Despite the complex role of baryonic feedback processes in galaxy formation, galaxies exhibit a surprising degree of regularity, with striking correlations among their physical properties. Disk galaxies display a remarkably tight correlation between the total baryonic mass (stellar + gas) and the gravitational acceleration in galaxies~\cite{McGaugh:2016leg,Lelli:2016cui}. This apparent conspiracy, known as the mass discrepancy acceleration relation (MDAR), states that the acceleration experienced by a baryonic particle (irrespective of whether it is due to DM, new long-range forces, or both) can be uniquely predicted from the baryon density profile. At large distances, the MDAR implies the baryonic Tully-Fisher relation~\cite{McGaugh:2011ac}, which relates the total baryonic mass to the asymptotic/flat rotation velocity as~$M_{\rm b} \sim v^4_{\rm f}$. This relation holds over five decades in mass, with remarkably small scatter.  Another scaling relation is the correlation between the central stellar and dynamical surface densities in disk galaxies~\cite{Lelli:2016uea}. 

Much work has been done to understand how the MDAR arises within~$\Lambda$CDM. Hydrodynamical simulations that generate galaxies whose properties broadly match observations also tend to reproduce the MDAR ({\it e.g.},~\cite{Keller_2017}), although with some discrepancy in the trend and scatter~\cite{Ludlow:2016qzh,Tenneti_2018}. Starting from the early study by~\cite{vandenBosch_2000}, various semi-analytical arguments have been put forth in recent years to explain the MDAR in $\Lambda$CDM~\cite{Desmond:2015nja,Desmond:2016azy,Navarro:2016bfs,Wheeler_2019,Grudic_2020,Paranjape:2021uyu}.

The MDAR was predicted over forty year ago by Milgrom~\cite{Milgrom:1983ca} with MOdified Newtonian Dynamics (MOND) (see Refs.~\cite{Famaey:2011kh, Sanders:2018jiv, McGaugh:2020ppt} for recent reviews). The MOND law states that the gravitational acceleration~$a$ is related to the baryonic acceleration~$a_{\rm b}$ ({\it i.e.}, the Newtonian acceleration due to ordinary matter alone) via
\begin{equation}
a =\left\{\begin{array}{cl} 
a_{\rm b} \qquad ~~~~a_{\rm b} \gg a_0\\[.5cm]
\sqrt{a_{\rm b}a_0} \qquad a_{\rm b} \ll a_0 \,,
\end{array}\right.
\label{MDAR}
\end{equation}
where~$a_0$ is a characteristic acceleration scale. Its best-fit value is intriguingly of order the speed of light times the Hubble constant~$H_0$:
\begin{equation}
a_0 \simeq \frac{1}{6} H_0 \simeq 1.2\times 10^{-8}~{\rm cm}/{\rm s}^2\,.
\label{a0}
\end{equation}
This empirical force law has been shown to fit both galaxy rotation curves~\cite{Famaey:2005fd, Gentile:2010xt}, as well as elliptical galaxies~\cite{Richtler:2011ec, Dabringhausen:2016klu, Chae:2017bhk}. All of the aforementioned galactic scaling relations follow from Milgrom's empirical relation~\eqref{MDAR}. For example, the baryonic Tully-Fisher relation arises naturally in the MONDian regime, since a test particle orbiting around an isolated spherically-symmetric source would satisfy~$\frac{v_{\rm f}^2}{r} = \sqrt{\frac{G M_{\rm b}a_0}{r^2}}$, which implies~$Ga_0 M_{\rm b} = v_{\rm f}^4$. 

Although originally promoted as an alternative to DM, the possibility of completely eliminating the need for DM now seems highly unlikely. The observational case for an additional component behaving as a non-relativistic, collisionless fluid on large scales seems incontrovertible (in this regard, see~\cite{Pardo:2020epc}). Furthermore, from a theoretical standpoint, postulating the existence of a new particle offers a simpler and more tractable explanation for observations than introducing contrived (and possibly non-local~\cite{Deffayet:2014lba,Deffayet:2024ciu}) modifications to Einstein’s theory of gravity, particularly since all known (to us) proposals for MOND exhibit some form of pathology, whereas there exist a plethora of self-consistent and well-motivated DM candidates. Nevertheless, as an empirical fact the success of the MOND law at fitting galactic properties is clear, especially in rotationally-supported systems. Even if DM consists of conventional weakly interacting massive particles, its density distribution within galaxies should ultimately reproduce~\eqref{MDAR}.

It is instructive to explore how Eq.~\eqref{MDAR} can arise from a modification of gravity. A straightforward approach is to consider a scalar field, with a non-relativistic action of the form~\cite{Bekenstein:1984tv}
\begin{equation} 
{\cal L} = - \frac{1}{12\pi G a_0} \left(\big(\vec{\nabla} \phi\big)^2\right)^{3/2} + \phi\rho_{\rm b} \,,
\label{MOND Lag}
\end{equation}
in terms of the baryonic mass density~$\rho_{\rm b}$. The corresponding equation of motion is a non-linear Poisson equation:
\begin{equation}
\vec{\nabla}\cdot \left(\frac{\big\vert\vec{\nabla}\phi\big\vert}{a_0} \vec{\nabla}\phi \right) = 4\pi G \rho_{\rm b}\,.
\end{equation}
Ignoring a homogeneous curl term, which vanishes for spherical symmetry, the solution is~$a_\phi = \big\vert\vec{\nabla}\phi\big\vert = \sqrt{a_{\rm b}a_0}$. The total acceleration,~$a = a_{\rm b} + a_\phi$, is consistent with Eq.~\eqref{MDAR}.

As a theory of a fundamental scalar field, the non-analytic form of the kinetic term in Eq.~\eqref{MOND Lag} is somewhat unpalatable. As first noted in~\cite{Berezhiani:2015bqa,Berezhiani:2015pia,Khoury:2016ehj,Berezhiani:2017tth}, however, the similarity between the non-standard kinetic term in Eq.~\eqref{MOND Lag} and the effective theory of superfluid phonons tantalizingly suggests that the MDAR could arise effectively from a phonon-mediated long-range force, as the result of a direct coupling between SDM and baryonic components. See~\cite{Blanchet:2006yt,Bruneton:2008fk,Blanchet:2008fj,Li:2009zzh,Ho:2010ca, Ho:2011xc, Ho:2012ar, Edmonds:2013hba, Ng:2016qvh, Blanchet:2015sra,Blanchet:2015bia,Cai:2015rns, Verlinde:2016toy,Salucci:2017cet,Skordis:2020eui,Skordis:2021mry} for other DM-MOND hybrid ideas explored in the literature.

In this Section we consider the general possibility that the MDAR is an environmentally emergent phenomenon, from the DM microscopic properties and interactions with baryons. We discuss some of the features of the required SDM for realizing this idea. Not surprisingly, we will see that this requires a somewhat more exotic superfluid model compared to the vanilla SDM discussed in previous sections. We then discuss the phenomenology of such SDM theories, including the observational constraints and some of the challenges of this paradigm. Finally, to further motivate the approach and generalize the results beyond the MDAR application, we discuss a prototype SDM scenario with an emergent effective long-range interaction between external baryonic perturbations, submerged within the DM condensate. 
These results will lay the foundation for further studies on the interplay between DM superfluids and baryons in galaxies.

It is worth stressing that the desired DM profile may naturally emerge also as the equilibrium configuration from DM-baryon, short-range, collisional interactions, whose necessary properties can be reverse-engineered from observational insights related to the MDAR~\cite{Famaey:2017xou,Famaey:2019baq}. In particular, one needs a relaxation time of the order of the dynamical time, and a DM-baryon cross section which is inversely proportional to the DM number density and velocity squared,~$\sigma \propto 1/n v^2$. The velocity dependence can arise with Sommerfeld enhancements or charge-dipole interactions, while the density dependence may arise from an environmental dependence of DM-baryon couplings.

\subsection{Effective field theory of phonon-mediated force}

As already discussed, the low-energy effective theory of zero-temperature superfluids, in the presence of a Newtonian potential~$\Phi_\text{\tiny N}$, is given by the phonon Lagrangian
\be
{\cal L} = P(X)\,;\qquad X = \mu_\text{\tiny NR} - m \Phi_\text{\tiny N} + \dot{\pi} - \frac{(\vec{\nabla} \pi)^2}{2m}\,,
\ee
where~$\mu_\text{\tiny NR}$ is the non-relativistic chemical potential. The functional form of~$P(X)$ specifies the superfluid equation of state. To reproduce Eq.~\eqref{MOND Lag} in the static limit, we consider the example~\cite{Berezhiani:2015bqa, Berezhiani:2015pia}
\begin{equation}
{\cal L} = \frac{2\Lambda(2m)^{3/2}}{3} X\sqrt{|X|} - \alpha  \frac{\Lambda}{M_{\rm Pl}} \pi \rho_{\rm b}\,,
\label{LMOND}
\end{equation}
where the scale~$\Lambda$ will be fixed shortly. Let us make a few remarks:

\begin{itemize}

\item The appearance of the fractional power of~$X$ is not {\it a priori} an issue. For instance, a superfluid comprised of massive particles with repulsive $3$-body contact interaction, {\it i.e.}, with~$\Phi^6$ potential, is described by~$\mathcal{L}\propto X^{3/2}$. This corresponds to a polytropic equation of state~$P\propto \rho^3$. Thus, despite the non-analytic dependence on~$X$, the equation of state is analytic in~$\rho$. A well-known example of a theory with fractional power in cold atom systems is the Unitary Fermi Gas~\cite{Giorgini:2008zz,Randeria:2013kda}, which describes fermionic atoms tuned at unitary. The superfluid action in this case is fixed by non-relativistic scale invariance to the non-analytic form ${\cal L}_{\rm UFG}(X) \sim X^{5/2}$~\cite{Son:2005rv}. 

\item The appearance of the absolute value under the square root is less trivial to generate and is due to the necessity to source a real-valued phonon configuration in the limit of large gradients, {\it i.e.}, for $|\vec{\nabla} \pi|/m\gg c_s$. A key challenge is to find a  microscopic theory of DM particles, from which one would have a tractable derivation of \eqref{LMOND} (see Refs.~\cite{Addazi:2018ivg, Hertzberg:2021fvu} for relevant work in this direction).

\item The phonon-baryon coupling, with dimensionless parameter~$\alpha$, is introduced to generate a phonon-mediated long-range force between ordinary matter. This~$M_{\rm Pl}$-suppressed operator explicitly breaks the shift symmetry of the theory, albeit weakly, making the phonon~$\pi$ a pseudo-Goldstone boson.  
It could arise, for instance, if the superfluid includes two components coupled through a Rabi-Josephson interaction~\cite{Ferreira:2018wup}. (See  also~\cite{Mistele:2019byy} for a discussion on the interplay between the superfluid chemical potential and the explicit coupling to baryons.)

\end{itemize}

The action~\eqref{LMOND} reproduces the low acceleration limit of \eqref{MDAR} in the limit~$(\vec{\nabla} \pi)^2 \gg 2 m \hat{\mu}_\text{\tiny NR}$, with $\hat{\mu}_\text{\tiny NR} \equiv \mu_\text{\tiny NR} - m \Phi_\text{\tiny N}$. Indeed, up to a homogeneous curl term, we have $\vec{\nabla} \pi |\vec{\nabla} \pi| \simeq \alpha M_{\rm Pl} \vec{a}_{\rm b}$, in terms of the Newtonian acceleration~$\vec{a}_{\rm b}$ due to baryons. Using the fact that the $\pi$-mediated acceleration from the interaction term is~$\vec{a}_\pi = \alpha \Lambda \vec{\nabla} \pi/ M_{\rm Pl}$, one reproduces the deep MONDian form
\begin{equation}
a_\pi = \sqrt{\frac{\alpha^3 \Lambda^2}{M_{\rm Pl}} a_{\rm b}} \equiv \sqrt{a_0 a_{\rm b}}\,, \qquad a_0 =  \frac{\alpha^3 \Lambda^2}{M_{\rm Pl}}\,.
\label{a0 expression}
\end{equation}
With~$\alpha \sim {\cal O}(1)$, we see that~$\Lambda\sim {\rm meV}$ in order to reproduce the critical acceleration~\eqref{a0}.
A key difference compared to the standard results of MOND is that the DM halo itself contributes as a gravitational acceleration~$a_\DM$ on baryons, such that the  total acceleration reads
\begin{equation}
\vec{a} = \vec{a}_{\rm b} + \vec{a}_\DM + \vec{a}_\pi\,.
\label{a total}
\end{equation}
The new contribution~$a_\DM$ is subdominant at distances probed by galactic rotation curves, but becomes comparable to the phonon force at distances approximately equal to the size of the superfluid core.

While the zero-temperature action~\eqref{LMOND} reproduces the MDAR for spherically-symmetric profiles, linearized perturbations around such backgrounds are unstable~\cite{Berezhiani:2015pia}. However, it was noted that these instabilities can be cured by finite-temperature effects (see Sec.~\ref{sec: finitetemperature}). A simple finite-temperature modification of~\eqref{LMOND} that does not suffer from such pathology is
\begin{equation}
\label{LagrMONDT}
{\cal L} = \frac{2\Lambda(2m)^{3/2}}{3} X\sqrt{|X - \beta Y|} + \alpha \frac{\Lambda}{M_{\rm Pl}} \pi \rho_{\rm b}\,,
\end{equation}
where~$Y = \hat{\mu}_\text{\tiny NR} + \dot{\pi} + \vec{v} \cdot \vec{\nabla} \pi$  depends on the velocity $\vec{v}$ of the normal fluid component. The coefficient~$\beta$ parametrizes the size of temperature-dependent effects, with~$\beta \to 0$ to recover the original action of Eq.~\eqref{LMOND}. This parameter should implicitly depend on~$T/T_{\rm c}$, even though for simplicity we treat it as a constant. As shown in Refs.~\cite{Berezhiani:2015bqa, Berezhiani:2015pia},~$\beta \geq 3/2$ is necessary to reproduce the MDAR with stable perturbations.

The static equation of motion for phonons is~\cite{Berezhiani:2015bqa, Berezhiani:2015pia} 
\begin{equation}
\vec{\nabla} \cdot \left( \frac{\big(\vec{\nabla} \pi\big)^2 + 2 m \left(\frac{2\beta}{3} - 1\right) \hat{\mu}_\text{\tiny NR}}{\sqrt{\big(\vec{\nabla} \pi\big)^2 + 2 m (\beta - 1) \hat{\mu}_\text{\tiny NR}}} \vec{\nabla} \pi \right) = \frac{\alpha \rho_{\rm b}}{2 M_{\rm Pl}}\,.  
\end{equation}
Assuming spherical symmetry, this can be readily integrated to give 
\begin{equation}
\frac{\pi'^{\,2} + 2 m \left(\frac{2\beta}{3} - 1\right) \hat{\mu}_\text{\tiny NR}}{\sqrt{\pi'^{\,2} + 2 m (\beta - 1) \hat{\mu}_\text{\tiny NR}}}\pi'   = \alpha M_{\rm Pl} a_{\rm b}(r)\,, 
\end{equation}
where primes denote radial derivatives, and~$a_{\rm b}(r) = \frac{GM_{\rm b}(r)}{r^2}$. 
This implies a cubic equation for~$\pi'^{\,2}$, whose real root does not have a particularly illuminating
analytic form. Instead, it is useful to consider different limiting regimes, following~\cite{Berezhiani:2015bqa, Berezhiani:2015pia}. Sufficiently close to the baryon source ({\it i.e.}, in the central regions of galaxies), one has~$\pi'^{\,2} \gg 2 m \hat{\mu}_\text{\tiny NR}$, such that the desired behavior is obtained,~$\pi' = \sqrt{\alpha M_{\rm Pl} a_{\rm b}(r)}$. In the opposite regime far from baryons, $\pi'^{\,2} \ll 2 m \hat{\mu}_\text{\tiny NR}$, the solution instead scales like~$\pi' = \sqrt{\frac{3}{2 m \hat{\mu}_\text{\tiny NR}}} \sqrt{\frac{\beta-1}{2\beta-3}} \alpha M_{\rm Pl} a_{\rm b}(r)$ for approximately constant~$\hat{\mu}_\text{\tiny NR}$. Thus the~$\pi$-mediated acceleration takes the form
\begin{equation}
a_\pi(r) \simeq \begin{cases}
   \sqrt{a_0 a_{\rm b}(r)} \,, \qquad~~~~~~~~~~~~~~~~~~~ \text{phonon-dominated regime}\,;  \\
   \sqrt{\frac{3}{2 m \hat{\mu}_\text{\tiny NR}}} \sqrt{\frac{\beta-1}{2\beta-3}} \alpha^2 \Lambda a_{\rm b}(r)\,, \quad ~~~\text{DM halo-dominated regime}\,,
\end{cases}
\label{piacc}
\end{equation}
where we have used the expression~\eqref{a0 expression} for~$a_0$. In particular, outside of the baryonic source~($M_{\rm b} = {\rm const.}$), the~$\pi$-mediated acceleration scales as~$1/r$ and~$1/r^2$ respectively.

To compute the total acceleration~\eqref{a total}, one must also calculate the DM contribution~$a_\DM$. The latter can be derived from the DM density profile~$\rho_\DM(r)$, obtained by varying~\eqref{LagrMONDT} with respect~$\Phi_\text{\tiny N}$ as~\cite{Berezhiani:2017tth}:
\begin{equation}
 \rho_\DM(r) = \frac{2 \sqrt{2} m^{5/2} \Lambda \left( 3 (\beta-1)\hat{\mu}_\text{\tiny NR} + (3-\beta) \frac{\pi'^{\,2}}{2m} \right)}{3 \sqrt{(\beta-1)\hat{\mu}_\text{\tiny NR} + \frac{\pi'^{\,2}}{2m}} }\,.
\end{equation}
Working outside the baryon source, for simplicity, the DM acceleration reads
\begin{equation}
a_\DM \simeq \begin{cases}
   \frac{m^2 \Lambda}{3 M_{\rm Pl}^2} (3 - \beta)\sqrt{\alpha M_{\rm Pl}  a_{\rm b} r^2} \sim r^0\,, \quad~~~\, \text{phonon-dominated regime}\,;  \\
   \frac{\sqrt{2} m^{5/2} \Lambda}{M_{\rm Pl}^2} \sqrt{(\beta - 1)\hat{\mu}_\text{\tiny NR}r^2} \sim r \,, \quad~~~~~ \text{DM halo-dominated regime}\,.
\end{cases}
\end{equation}
When compared to the $\pi$-mediated acceleration of~\eqref{piacc}, this confirms the hierarchy between the two contributions in the total acceleration.

A more precise calculation of the accelerations in the two regimes is described in detail in Ref.~\cite{Berezhiani:2017tth}. There, as an example, the corresponding density profiles were fitted to two representative galaxies, a low surface brightness galaxy IC~2574 and a high surface brightness galaxy UGC~2953, providing an excellent fit to their rotation curves. Phenomenological implications for other astrophysical systems were also discussed in Refs.~\cite{Berezhiani:2017tth,Hodson:2016rck,Hossenfelder:2018iym,Berezhiani:2019pzd}.\footnote{See also Ref.~\cite{Banerjee:2020obz} for simulations of structure formation in MONDian SDM models, where strong DM–baryon interactions lead to faster structure growth compared to~$\Lambda$CDM, potentially allowing constraints to be placed on the coupling strength to baryons.}

\begin{figure}[t!]
	\centering
	\includegraphics[width=0.39\textwidth]{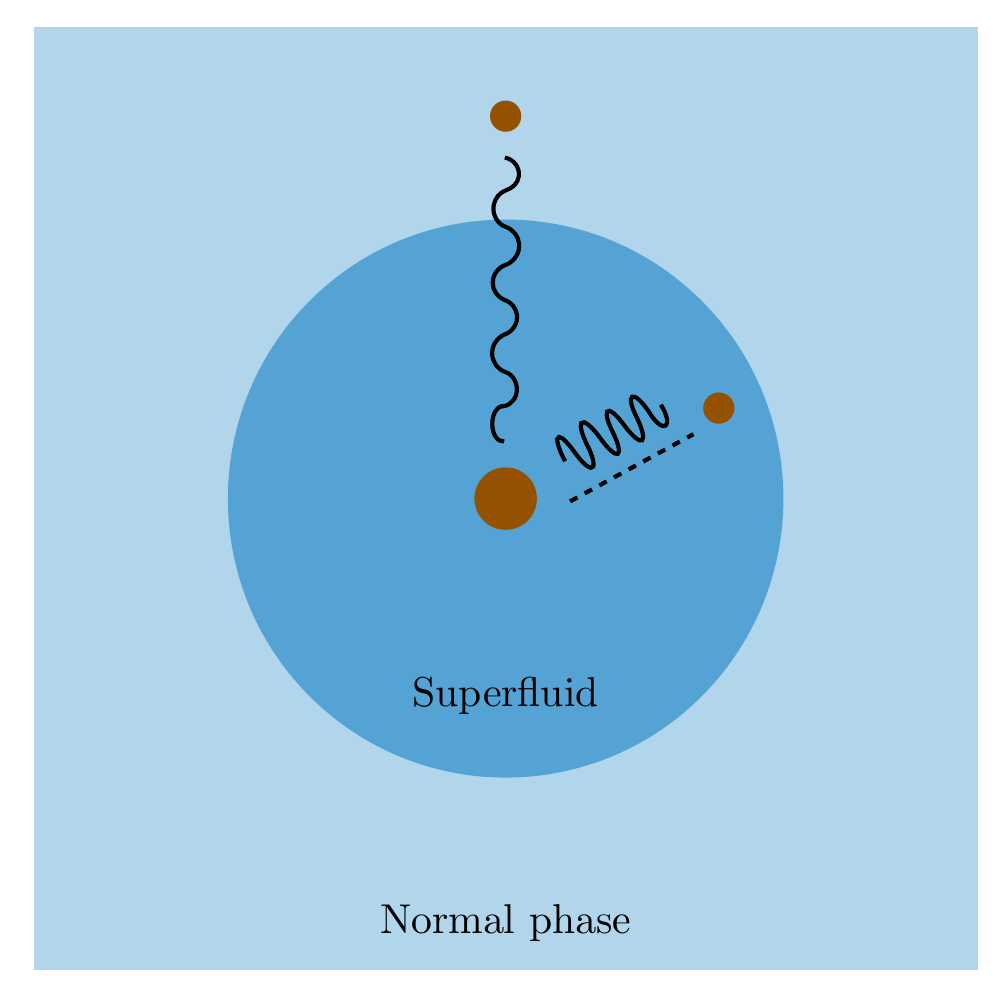}
 \hspace{2cm}
   \includegraphics[width=0.47\textwidth]{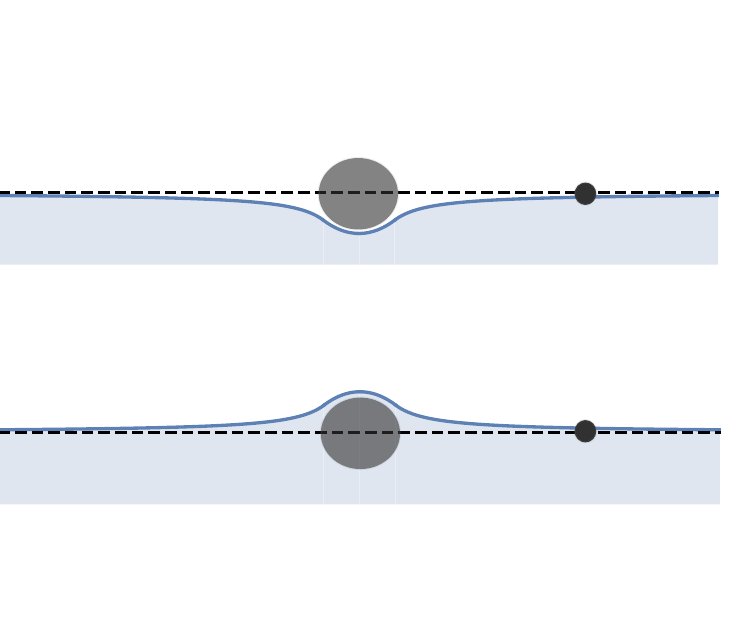}
	\caption{{\it Left panel}: Galactic forces between baryons. The light blue region depicts the envelope of the normal DM component, enclosing the superfluid core (blue area), while brown disks depict baryons. Wavy lines correspond to the gravitational interaction sourced both by baryons and DM, while the dashed line depicts the phonon-mediated force. {\it Right panel}: Physical picture of environmentally emergent forces between baryons (gray and black disks) as a result of the displacement of the DM condensate (light blue).}
	\label{figforces}
\end{figure}

Compared to the ``vanilla'' SDM scenario discussed in earlier sections, the twist of the SDM scenario proposed in~\cite{Berezhiani:2015bqa,Berezhiani:2015pia,Berezhiani:2017tth} is to posit the environmentally emergent, phonon-mediated long-range force within the coherence length of the superfluid core. The left panel of Fig.~\ref{figforces} summarizes the qualitative picture. Baryons submerged within the superfluid core would experience both gravitational forces and the phonon-mediated force. However, the latter would be lost outside the core due to the loss of coherence.

Although the idea of a phonon-mediated force may at first seem somewhat exotic, it has a clear physical meaning in the language of fluid mechanics. The idea is captured by the right panel of Fig.~\ref{figforces}, which shows the redistribution of DM condensate in the presence of a central baryon lump (large gray disk). There are two possible scenarios, depending on the sign of the coupling between baryons and DM, which determines whether DM is expelled by or accreted onto the baryons. The far away probe baryon (small black disk) can feel the redistribution of the DM condensate. Furthermore, if it couples to DM in the same way as the central lump ({\it e.g.}, if both are comprised of the same particles) then it will be pulled towards it in both pictures. 

\subsection{Phenomenology and bounds}

In the following, we summarize some of the main phenomenological predictions and implications of MONDian superfluids, discussing possible bounds on the parameters of the theory from various observational constraints.

\begin{itemize}

\item {\bf External field effect:} A key aspect of MONDian phenomenology is the external field effect. Consider a subsystem with small internal acceleration~$a_\text{\tiny int} \ll a_0$ in the background of a large homogeneous external acceleration~$a_\text{\tiny ext} \gg a_0$. 
In General Relativity, the external acceleration can be removed by a coordinate transformation (going to the freely-falling frame) and does not have physical consequences for the internal dynamics. In MOND, however, it is physical and renders the subsystem Newtonian.

The most notable observable consequence of the external field effect is that isolated galaxies should exhibit asymptotically flat rotation curves, whereas galaxies in dense environments are expected to show declining curves at larger radii. Refs.~\cite{Chae:2020omu,Chae:2021dzt} have presented evidence of a statistical detection of this effect in galaxies from the SPARC sample. Interestingly, it has been argued in~\cite{Paranjape:2021eng} that an effective external field effect is expected statistically in CDM models with realistic galaxy rotation curves.

In field theory, the external field phenomenon arises in scalar field theories with kinetic interactions, and is an example of the kinetic screening mechanisms~\cite{Babichev:2009ee,Babichev:2011kq,Brax:2012jr, Burrage:2014uwa}. For example, in~$P(X)$ theories, non-linearities in~$\vec {\nabla} \pi$ can result in the suppression of scalar field effects locally and the recovery of standard gravity~\cite{Joyce:2014kja}.

In case of phonon-mediated forces in DM superfluids, the long-range force between two bodies only applies when both bodies are within the superfluid region. The same requirement also applies to the external field effect~\cite{Berezhiani:2017tth}. For example, within the superfluid core of the Milky Way, we expect both satellite galaxies and globular clusters to be affected by the external field effect. Dwarf spheroidal galaxies would be affected by this phenomenon only if they reside within sufficiently large superfluid cores. In contrast, tidal dwarf galaxies---formed from the interactions of massive spiral galaxies---are not expected to host substantial DM halos and should therefore exhibit Newtonian dynamics~\cite{Lelli:2015rba, Lelli:2015yle}. Lastly, galaxies within a galaxy cluster, which should have a small superfluid core, are not expected to experience any external field effect since they are orbiting outside of the cluster superfluid core.

\item {\bf Solar system tests:} Although typical accelerations in the solar system are large compared to~$a_0$, post-Newtonian tests are sensitive to small corrections to Newtonian gravity and require further suppression of the photon-mediated force at short distances. The superfluid framework offers an elegant explanation. As argued in~\cite{Berezhiani:2015pia,Berezhiani:2015bqa}, the large phonon gradient induced by an individual star results in a breakdown of superfluidity in its vicinity. Specifically, the local superfluid velocity~$|\vec{\nabla}\pi|/m$ sourced by the Sun exceeds the Bose-Einstein condensate critical velocity $\sim (\rho/m)^{1/3}$ in our Milky Way neighborhood within a distance of order~${\cal O}\big(10^2\big)$~AU from the Sun. Thus coherence is lost within the solar system. Individual DM particles still interact with baryons, but no long-range force can be mediated. Other possibilities include modifying~$P(X)$ at large~$X$~\cite{Bruneton:2007si} and/or introducing higher-derivative operators to achieve local screening~\cite{Babichev:2011kq}.

\item {\bf Cherenkov radiation:} As pointed out in~\cite{Mistele:2021qvz,Mistele:2022dfj}, the presence of a direct coupling between phonons and baryons can lead to orbital decay for supersonic objects, such as stars, through phonon Cherenkov radiation. It was demonstrated that certain orbits are significantly affected by this channel of energy dissipation, leading to lifetimes $\lesssim 10~{\rm Gyr}$, for the realization of SDM given by~\eqref{LagrMONDT}. This effect, while technically correct, does rely on the validity of the superfluid EFT on small scales, which as argued above must break down in the vicinity of stars.\footnote{As yet another possibility,~\cite{Mistele:2020qha, Mistele:2022dfj} considered a two-fluid model of SDM, where one phonon field couples to matter and is responsible for mediating a MOND-like force, while the other phonon field carries the superfluid energy density. Since only this component has a non-relativistic sound speed and can thus be potentially radiated away as Cherenkov radiation, the suppressed coupling between the fields allows for only a small amount of radiation emitted in the process, avoiding the corresponding constraint~\cite{Mistele:2022dfj}.}

\item {\bf Kinematic observables:} As discussed above, MOND-like features of SDM are confined within the superfluid core, where the phonon force is active. The DM envelope engulfing the soliton, on the other hand, is expected to be in the form of the NFW profile without phonon-mediated force. Therefore, the SDM predictions for kinematic observables within the envelope are expected to be similar to CDM with cored NFW density profile~\cite{Berezhiani:2017tth}; see also~\cite{Mistele_2023}. A detailed numerical study of the distribution of DM in the envelope is thus required before a detailed fit with kinematic observables can be performed outside the core.

It was argued in~\cite{Lisanti:2018qam,Lisanti:2019nmn} that the phonon-mediated force poses a problem in simultaneously explaining the galaxy rotation curve and vertical velocity dispersion data in the Solar vicinity~\cite{Lisanti:2018qam}. Specifically, the acceleration in the direction perpendicular to the disk plane may over-predict vertical velocity dispersions, providing a source of tension with local Milky Way data~\cite{Zhang:2012rsb, deSalas:2019pee}. However, central to the Jeans analysis of~\cite{Lisanti:2018qam,Lisanti:2019nmn} is the assumption that the Milky Way disk is in steady state, which is at odds with the observed disequilibrium of disk stars~\cite{Widrow:2012wu,Bennett_2018}. Indeed, as can be clearly seen in Fig.~1 of~\cite{Lisanti:2019nmn}, neither the superfluid or CDM model assuming steady state offers a particularly good fit to the observed vertical velocity dispersions.  

Lastly, weak gravitational lensing may also place constraints on the model. As pointed out in~\cite{Berezhiani:2017tth}, the phonon-force within the superfluid core should not affect photons in order to avoid constraints from GW observations like GW170817, where the gravitational signal arrives almost simultaneously as the electromagnetic one; see also~\cite{Sanders:2018jiv, Boran:2017rdn, Hossenfelder:2018iym}.\footnote{However, let us notice that the presence of Bose-Einstein condensates along the line-of-sight of propagating GWs may in principle induce a deviation, albeit tiny, in the GW speed, due to the change in the effective refractive index associated to phonon excitations~\cite{Dev:2016hxv, Cai:2017buj, Banerjee:2022zii}.} This implies that gravitational lensing in SDM should work similar to CDM with cored-NFW profile (see also~\cite{Mistele_2023}). 

\end{itemize}

\subsection{Emergent long-range interaction from~$U(1)$-invariant coupling}

The phonon-mediated force considered above hinges on explicitly breaking (albeit weakly) the global~$U(1)$ symmetry associated with particle number conservation. We will now show the emergence of a long-range force, which arises from the~$U(1)$-invariant coupling of the modulus of the complex scalar field to baryons~\cite{Berezhiani:2018oxf}. Based on the intuition from the Higgs mechanism, one would expect that the modulus is a heavy mode. However, because the superfluid state breaks Lorentz invariance spontaneously, both heavy and light modes are linear combinations of the phase and the modulus of the complex scalar field, such that the range of the force mediated by the modulus can be much longer than the Compton wavelength of the heavy mode. 

Concretely, we will see that in a theory with a mass gap and collisional particle-number-preserving interactions only, the formation of a Bose-Einstein condensate leads to the appearance of forces with range set by the healing length of the condensate. This effect had been studied previously in the absence of self-interactions, corresponding to infinite healing length~\cite{Ferrer:2000hm}. In that case, in the presence of a heat bath of ideal gas, finite-temperature/density corrections to the propagator of the massive particles result in the emergence of a long-range force below the critical temperature. This phenomenon is the bosonic counterpart of the Kohn-Luttinger effect~\cite{Kohn:1965zz}, which transforms repulsive contact interactions of fermions into long-range attractive forces in the presence of a Fermi sea.

Consider a complex scalar field with quartic self-interactions and $U(1)$-invariant coupling to a source $J$,
\beq
\mathcal{L}=-|\partial\Phi|^2-m^2|\Phi|^2-\frac{\lambda_4}{2}|\Phi|^4-\frac{1}{\Lambda^2}|\Phi|^2J\,.
\label{lag}
\eeq
Here, $\Lambda$ denotes the scale associated with the coupling to the source $J$. The presence of a mass gap in the theory implies the absence of long-range correlations in vacuum, {\it i.e.}, the $\Phi$-mediated force between two sources due to their contact interactions is necessarily short-range. The situation changes, however, if we submerge the sources within a condensate of~$\Phi$ particles, given by Eq.~\eqref{sect2:fieldconf}. In this case, the source will distort the homogeneous field configuration, as in Eq.~\eqref{perturbedSol}, such that the Lagrangian becomes 
\begin{align}
\label{pertlag}
\mathcal{L}=&~\dot h^2-\big(\vec{\nabla}h\big)^2+(v+h)^2\Big( \lambda_4 v^2 +2\mu\dot{\pi}+\dot{\pi}^2-\big(\vec{\nabla}\pi\big)^2\Big)-\frac{\lambda_4}{2}(v+h)^4-\frac{(v+h)^2}{\Lambda^2}J\,.
\end{align}
Importantly, even though the phase of~$\Phi$ is decoupled from~$J$ due to the $U(1)$ invariance in Eq.~\eqref{lag}, the spontaneous breaking of Lorentz invariance by the background~$v$ induces a mixing between the gapless mode~$\pi$ and the modulus~$h$. This is key to obtain a force with a range much longer than the Compton wavelength~$(2m)^{-1}$ of the heavy mode.

To derive the force law, let us focus on static sources, setting~$\dot{h}= \dot{\pi} = 0$.\footnote{For a real scalar field undergoing Bose-Einstein condensation, the long-range force would be characterized by an additional rapid oscillatory prefactor~\cite{dePireySaintAlby:2017lwc} that can be removed by averaging over time.} The Lagrangian reduces to
\begin{align}
\mathcal{L}_\text{\tiny static}=&-\big(\vec{\nabla}h\big)^2+(v+h)^2\Big( \lambda_4 v^2-\big(\vec{\nabla}\pi\big)^2 \Big)-\frac{\lambda_4}{2}(v+h)^4-\frac{(v+h)^2}{\Lambda^2}J\,.
\label{staticlag}
\end{align}
Since~$\pi$ does not couple to the source directly, it can be set to zero without contradicting the classical equations of motion.
Assuming small deviations from the superfluid state,~$h\ll v$, one obtains the following linear equation of motion for~$h$
\beq
\Big(\Delta-2\lambda_4 v^2\Big)h=\frac{v}{\Lambda^2}J\,.
\label{h eom linear}
\eeq
A point-like source,~$J=M\delta^{3}(\vec{x})$, gives rise to the Yukawa profile
\beq
h(r)=-\frac{v}{\Lambda^2}\frac{M}{4\pi r}{\rm e}^{-r/\ell}\,; \qquad \ell=\frac{1}{\sqrt{2\lambda_4 v^2}} \simeq \frac{1}{2 m c_s} \gg \frac{1}{2m}\,.
\label{hlinearprofile}
\eeq
Evidently, the force is an inverse-square law $\sim 1/r^2$ for $r\ll \ell$, it is exponentially suppressed at distances greater than the healing length, where self-interactions start to smooth out the perturbation induced in the condensate by the localized source,
and it becomes long-range~($\ell\rightarrow\infty$) in the limit of vanishing self-interaction,~$\lambda_4\rightarrow 0$. 

While a~$1/r^2$ force is usually associated with the existence of a massless mediator with linear dispersion relation~$\omega_k\sim k$, here it is associated in the limit~$\lambda_4\rightarrow 0$ with a quadratic gapless dispersion relation,~$\omega_k=k^2/2m$. This is due to the kinetic mixing term with which the gapless mode enters the amplitude of the complex scalar field, as discussed in the following inset.

\begin{framed}
{\small
\vspace{-.15cm}
\noindent 
To clarify the role of kinetic mixing, consider the Lagrangian~\eqref{pertlag} in the absence of self-interactions~($\lambda_4 = 0$), at leading order in perturbations:
\beq
\mathcal{L}_{\lambda_4=0}=\dot h^2-\big(\vec{\nabla} h\big)^2+\dot{\tilde{\pi}}^2-\big(\vec{\nabla}\tilde{\pi}\big)^2+4mh\dot{\tilde{\pi}}
-2\frac{v}{\Lambda^2}hJ\,,
\label{linth}
\eeq
where~$\tilde{\pi}\equiv\pi v$. Notice that the kinetic part is symmetric in~$h$ and~$\tilde{\pi}$ up to a mixing term~$h\dot{\tilde{\pi}}$, proportional to the mass, without which there would be two gapless modes in the spectrum. The observation that~$h$ becomes massless in the absence of kinetic mixing is precisely why it can mediate a long-range force between static sources. 

To make this more transparent, one can integrate out~$\tilde{\pi}$ by substituting the solution to its equation of motion,~$\tilde{\pi}=\frac{2m}{\Box}\dot{h}$. One obtains
\beq
\mathcal{L}^h_{\lambda_4=0}=\dot{h} \frac{(2m)^2}{-\Box} \dot{h}+h\Box h-2\frac{v}{\Lambda^2}hJ\,.
\eeq
Performing the field redefinition $h = \frac{\sqrt{-\Box}}{2m} h_c$, and taking the non-relativistic limit $k\ll m$, which implies that the frequency of the gapless mode satisfies~$\omega_k \ll m$ and thus freezing out the gapped mode (since its on-shell production is energetically forbidden), one gets 
\beq
\mathcal{L}^{h_c}_{\lambda_4=0}=\dot{h}_c^2-h_c\frac{\Delta^2}{4 m^2} h_c-2\frac{v}{\Lambda^2} J \frac{\sqrt{-\Delta}}{2m}h_c \,.
\label{hlin}
\eeq
This formulation shows manifestly that the only dynamical degree of freedom is a phonon with dispersion relation $\omega_k=k^2/2m$, which couples to
the source with a momentum-dependent form factor. The combination of these insights results in the gapless mode mediating a $1/r^2$ force. 

With self-interactions turned on, phonons have a non-vanishing sound speed~$c_s^2\equiv \lambda_4 v^2/2m^2$, such that the Lagrangian becomes
\beq
\mathcal{L}^{h_c}_{\lambda_4>0} = \dot{h}_c^2-h_c\left(-c_s^2\Delta+\frac{\Delta^2}{4 m^2} \right)h_c-2\frac{v}{\Lambda^2}J \frac{\sqrt{-\Delta}}{2m}h_c\,,
\label{hlinlambda}
\eeq
The range of the force shrinks to~$\ell=(2mc_s)^{-1}$, as shown in Eq.~\eqref{hlinearprofile}, even though the mediator remains gapless.
The qualitative understanding of this behavior readily follows from Fig.~\ref{figforces}. For finite sound speed, the healing length of the condensate is finite, which in turn sets the distance beyond which self-interactions begin to erase the dent created in the condensate by the localized source. 
}
\end{framed}

\begin{figure}[t!]
	\centering
	\includegraphics[width=0.43\textwidth]{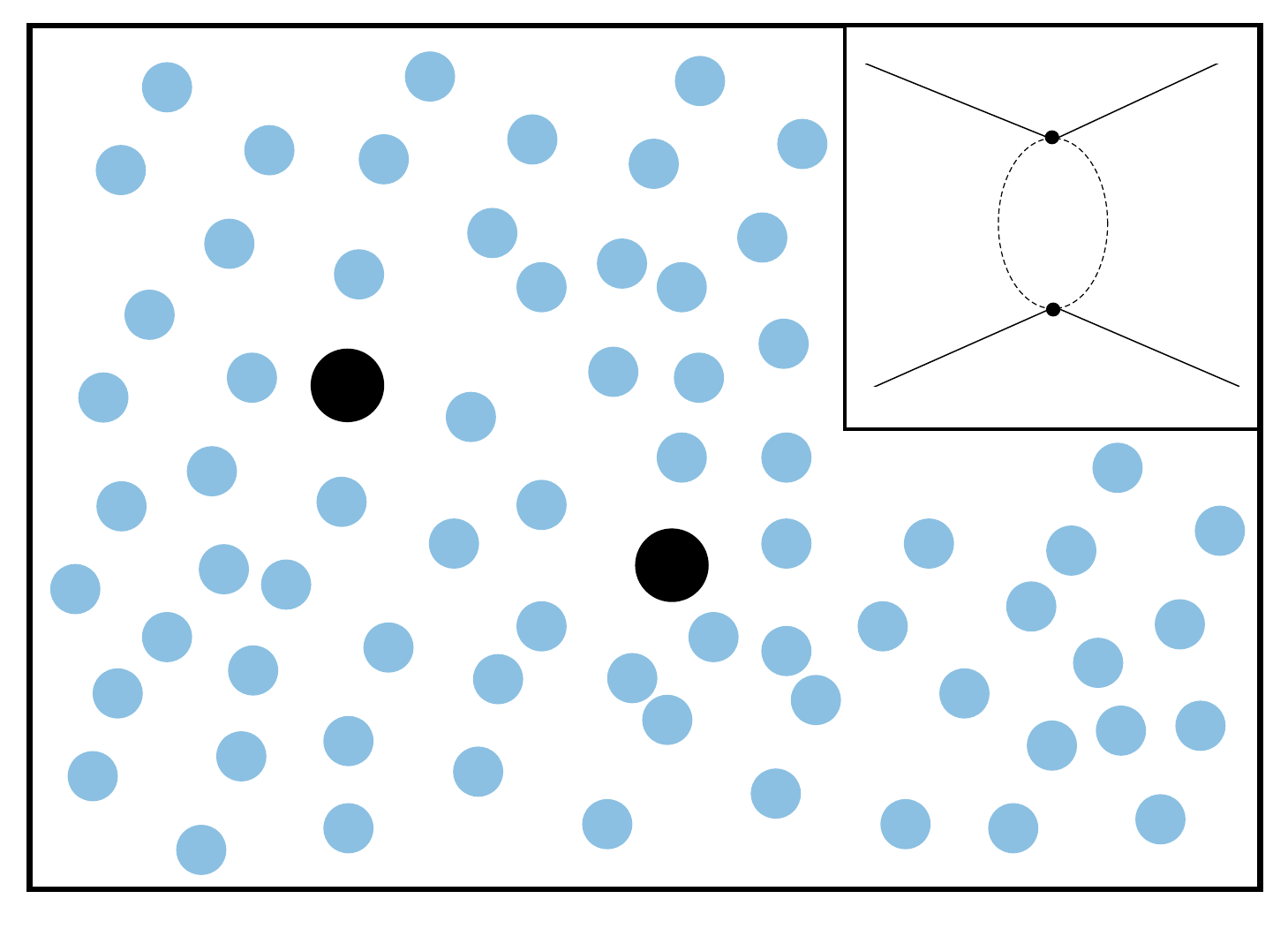} \hskip 20pt \includegraphics[width=0.424\textwidth]{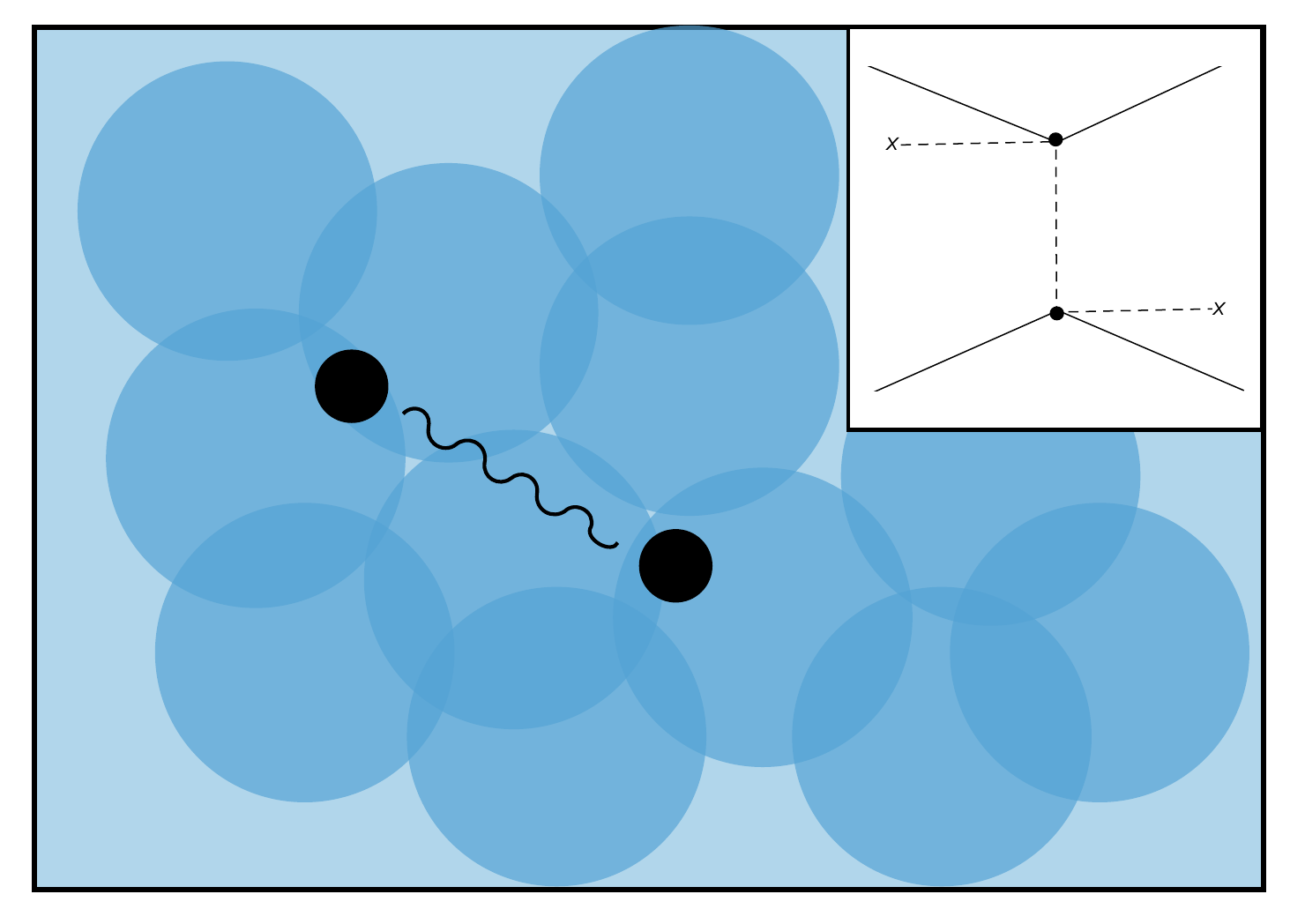}\\
	\caption{The environmentally emergent force as a result of the exchange of the condensate particle. \textit{Left panel:} Two sources (black disks) submerged in a non-degenerate gas of~$\Phi$ particles (blue disks, with the diameter representing the de Broglie wavelength). The Feynman diagram depicts the leading interaction between sources in the vacuum, {\it i.e.}, in the absence of the condensate. \textit{Right panel:} The picture corresponds to subcritical temperatures when the de Broglie wavelengths begin to overlap significantly, while the Feynman diagram depicts the leading-order interaction between sources in the presence of the condensate.}
	\label{figmicro}
\end{figure}

The mechanism of this emergent force is sketched in Fig.~\ref{figmicro}. In the left panel, the system is above the critical temperature and is in the form of the gas of~$\Phi$ particles with non-overlapping de Broglie wavelength. In this case, sources can interact with the exchange of the loop of massive particles, which is inevitably short-range because of the absence of coherence. In the right panel, the system is at subcritical temperature, such as the condensate is formed due to the significantly overlapping wavefunctions. In this case, the thermal loop correction at large distances is dominated by the condensate insertions, depicted as crosses in the diagram in the top-right inset. In other words, a constituent of the condensate, whose wavefunction is delocalized on scales larger than the separation between sources, can be virtually borrowed from one source and then deposited back to another after the exchange. The range of such force is determined by the off-shellness of the exchanged particles. In the Feynman diagram of the left panel, sources exchange the quantum loop of a massive particle (replaced by a single massive propagator in the case of a linear coupling),  with the mass setting just how off-shell virtual particles must be. On the other hand, in the presence of a condensate, the particles are already present, and therefore there is no mass-gap cost (as the virtual energy is kinetic).

\subsubsection{Distorted Bose-Einstein condensates}

Inspection of Eq.~\eqref{hlinearprofile} reveals that the weak-field approximation~$h\ll v$ is valid only on sufficiently large scales, $r \gg r_* \equiv \frac{M}{4\pi\Lambda^2}$.\footnote{Notice that, since the interaction between the condensate degrees of freedom and the source is a higher-dimensional operator suppressed by $\Lambda$, to neglect higher dimensional operators in an EFT approach one should require $v\ll \Lambda$.} 
On smaller scales, the superfluid profile can be significantly distorted by the source. In general, if the point-like source is replaced by an object of homogeneous density~$\rho$, then its radius~$R$ should be greater than~$r_*$ in order for the superfluid condensate to be only marginally distorted and remain in the linear regime, giving rise to the bound~$\rho R^2 < \Lambda^2$. When the source is dense and/or large enough to violate this bound, one must solve for the full non-linear profile for~$h$. 

In the~$\lambda_4\rightarrow 0$ limit, the nonlinear generalization of Eq.~\eqref{h eom linear} is
\begin{equation}
\Delta h=\frac{v+h}{\Lambda^2}J \,.
\end{equation}
Assuming a homogeneous and constant density source $J=\rho$,  a spherically-symmetric classical solution can be obtained, which matches the decaying configuration at infinity and thereby restores the unperturbed condensate asymptotically. Explicitly, within the source~($r\leq R$), imposing regularity at the origin gives the solution 
\beq
h_{\rm in}(r)=v \left( -1+ \frac{{\rm sech} \, x}{x}  \, \frac{R}{r}\sinh{\frac{r x}{R}} \right)\,; \qquad x\equiv \frac{R\sqrt{\rho}}{\Lambda}\,.
\eeq
Outside the source, the solution is of the long-range form
\begin{equation}
h_{\rm out} (r) =-\frac{v}{\Lambda^2}\frac{M_\text{\tiny eff}}{4\pi r}\,, 
\end{equation}
in terms of the effective mass
\begin{equation}
M_{\rm eff}\equiv \frac{4\pi}{3}\rho R^3 \times \frac{3}{x^2}\left( 1-\frac{\tanh x}{x} \right) \,.
\end{equation}
Note that~$M_{\rm eff}$ depends on the density and size of the source. In particular, for a small/low-density source,~$x\ll 1$, the effective mass experienced by a probe outside the source is very close to the gravitational mass, {\it i.e.},
\beq
\left.M_{\rm eff}\right\vert_{x\ll 1}\simeq \frac{4\pi}{3}\rho R^3 \equiv M \qquad  \longrightarrow \qquad h_{\rm out}(r)=-\frac{v}{\Lambda^2}\frac{M}{4 \pi r } \qquad (\text{low-density/unscreened}) \,.
\label{lowxmeff}
\eeq
For a large/high-density source,~$x\gg 1$, on the other hand, the effective mass gets significantly screened compared to the gravitational mass, 
\beq
\left.M_{\rm eff}\right\vert_{x\gg 1}\simeq 4\pi \Lambda^2 R \qquad \longrightarrow \qquad h_{\rm out}(r)=-v\frac{R}{r} \qquad (\text{high-density/screened}) \,,
\label{highxmeff}
\eeq
with $h$ being independent of the object density and scale $\Lambda$.
In other words,~$h$ acquires an effective mass for high-density sources, which drives symmetry restoration~$\Phi=0$ at their center. This is similar, albeit in a Lorentz-violating background, to the symmetron~\cite{Hinterbichler:2010es} and chameleon~\cite{Khoury:2003aq,Khoury:2003rn} screening mechanisms, where screening is achieved by making the scalar mediator respectively heavy or weakly-coupled inside dense sources. See~\cite{Joyce:2014kja} for a review of screening mechanisms. 

At this point, we can provide various cases for the emergent force between two sources submerged within the condensate, depending on their respective size/density:
\begin{align}
F_h = \begin{cases}
\frac{v^2}{4 \pi\Lambda^4}\frac{M_1 M_2}{r^2}\,, \qquad  \text{unscreened source and probe}\,; \nonumber \\
4\pi v^2\frac{R_1 R_2}{r^2}\,, \qquad  \,\text{screened source and probe}\,; \nonumber \\
\frac{v^2}{\Lambda^2}\frac{R_1M_2}{r^2}\,, \qquad \,\,\,\,\,\,\text{screened source 1 and unscreened probe 2}\,.
\end{cases}
\end{align}
These results agree with the wake force formalism at zero momentum discussed in Ref.~\cite{VanTilburg:2024xib}. This formalism is based on quadratic interactions between the source/target particles and the mediating ones, and can be understood as source particles causing a disturbance (a wake) in the background waves, which subsequently interacts with the target particles. 

Before closing this Section, let us briefly comment on the sign of the interaction between the source~$J$ and~$\Phi$. While the previous derivation assumed a repulsive interaction, setting instead $J=-\rho$ would correspond to an attractive interaction. In this case, in the strongly-distorted regime, a gradient instability emerges. This can be seen by looking at the Goldstone boson dispersion relation inside the source
\beq
\label{attrdisp}
\omega_k^2= \pm \frac{\rho}{4\Lambda^2 m^2}k^2+\frac{k^4}{4m^2} \qquad \text{for}~J = \pm \rho\,.
\eeq
It follows that perturbations with wavenumber~$k<k_*\equiv \sqrt{\rho}/\Lambda$ are unstable for~$J = - \rho$. Since we can talk about such soft modes as long as the source radius $R$ is larger than $k_*^{-1}$, {\it i.e.} if $x > 1$ (see discussion above), then the instability would kick in whenever the condensate distortions become non-linear. In other words, in the attractive case, increasing the density of the source beyond a critical threshold destabilizes the condensate. Conversely, in the repulsive case, the condensate retains long-range coherence and exhibits the mass screening phenomenon discussed earlier.

\subsubsection{Bounds on emergent forces: low-surface brightness galaxies}

The long-range effect discussed above could have intriguing implications for scalar DM models where there is a contact interaction between the ultralight (and weakly self-interacting) DM particles and baryons.

Suppose that~$\Phi$ describes DM, while~$J$ is the trace of the baryonic energy-momentum tensor, $J=-T^\mu_{\,\mu}$.\footnote{The scalar profile created by an extended distribution of baryons was studied in Ref.~\cite{Hees:2018fpg}, where it was shown that their density could get significantly screened or amplified depending on the sign of the coupling between the scalar field and baryons.} If DM forms a Bose-Einstein condensate in the central regions of galaxies, such that it encompasses part or all of the baryonic disk, then baryons within this core will experience an additional force mediated by~$h$, alongside gravity. Since large-scale structure observations imply a lower bound of~$\gtrsim 10^{-22}$~eV on the DM mass, in the limit $\lambda_4\rightarrow 0$, the emergent force can extend up to approximately $\ell_\text{\tiny deg} \simeq \big(G \rho_\text{\tiny nfw} m^2\big)^{-1/4} \simeq 10$~kpc from the galactic center (see also Eq.~\eqref{JeansscaleQ}). Therefore, while the new force should drop significantly in the flat part of the rotation curves of 
high surface brightness (Milky Way-like) galaxies, which extend out to~$\sim 50$~kpc, it could have an important impact on the dynamics of low-surface brightness galaxies, whose rotation curves typically extend out to at most~$10$~kpc. This coincidence is notable, as these galaxies arguably present a more significant challenge to the standard CDM paradigm. 

To quantify the strength of this force, one may consider a fiducial low surface brightness galaxy, IC 2574, with total baryonic mass $M_\text{\tiny LSB}\simeq 10^9\,M_{\odot}$~\cite{Lelli:2016zqa} and baryonic density profile modeled as a uniform-density sphere of radius $R_\text{\tiny LSB}\simeq 8~{\rm kpc}$. As discussed in Ref.~\cite{Berezhiani:2018oxf}, in the linear/unscreened regime, valid for $\Lambda \gtrsim 10^{-4} \mpl$, the force $F_h$ will be comparable or stronger than gravity if the DM mass satisfies the bound
\beq
m\lesssim 3\times 10^{-23}~{\rm eV}\cdot \left(\frac{\Lambda}{10^{-4}\mpl} \right)^{-2}\,; \qquad \text{linear/unscreened regime}\,,
\eeq
which is problematic for large-scale structure observations~\cite{Hui:2016ltb}.
On the other hand, in the non-linear/screened regime, valid for~$\Lambda \lesssim 10^{-4} \mpl$, the bound becomes
\beq
m\lesssim 2\times 10^{-23}~{\rm eV}\cdot \left( \frac{\Lambda}{10^{-4}\mpl} \right)^{-1}\,; \qquad \text{non-linear/screened regime}\,.
\label{mbound}
\eeq
Thus, by lowering~$\Lambda$ this allows DM candidates heavier than~$10^{-22}~{\rm eV}$, compatible with large-scale structure constraints. 
To summarize, in the regime where low surface brightness galaxies are screened, it is possible for the long-range force to be strong enough to compete with the baryonic gravitational force near the edge of the baryonic distribution (despite being suppressed at its center because of the lower DM density), while having DM masses large enough to satisfy constraints from structure formation.

\section{Discussions and conclusions}
\label{sec:conclusions}
In this review, we explored an alternative class of dark matter models based on the existence of sub-eV bosonic particles with repulsive self-interactions, capable of forming a superfluid state on galactic scales. This framework has emerged as a compelling avenue in modern cosmology, offering novel mechanisms to explain a range of astrophysical phenomena that challenge the standard paradigm.

In Sec.~\ref{Sec: Superfluidity}, we reviewed the theoretical underpinnings of superfluidity, presenting both its field-theoretic effective description via Bose-Einstein condensation and its non-relativistic hydrodynamical formulation. We then examined the coupling of the superfluid to gravity, showing how its low-energy excitations---phonons, associated to the spontaneous breaking of the global~$U(1)$ symmetry---mediate distinctive gravitational dynamics.

Building on this foundation, Secs.~\ref{Sec: SDMgalaxies} and~\ref{sec: cosmo} addressed the cosmological evolution of the model and the conditions under which superfluid soliton states can emerge in galaxies. Despite the lack of a concrete UV completion of the proposed scalar field model, it is plausible to assume that sub-eV DM particles could be produced non-thermally via the vacuum misalignment mechanism, which would generically lead to a highly homogeneous and coherent scalar field profile in the early universe. As the Hubble parameter redshifts to the scale of the DM mass, the field begins to behave like a condensate of massive particles, eventually forming halos through virialization. In the early stages of virialization, when self-interactions are negligible, the system gradually loses its initial coherence and behaves as cold, pressureless matter. However, as self-interactions become important, they enable re-equilibration and the restoration of coherence, culminating in the formation of a superfluid phase within galaxies. This state evolves through gravitational interactions, eventually yielding a dense central core surrounded by streams of superfluid debris.

The formation of superfluid cores in galaxies has profound implications for their structure and dynamics, as discussed in Secs.~\ref{sec: pheno}$-$\ref{sec:Forces}. These include modified cluster merger dynamics---leading to a revised interpretation of the Bullet Cluster bound---and the formation of vortex lattices that preserve angular momentum through the phase transition. Additionally, while various dark matter models produce density spikes around supermassive black holes in galactic centers, superfluid dark matter generates a characteristic spike with distinctive observational features that set it apart from other DM candidates.
We also examined the behavior of binary systems embedded in a superfluid background, deriving predictions for dynamical friction and assessing prospects for detection in future gravitational wave experiments. Finally, the interaction of phonons with external sources, such as baryons, gives rise to long-range forces which, in some scenarios, reproduce MOND-like phenomenology and the associated galactic scaling relations.

In a nutshell, the superfluid dark matter scenario sits at the crossroads of other paradigms, such as cold, warm, self-interacting, and fuzzy dark matter models. It attempts to merge several of their key features under a single framework. On cosmological scales, it reproduces the successful predictions of CDM, behaving effectively as a collisionless fluid and preserving the observed large-scale structure. However, on smaller, galactic scales, the coherence is lost outside the central superfluid soliton due to the combination of the Jeans instability and the tidal disruption of would-be superfluid solitons. Within the central soliton, the superfluid phase transition gives rise to collective excitations that could mediate additional long-range forces between baryons, leading to cored halo profiles and scaling relations that could match galactic observations. Because of the finite Jeans scale, this scenario erases substructures below that characteristic length, in analogy with warm and fuzzy dark matter, which suppress structure below the de Broglie wavelength inside galaxies.

At the same time, this suppression of small-scale power introduces both predictive advantages and potential challenges. The erasure of substructures below the Jeans scale can help mitigate long-standing small-scale issues of CDM, such as the overabundance of satellites and overly dense halo centers. Yet, if the Jeans length is too large, it risks conflicting with the observed abundance and internal dynamics of dwarf galaxies, requiring fine-tuning of the boson mass and self-interaction strength accordingly. However, the physical origin and extent of this suppression differ substantially from other models. In fuzzy dark matter, and to some extent in warm dark matter, the de Broglie wavelength of the particles sets a universal cutoff, preventing the formation of substructures below that scale throughout the halo. In contrast, for superfluid dark matter, this suppression applies only within the thermalized superfluid core, where condensation and collective behavior occur. The outer regions of halos, shaped by tidal disruption of would-be superfluid solitons, consist of virialized superfluid streams that never fully re-thermalize due to the absence of a dense, gaseous dark matter phase with a highly occupied phase space. As a result, these non-thermalized envelopes may be able to form smaller substructures than the Jeans length. However, this qualitative discussion needs to be substantiated by numerical simulations. This dual structure, superfluid core plus virialized outer halo, places superfluid dark matter in a unique middle ground between collisionless and wave-like models.

In summary, the study of superfluid dark matter is a rich and rapidly evolving research program, with numerous theoretical, numerical, and observational opportunities for further development. Foremost among these is the necessity of conducting sophisticated numerical simulations to fully understand the formation and evolution of superfluid structures across cosmic time. Such simulations are essential to accurately compare the model to data, and may reveal novel observational signatures of this model on both small and large scales. Equally important is constructing a consistent particle physics framework for these ultralight bosons, encompassing their cosmological origin and thermal history. Lastly, a systematic evaluation of the model’s ability to address the small-scale challenges to the~$\Lambda$CDM model is crucial for assessing its viability. We hope this review provides a useful overview of the current state of the field and offers a springboard for future investigations.

\subsubsection*{Acknowledgments}
We would like to thank Lam Hui for valuable discussions. G.C. is supported by the French
government under the France 2030 investment plan, as part of the Initiative d’Excellence
d’Aix-Marseille Université - A*MIDEX (AMX-19-IET-012), and by the  ``action th\'ematique'' Cosmology-Galaxies (ATCG) of the CNRS/INSU PN Astro. 
V.DL. is supported by funds provided by the Center for Particle Cosmology at the University of Pennsylvania. 
The work of J.K. is supported in part by the DOE (HEP) Award DE-SC0013528.

\appendix
\section{More on the effective field theory of superfluidity}
\label{sec:Scattering}
In this Appendix we provide further details on the connection between phonons and constituent particles of the condensate, showing more quantitative analyses of certain aspects we discussed in the main body of the review.

\subsection{Motion of a probe particle in the superfluid bulk}

We wish to study processes involving the coupling between an external probe and the superfluid phonons, and point out that dissipative channels are kinematically allowed only if the probe moves supersonically, in agreement with Landau's criterion~\cite{Landau:1941vsj}. Furthermore, we prove that higher gradient corrections are important to characterize dissipative processes in which the transferred momentum exceeds the scale~$2 m c_s^2$, which is relevant if the perturber mass~$M$ greatly exceeds the mass~$m$ of the constituents.   
Finally, we demonstrate that, in the same limit, the energy dissipation experienced by the probe due to phonon radiation  is almost equivalent to the energy loss resulting from scattering off an isolated constituent.

Let us focus on the theory of a scalar field with contact two-body interactions, described by the Lagrangian \eqref{eq:LagrPhi4}, with the addition of a coupling to an external real scalar field~$\chi$
\begin{equation}
    \mathcal{L}=-\partial_\mu \Phi^* \partial^\mu \Phi-m^2|\Phi|^2-\frac{\lambda_4}{2}|\Phi|^4-\frac{g_\chi}{2}|\Phi|^2\chi^2\,.
\end{equation}
In this case, the external perturbation~$\chi$ does not carry an expectation value, and it represents an external perturbation that couples directly to the constituents of the superfluid. Following the EFT approach discussed in Sec.~\ref{sec:HigherOrder}, and  integrating out the radial field~$h$, the spectator field~$\chi$ becomes directly coupled to the phonon field according to the interaction Lagrangian
\begin{flalign}
\mathcal{L}=\frac{1}{2}\left(\dot{\pi}^2 + c_s^2\,\pi\Delta\pi-\frac{\left({\Delta}\pi\right)^2}{4 m^2}\right)
-\frac{g_\chi v}{\sqrt{2}}\frac{\dot{\pi}}{\sqrt{-\Delta+4m^2 c_s^2}}\chi^2
+ \dots
\label{eq:Lagr1gx}
\end{flalign}
in terms of the relative velocity $v$ between the probe and the superfluid.
This coupling allows a probe propagating in the superfluid bulk to dissipate energy by radiating phonons. In the cubic Lagrangian, higher order corrections in spatial derivatives enter due to the spatial Laplacian~$\Delta$ that appears inside the square root in the denominator of the cubic vertex. These corrections are the ones that correctly capture the high-$k$ behavior of the theory. However, an interesting property of the~$\chi^2 \dot{\pi}$ vertex is that the amplitude of the process 
\begin{equation}
    \chi(p_i) \rightarrow \chi(p_f)+\pi(k)
\label{eq:processChiPi}
\end{equation}
is independent of the higher order corrections in~$k$, once all particles have been put on-shell. To be more explicit, the amplitude of the process \eqref{eq:processChiPi} reads
\begin{equation}
\mathcal{A}\left(\chi \rightarrow \chi+\pi\right)=\frac{g_\chi v}{2\sqrt{2}m} k\,,
\end{equation}
independently of whether we study the theory at the leading-order in derivatives or if include corrections. 
Therefore, the contribution of these corrections appears only once the amplitude is integrated over the phase space to get the interaction rate. This happens because higher order corrections to the quadratic Lagrangian, through the modified phonon dispersion relation, will affect the delta function responsible for energy conservation in the high-momentum limit~\cite{Berezhiani:2020umi}. 

At this point, we can evaluate the interaction rate of the process \eqref{eq:processChiPi}
by integrating the differential rate
\begin{equation}
    {\rm d}\Gamma_{\chi^2 \pi}=\frac{{\rm d}^3 k \, {\rm d}^3 p_f}{32 \pi^2 \omega_k \omega_{p_f}\omega_{p_{i}}} |\mathcal{A}(\chi\rightarrow \chi+\pi)|^2 \delta^{4}\left(p_i^\mu-k^\mu-p_f^\mu\right)\,,
\end{equation}
as shown in~\cite{Berezhiani:2020umi}. Once we integrate the delta function over the three-momentum using the~$p_f$ integral, we find that energy conservation implies that we have to integrate over phonon momenta satisfying the 
relation
\begin{equation}
\cos\theta=\frac{k}{2M v_i}+\sqrt{\frac{c_s^2}{v_i^2} +\frac{k^2}{4m^2v_i^2}}\,,
\label{eq:thetaint}
\end{equation}
where~$\theta$ is the scattering angle and we have used the non-relativistic momentum~$p_i=M v_i$. If we integrate over~$\cos \theta < 1$, the integration range in~$k$ is restricted by the relation~\eqref{eq:thetaint}, showing that the integral over~$\theta$ vanishes if~$v_i<c_s$. Notice that this relation only involves velocities and not momenta.

By performing the integral, one finds that the integrated rate reads
\begin{equation}
    \Gamma_{\chi^2 \pi}=\frac{g^2_\chi \rho}{128 \pi m^3 M^2 v_i} \left\{\gamma\sqrt{\gamma^2+4 m^2 c_s^2}-4 m^2 c_s^2 \log\left(\frac{2 m c_s}{-\gamma+\sqrt{\gamma^2+4m^2 c_s^2}}\right)\right\}, \quad v_i> c_s\,,
\end{equation}
in terms of the superfluid density~$\rho$, and the momentum~$\gamma$ which solves Eq. \eqref{eq:thetaint} for~$\cos\theta =1$,
\begin{equation}
    \gamma= \frac{2 m M}{M^2-m^2}\left(-m v_i+\sqrt{m^2c_s^2+M^2 \left(v_i^2-c_s^2\right)}\right)\,,
\end{equation}
where we have assumed that~$M>m$ for convenience. Notice the non-linear nature of the solution in~$c_s$. One finds that $\gamma$ vanishes for~$v_i=c_s$, implying that the interaction rate vanishes smoothly in the subsonic regime.
In other words, the interaction rate is non-vanishing only for supersonic perturbers, which is a manifestation of Landau's criterion for superfluidity.

For a much heavier perturber~$M \gg m$, the interaction rate reduces to 
\begin{equation}
    \Gamma_{\chi^2 \pi}=\frac{g^2_\chi \rho}{32 \pi m M^2}v_i \left\{\sqrt{1-\frac{c_s^2}{v_i^2}}-\frac{c_s^2}{v_i^2} \log \left(\frac{v_i}{c_s}-v_i\sqrt{1-\frac{c_s^2}{v_i^2}}\right)\right\}, \quad v_i> c_s\,.
\end{equation}
The overall factor exactly corresponds to the non-relativistic two-body scattering rate of two particles of mass~$M$ and~$m$ in the vacuum, in the limit that the former is infinitely more massive than the latter. The factor inside the brackets approaches unity for an incoming particle with~$c_s\ll v_i$.  The deviation from unity is quadratically sensitive to the ratio~$c_s/v_i$, resulting in a rapid saturation.

Therefore, in the large velocity limit, the interaction rate exactly reduces to the two-body scattering rate of particles, in agreement with the interpretation of hard phonons as single-particle excitations. This happens because the internal self-interactions of the fluid become irrelevant in this limit,\footnote{The limit~$v_i\gg c_s$ coincides with the regime where self-interactions between constituents are set to zero. Therefore, in this kinematic regime, the perturber sees the superfluid as a noninteracting Bose-Einstein condensate.} which can be interpreted as the fact that the external perturbation is transferring enough momentum on a constituent particle to kick it out from the condensed phase, without disturbing neighboring constituents.

\subsection{Phonon self-interactions as the scattering of constituents}

As an alternative scenario for the scattering of probes in superfluids, consider replacing the external perturber~$\chi$ with a constituent particle~$\Phi$. 
While for distinguishable particles the frictionless behavior arises from energy-momentum conservation, we will see that for indistinguishable particles injected into the condensate it results from a backreaction effect.

Let us start by considering a particle~$\Phi$ shot in the superfluid. Following the description of the previous subsection using propagating wavepacket of phonons, the interaction between the propagating constituent and the superfluid bulk is described in terms of phonon self-interactions, with the leading process determined by
\begin{equation}
\pi(k_1)\rightarrow \pi(k_2)+\pi(k_3)\,.
\label{eq:phonon3}
\end{equation}
Processes with more phonons in the initial or final state may contribute as well, but are suppressed by additional powers of~$c^2_s$. To bring an asymptotic constituent inside the condensate and win over repulsive collective self-interactions, the constituent must have a minimum energy given by the non-relativistic chemical potential~$\mu_\text{\tiny NR} = m c_s^2$, {\it i.e.}, it must satisfy the condition for non-relativistic particles
\begin{equation}
    E_\Phi=\frac{1}{2}m v_i^2 \geq m c_s^2 \qquad \Longrightarrow \qquad v_i >  \sqrt{2} c_s\,.
\end{equation}
In other words, the asymptotic constituent should have a velocity larger than the sound speed of the condensate in order to enter the superfluid and be matched to a wavepacket of phonons. This condition is slightly stronger than Eq.~\eqref{eq:LandauCr}.

In the high momentum regime we may use the identification one phonon~$\sim$ one particle, but this interpretation fails when applied to phonons of arbitrary soft momentum. To approach the problem, it is convenient to resort to Bogoliubov's theory of superfluidity~\cite{Bogolyubov:1947zz}. In that context, the phonon field~${\pi}$ is obtained from the fundamental field~$\Phi$ by the Bogoliubov transformation of parameters
\begin{equation}
b_k=\sinh\frac{\theta_k}{2} a_k+\cosh\frac{\theta_k}{2} a^\dagger_k\,,\qquad \text{with}\qquad \tanh \theta_k= \
    \frac{1}{1+\frac{k^2}{8 m^2 c_s^2}}\,,
\label{sec2:Bogoliubov}
\end{equation}
where~$a_k^\dagger$ and~$a_k$ are the creation and annihilation operators of the~$\Phi$ field, 
while~$b_k$ and~$b_k^\dagger$ correspond to the ladder operators of the phonon field.\footnote{The starting point of Bogoliubov's theory is the non-relativistic Hamiltonian of the system
\begin{equation}
\mathcal{H}_\text{\tiny eff}=\mathcal{H}-\mu_\text{\tiny NR}N=\frac{\vec \nabla \Phi \vec \nabla \Phi^\dagger}{2m }+\frac{\lambda_4}{8 m^2}|\Phi|^4-4 m c_s^2|\Phi|^2\,,
\label{BogHam}
\end{equation}
where the finite density of the system is enforced by the Lagrange multiplier $\mu_\text{\tiny NR}$, and the parameters coincide with those of the relativistic theory of Eq.~\eqref{eq:LagrPhi4}. Finite density effects in fluctuations are introduced by expanding~$\Phi$ in terms of the ladder operators~$a_k^\dagger=\sqrt{N}+a^\dagger_{k\neq 0}$, where~$N$ describes the high occupancy of the zero-momentum state. Using the Bogoliubov transformations of Eq.~\eqref{sec2:Bogoliubov}, one can  rotate the massive field~$\Phi$ into the gapless phonons~$\pi$.} Notice that the Bogoliubov transformation~\eqref{sec2:Bogoliubov} becomes trivial in the high-momentum limit, which proves that a single hard phonon becomes equivalent to a propagating constituent. However, this is not true for momenta smaller than the non-relativistic chemical potential $\mu_\text{\tiny NR} = m c_s^2$. 

To understand this, let us introduce the state~$|1_{\rm p}\rangle=a_{\rm p}^\dagger |\mu_\text{\tiny NR}\rangle$, with~$|\mu_\text{\tiny NR}\rangle$ denoting the vacuum state of phonons.
The state $|1_{\rm p}\rangle$ corresponds to a constituent particle with momentum~$p =  m v_i$  propagating in the superfluid bulk. By inverting the Bogoliubov transformation, we find that the number of phonons associated with this state reads
\begin{equation}
\text{N}_\text{\tiny ph}=\langle
1_\text{p}|\hat{N}_\text{\tiny ph}|1_\text{p}\rangle=\cosh^{2}\left(\frac{1}{2}\tanh^{-1}\frac{1}{1+\frac{v_i^2}{8 c_s^2}}\right)\,,
\label{eq:Nph}
\end{equation}
where~$\hat{N}_\text{\tiny ph}=\sum _{k\neq 0} b^\dagger_k b_k$ is the phonon number operator. In the limit of supersonic constituents, a phonon corresponds to a single constituent particle. However, if the constituent particle propagates in the superfluid bulk with a small velocity, then we have~$\lim_{v_i/c_s\to 0} \text{N}_\text{\tiny ph}= c_s/v_i$, which resembles the propagation of a coherent beam of phonons, each with a typical momentum~$m v_i$. 
We can also check the energy of this configuration by writing the energy operator in terms of the phonon energies. By evaluating it on the state~$|1_\text{p}\rangle$ we find, at the leading order in~$v_i/c_s$,
\begin{equation}
    E_{\Phi_1}=\langle 1_\text{p}|\hat{E}_\text{\tiny ph}|1_\text{p}\rangle  \sim m c_s^2+...\qquad \text{with} \qquad \hat{E}_\text{\tiny ph}=\sum _{k\neq 0} \omega_k b^\dagger_k b_k\,.
\end{equation}
This shows that a particle~$\Phi$ can propagate in a superfluid bulk, provided that it has a minimum energy comparable to the chemical potential. Without paying the cost of this additional gap, it is not possible to describe the particle with a wavepacket of phonons.

At this point we can evaluate the interaction rate of the constituent particle with the superfluid bulk, which reads
\begin{equation}
  \Gamma_{\Phi-\text{\tiny bulk}}=\text{N}_\text{\tiny ph}\sum_{i=2}  \Gamma_{\pi}^i+\text{N}_\text{\tiny ph}(\text{N}_\text{\tiny ph}-1)\sum_{i=1}  \Gamma_{\pi^2}^i+...\qquad (v_i>\sqrt{2}c_s)\,.
  \label{eq:Phisfsc}
\end{equation}
Here,~$\Gamma_{\pi^n}^i$ is the interaction rate of a process with~$n$ incoming phonon legs and~$i$ outgoing legs. In other words, the interaction of a single constituent~$\Phi$ with velocity~$v_i$ is recast as the sum of all possible self-interactions between~$\text{N}_\text{\tiny ph}$ phonons of individual momentum~$m v_i$, with~$\text{N}_\text{\tiny ph}$ determined according to Eq.~\eqref{eq:Nph}.  

In the high energy limit, the process \eqref{eq:phonon3} provides the main contribution to Eq.~\eqref{eq:Phisfsc}, whose amplitude reads~\cite{Berezhiani:2020umi}\footnote{To evaluate it, we need the cubic Lagrangian in the phonon field, which reads~\cite{Berezhiani:2020umi}
\begin{equation}
\mathcal{L}_{\pi^3}=\frac{m
}{2\sqrt{\rho}}\left\{\left(\frac{\dot{\pi}}{\sqrt{-\Delta+4m^2 c_s^2}}\right)^2\frac{-\Delta}{\sqrt{-\Delta+4m^2 c_s^2}}\dot{\pi}-\frac{1}{2m^2}\left(\sqrt{-\Delta+4m^2 c_s^2}\,\vec{\nabla} \pi\right)^2
    \frac{\dot{\pi}}{\sqrt{-\Delta+4m^2 c_s^2}}\right\}\,.
\end{equation}}
\begin{flalign}
\mathcal{A}\left(\pi\rightarrow \pi+\pi\right)=\left(\frac{m}{ \sqrt{\rho}}\right)\left\{\frac{k_1}{2m} \frac{k_2}{2m} \frac{k_3}{2m}\right.&(k_1^2+k_2^2+k_3^2) -\frac{k_3}{m}\left({\vec k_1}\cdot {\vec k_2}\right)\frac{\omega_{k_1}}{k_1}\frac{\omega_{k_2}}{k_2}\nonumber\\&\left.-\frac{k_2}{m} \left({\vec k_1}\cdot {\vec k_3}\right)\frac{\omega_{k_1}}{k_1}\frac{\omega_{k_3}}{k_3}+\frac{k_1}{m} \left({\vec k_2}\cdot {\vec k_3}\right)\frac{\omega_{k_2}}{k_2}\frac{\omega_{k_3}}{k_3}\right\}\,,
\label{eq:phonon3ampl}
\end{flalign}
where all phonon legs have been put on-shell. The first term in the amplitude would have been missing in the leading order approximation in derivatives. 

In a Taylor series in the sound speed~$c_s/v_i\ll1$, the~${\cal O}(c_s^0)$ term vanishes by energy and momentum conservation, while the first non-vanishing correction to the amplitude appears at order~$c^2_s$. By integrating the differential interaction rate using the dispersion relation of phonons with zero sound speed, one gets
\begin{equation}
\lim_{v_i\gg c_s}\Gamma_{\Phi-\text{\tiny bulk}}=
\frac{9\lambda_4^2\rho}{128 \pi m^3} v_i\,.
\end{equation}
In other words, we recover the same result of the analysis of the scattering of an external perturbation~$\chi$ on the superfluid with $g_\chi = 3 \lambda_4/2$. A supersonic constituent scatters on the superfluid bulk as if it is scattering on a single constituent at rest, in the vacuum.

\bibliographystyle{JHEP}
\bibliography{Bib.bib}

\end{document}